\documentclass[%
 aip,
 amsmath,amssymb,
 preprint,%
]{revtex4-1}

\usepackage{graphicx}
\usepackage{dcolumn}
\usepackage{bm}
\usepackage{rotating}
\usepackage[utf8]{inputenc}
\usepackage[T1]{fontenc}
\usepackage{mathptmx}
\usepackage{booktabs}
\usepackage{siunitx}
\usepackage{threeparttable}
\usepackage{xcolor}
\usepackage{multirow}
\usepackage{makecell}
\usepackage{hyperref}
\usepackage{setspace}
\usepackage{tikz}
\usepackage{subcaption}
\usepackage{tabularx}
\usepackage[normalem]{ulem}

\DeclareSIUnit\hartree{\text{\ensuremath{E_{\mathrm{h}}}}}

\newcommand{\ket}[1]{\vert #1 \rangle}

\newcommand{\bra}[1]{\langle #1 \vert}

\newcommand{\Tbraket}[3]{\langle #1 \hspace{.10em} \vert \hspace{.10em} #2 \hspace{.10em} \vert \hspace{.10em} #3 \rangle}

\newcommand{\simH}{\bar{H}}

\DeclareSIUnit{\byte}{B}

\newcommand{\eTtwo}{$e^{T}$ 2.0}
\newcommand{\eTone}{$e^{T}$ 1.0}
\newcommand{\eT}{$e^{T}$}

\newcommand{\sarai}[1]{\textcolor{black}{#1}}

\newcommand{\mrm}{\mathrm}

\newcolumntype{P}[1]{>{\centering\arraybackslash}p{#1}}
\begin{document}

\title{\eTtwo: {An} efficient open-source {molecular electronic structure} program}

\author{Sarai Dery Folkestad}
\email{sarai.d.folkestad@ntnu.no}
\thanks{Equal contributions}
\affiliation{Department of Chemistry, Norwegian University of Science and Technology, 7491 Trondheim, Norway}
\author{Eirik F.~Kj{\o}nstad}
\email{{eirik.f.kjonstad@gmail.com}}
\thanks{Equal contributions}
\affiliation{Department of Chemistry, Norwegian University of Science and Technology, 7491 Trondheim, Norway}
\author{Alexander C. Paul}
\thanks{Equal contributions}
\affiliation{Department of Chemistry, Norwegian University of Science and Technology, 7491 Trondheim, Norway}
\author{Rolf H.~Myhre }
\affiliation{Department of Chemistry, Norwegian University of Science and Technology, 7491 Trondheim, Norway}
\affiliation{Development Centre for Weather Forecasting, Norwegian Meteorological Institute, Oslo, Norway}
\author{{Riccardo Alessandro} }
\affiliation{Dipartimento di Chimica, Biologia e Biotecnologie, Universit{a`} Degli Studi di Perugia, 06123 Perugia, Italy;}
\author{Sara Angelico }
\affiliation{Department of Chemistry, Norwegian University of Science and Technology, 7491 Trondheim, Norway}
\author{Alice Balbi}
\affiliation{Department of Chemistry, Norwegian University of Science and Technology, 7491 Trondheim, Norway}
\author{Alberto Barlini}
\affiliation{Scuola Normale Superiore, Piazza dei Cavalieri, 7, 56126 Pisa PI, Italy}
\author{Andrea Bianchi}
\affiliation{Scuola Normale Superiore, Piazza dei Cavalieri, 7, 56126 Pisa PI, Italy}
\author{{Chiara Cappelli}}
\affiliation{Scuola Normale Superiore, Piazza dei Cavalieri, 7, 56126 Pisa PI, Italy}
\author{Matteo Castagnola }
\affiliation{Department of Chemistry, Norwegian University of Science and Technology, 7491 Trondheim, Norway}
\author{Sonia Coriani }
\affiliation{DTU Chemistry---Department of Chemistry, Technical University of Denmark, DK-2800 Kongens Lyngby, Denmark}
\author{Yassir El Moutaoukal}
\affiliation{Department of Chemistry, Norwegian University of Science and Technology, 7491 Trondheim, Norway}
\author{Tommaso Giovannini }
\affiliation{{Department of Physics, University of Rome Tor Vergata and INFN, Via della Ricerca Scientifica 1, 00133 Rome, Italy}}
\author{Linda Goletto }
\affiliation{Scuola Normale Superiore, Piazza dei Cavalieri, 7, 56126 Pisa PI, Italy}
\author{Tor S. Haugland}
\affiliation{Department of Chemistry, Norwegian University of Science and Technology, 7491 Trondheim, Norway}
\author{{Daniel Hollas} }
\affiliation{{Centre for Computational Chemistry, School of Chemistry, University of Bristol, Cantocks Close, Bristol BS8 1TS, United Kingdom}}
\author{Ida-Marie H{\o}yvik }
\affiliation{Department of Chemistry, Norwegian University of Science and Technology, 7491 Trondheim, Norway}
\author{Marcus T. Lexander}
\affiliation{Department of Chemistry, Norwegian University of Science and Technology, 7491 Trondheim, Norway}
\author{Doroteja Lipovec }
\affiliation{Department of Chemistry, Norwegian University of Science and Technology, 7491 Trondheim, Norway}
\author{Gioia Marrazzini }
\affiliation{Scuola Normale Superiore, Piazza dei Cavalieri, 7, 56126 Pisa PI, Italy}
\author{Torsha Moitra }
\affiliation{DTU Chemistry---Department of Chemistry, Technical University of Denmark, DK-2800 Kongens Lyngby, Denmark}
\affiliation{Department of Physical and Theoretical Chemistry, Faculty of Natural Sciences, Comenius University, SK-84215 Bratislava, Slovakia}
\author{Ylva Os}
\affiliation{Department of Chemistry, The Arctic University of Norway, 9037 Troms{\o}, Norway}
\author{Regina Paul}
\affiliation{Department of Chemistry, Norwegian University of Science and Technology, 7491 Trondheim, Norway}
\author{Jacob Pedersen }
\affiliation{DTU Chemistry---Department of Chemistry, Technical University of Denmark, DK-2800 Kongens Lyngby, Denmark}
\affiliation{Department of Chemistry, Norwegian University of Science and Technology, 7491 Trondheim, Norway}
\author{Matteo Rinaldi}
\affiliation{Scuola Normale Superiore, Piazza dei Cavalieri, 7, 56126 Pisa PI, Italy}
\author{Rosario R. Riso }
\affiliation{Department of Chemistry, Norwegian University of Science and Technology, 7491 Trondheim, Norway}
\author{Sander Roet}
\affiliation{Department of Chemistry, Norwegian University of Science and Technology, 7491 Trondheim, Norway}
\affiliation{Structural Biochemistry, Bijvoet Centre for Biomolecular Research, Utrecht University, 3584 CG Utrecht, The Netherlands}
\author{Enrico Ronca }
\affiliation{Dipartimento di Chimica, Biologia e Biotecnologie, Universit{a`} Degli Studi di Perugia, 06123 Perugia, Italy;}
\author{Federico Rossi }
\affiliation{Department of Chemistry, Norwegian University of Science and Technology, 7491 Trondheim, Norway}
\author{Bendik S. Sannes }
\affiliation{Department of Chemistry, Norwegian University of Science and Technology, 7491 Trondheim, Norway}
\author{Anna Kristina Schnack-Petersen }
\affiliation{DTU Chemistry---Department of Chemistry, Technical University of Denmark, DK-2800 Kongens Lyngby, Denmark}
\author{Andreas S. Skeidsvoll}
\affiliation{Department of Chemistry, Norwegian University of Science and Technology, 7491 Trondheim, Norway}
\affiliation{\sarai{Department of Physics and Technology, University of Bergen, Norway}}
\author{{Leo Stoll}}
\affiliation{Department of Chemistry, Norwegian University of Science and Technology, 7491 Trondheim, Norway}
\author{Guillaume Thiam }
\affiliation{Dipartimento di Chimica, Biologia e Biotecnologie, Universit{a`} Degli Studi di Perugia, 06123 Perugia, Italy;}
\author{Jan Haakon M. Trabski}
\affiliation{Department of Chemistry, Norwegian University of Science and Technology, 7491 Trondheim, Norway}
\author{Henrik Koch}
\email{henrik.koch@ntnu.no}
\affiliation{Department of Chemistry, Norwegian University of Science and Technology, 7491 Trondheim, Norway}

\date{\today}

\begin{abstract}
    The \eT~program is an open-source electronic structure program with 
    {emphasis}
    {on performance and modularity.}
    As its name suggests, 
    the program features extensive coupled cluster 
    capabilities,
    {performing well compared to} other electronic structure {programs,} {and,} in some cases{,} outperforming commercial alternatives.
    {However, \eT~is more than a coupled cluster program; 
    other {models based on} wave function {theory} (such as full {and reduced space} configuration interaction and a variety of self-consistent field models) and density functional theory {are} supported.}    
    The second major {release} of the program,
    \eTtwo, 
    {has}
specialized functionality for strong light-matter coupling conditions.
    {I}n addition{, 
    {it}
    includes} a wide range of optimizations and algorithmic improvements,
    {as well as}
    {new} capabilities 
    {for}
    exploring potential energy 
    {surfaces}
    and 
    {for}
    modeling 
    experiments {in the UV {and} X-ray regimes.}
    {Molecular} gradients {are now available} at the coupled cluster level{,} and 
    {high-accuracy} 
    {spectroscopic simulations are}
    {available at} {reduced} {computational} {cost} {within} {the} {multilevel coupled cluster and multiscale frameworks. } 
    {We} present the
    {modifications} 
    to the program since its first {major} release, \eTone, highlighting some 
    notable new features
    {and demonstrating} 
    {the} 
    performance 
    {of the new version}
    relative to 
    {the first release}
    and {to}
     other established {electronic} structure programs.
\end{abstract}

\maketitle

\section{Introduction}
    Research in electronic structure theory relies on efficient, reliable, and maintainable software.
    Over the past six decades, a large number of
    {programs}
    have emerged to answer this need, some focusing on specific
    {electronic structure methods and others adopting} a broad and ambitious scope, see for instance Refs.~\citenum{g16,cfour2020,pyscf2020,dalton2014,dirac2020,serenity2023,gamess2023,turbomole2020,turbomole2023,terachem2020,octopus,orca2025,qchem2021,nwchem2021,psi42020,molpro2020,mrcc2020,molcas2020,OpenMolcas2023, veloxchem2020,pybest2024}.
    {A massive amount of work has {been} invested in the development of these programs.}
    In many cases, what 
    {may seem} like simple calculations {(for example, evaluating some molecular property like the dipole moment)} {often {depends}} on
    thousands of lines of code.
    {This can make extending the software a daunting and time-consuming task, emphasizing the need for adherence to best practices for software development.}

    {To remain useful, software} must {also} continually adapt to new computer paradigms to 
    {follow}
    the evolution of computer hardware and architecture. Perhaps the most important shift in recent years has been from serial to parallel programming, where high performance {gains are made by} exploiting the large number of central processing unit (CPU) cores in modern computers, and more recently towards {accelerator-based}, heterogeneous architectures. 
    This shift is 
    {reshaping}
    high-performance computing (HPC) facilities around the world, driven to a large extent by developments in artificial intelligence and machine learning.
    Adapting (or rewriting) software to accommodate new hardware paradigms, such as a transition from single-node CPU algorithms towards multi-node CPU (with shared or distributed memory) or graphical processing unit (GPU) algorithms, is central to the productivity of the field of quantum chemistry.\citep{di2023perspective}

    The size of the electronic structure programs, combined with developments in hardware, highlights the need for
    {code bases} {that are easily adaptable.} 
    {This fact} has led some to argue that electronic structure software requires commercialization to meet the need of  software expertise and maintenance.\citep{Krylov2015} Others have emphasized the importance of open-source software\citep{lehtola2022free} {and the broad adoption of software best practices to minimize the technical debt (and thus the time taken away from research) that results from poorly written and poorly maintained code bases}.\citep{di2023perspective}

    {The present work is in line with {the} latter perspective.}
    The \eT\;program is an open-source electronic structure program {under the GNU General Public License v3.0 (GPLv3)} with a developer community that puts emphasis on code-quality, rigorous testing, and maintainability, reflecting our belief that, under these conditions, large code bases can be developed and successfully managed by the scientific community. While initially focusing on coupled cluster methods, \eT\;is today a full-fledged electronic structure program with a wide variety of features.

    The \eT~program is an object-oriented code, primarily written in Fortran 2018. The code is  especially optimized for modern CPU nodes with fast (shared) memory and input/output (I/O), parallelized through OpenMP and extensive use of the BLAS and LAPACK libraries.
    The code is hosted on GitLab (\url{www.gitlab.com/eT-program/eT}) and publicly developed, with both {code} review and a development version of the code available to the public, in addition to the latest stable release. The user manual, along with news about the community, is available on the website (\url{www.etprogram.org}). The \eT\;community is therefore fully open to collaboration and committed to 
    transparency.

    In this paper, we present the second major release of the program, \eTtwo, and highlight the improved performance of the program, compared to its first release, as well as {showcase} several significant features.

\section{Program structure and features}
    The \eTtwo\; program
    offers a wide range of features for electronic structure calculations. Its structure reflects the subtasks involved in such calculations, with dedicated modules (more precisely, classes {in object-oriented terminology}) with distinct and well-defined responsibilities. High-level examples include the classes referred to as \emph{engines} (which drive calculations by invoking a sequence of \emph{tasks}), \emph{solvers} (which {solve} model equations), and \emph{wave functions} (which {implement} the {particular} model equations). This modular structure makes it easy to extend the program and reuse existing code when designing new features. 
    It also facilitates interfacing with other codes, {which may invoke high-level modules to obtain the desired electronic structure information (see, e.g., Ref.~\citenum{kjonstad2024photoinduced}).}

    {Compared with its {first release},\cite{eTprog} the new version of \eT~includes various algorithmic improvements{, in addition to } 
    extensions in terms of which areas of chemistry can be explored.}
    {In particular, it has been shown that molecular} properties can be manipulated by placing molecules between closely spaced mirrors, resulting in a strong coupling of electrons and photons. These situations are described through cavity quantum electrodynamics (QED) {methods, which incorporate the coupling of the electronic structure to one\sarai{,} or 
    { several} photonic modes}. The \eT program now provides energies at both mean-field  and correlated levels of theory {for investigating such strong-coupling chemistry}.  \cite{,haugland2020coupled,haugland2021intermolecular,riso2022molecular,el2024toward,elmoutakal2025strong,angelico2023coupled,castagnola2024polaritonic} {The QED functionality has also been extended to use plasmonic modes,\cite{Fregoni2021,Romanelli2023} that are calculated separately for the nanostructure and imported into the \eT~program. This enables the study of strong coupling between localized surface plasmons and molecules.}

    {The new version also enables a broader set of applications in the areas of spectroscopy and photochemistry. For example,}
    ground and excited state equilibrium geometries, as well as vibrational frequencies, {calculated} using analytical nuclear gradients, can now be efficiently determined at the Hartree--Fock (HF) and coupled cluster singles and doubles (CCSD) (ground and excited states)\cite{schnack2022efficient} levels of theory. {Furthermore, molecular gradients at the QED-HF level\cite{lexander2025exploring, barlini2025} are also available.
    Hence, the program can be used to explore new regions of the potential energy landscapes, as well as for ground and excited state Born-Oppenheimer molecular dynamics calculations.\cite{lexander2024analytical,schnack2022efficient} %

    The new version enables fast simulation of absorption spectra with coupled cluster theory. These features have also been extended in new directions, such as the determination of triplet states and improved support for multilevel coupled cluster calculations.  
    In \eTtwo, excitation energies {and transition moments} can also be computed with time dependent Hartree--Fock (TDHF) theory. 
    The time-dependent coupled cluster (TDCC)\cite{skeidsvoll2020time} method has been extended with the time-dependent equation-of-motion coupled cluster (TD-EOM-CC) technique,\cite{skeidsvoll2020time,skeidsvoll2022} further extending the program's capabilities of simulating electron dynamics with explicit pulses. 

    {\eTtwo~also has extended} support for
    other established electronic structure models.
    The full configuration interaction {(FCI)} model, both {in {a}} full or {in an} active {orbital space,} {has} been added to the {program.} 
    Within coupled cluster theory, the CCSDT model is {now} available for both ground and excited state calculations. At the Hartree--Fock level, we have expanded \eT~to include {the restricted open-shell model}, implemented through the constrained unrestricted Hartree--Fock {approach}.\cite{tsuchimochi2010communication} Kohn-Sham density functional theory has 
    been added for ground state calculations using the local density approximation (LDA), generalized gradient approximation (GGA), and hybrid functionals.\cite{marrazzini2021multilevel,giovannini2023integrated}  The DFT grid is constructed using the widely employed Lebedev grid,\citep{lebedev1999quadrature} with the radial quadrature proposed in Ref.~\citenum{krack1998adaptive}.

    {The \eT~program now has}
    {broadened} 
    support for molecular properties. 
    {Dipole and quadrupole moments are available with the Hartree--Fock models, the coupled cluster models (except CCSDT), and with FCI and CASCI.}
    Static and frequency dependent (dipole) polarizabilities can be calculated at the Hartree--Fock and coupled cluster levels of theory, and nuclear magnetic resonance (NMR) shielding parameters are available with Hartree--Fock. 
    {Dipole oscillator strengths in various gauges (length, velocity and mixed) as well as 
    rotatory strengths of electronic circular dichroism, in both length and modified velocity gauges,}
    are available with the {CCSD} model{, and with restricted closed-shell Hartree--Fock and QED-Hartree--Fock.}

    Some of the program features rely 
    {on} {external}
    open-source software packages and libraries.
    Integrals are provided through one of three integral libraries; Libcint\cite{libcint} is the default integral library, but \eT can also be used with integrals from Libint\cite{libint} or PhasedInt\cite{bianchi2025} through a unified interface that allows for seamless exchange of integral {provider}. Optionally, polarizable continuum environments can be included through the PCMSolver library,\cite{di2019pcmsolver} and {the LibXC library\cite{libxc,marques2012libxc}  enables DFT calculations}. However, to use most of the features in \eT, only one integral library must be installed ({i.e., } Libcint, Libint, or PhasedInt).

    {On the algorithmic front, the}
    \eT program offers a range of solvers that ensure efficient and robust convergence {for} the variety of equations that must be solved. 
    Since the release of the first major version, the program has significantly improved in this regard.  The program now has support for a trust-region algorithm\cite{fletcher2013practical,Jorgensen:1983aa,hoyvik2012trust} that is currently used for quadratically convergent self-consistent field (SCF) optimizations,  occupied and virtual orbital localization,\cite{ER2022implementation} and for SC-QED-HF optimization.\cite{el2024toward} However, the trust-region solver can be used for any optimization problem, as long as the functional, its gradient, and the linear transformation by the (approximate) Hessian is implemented.  Similarly, the \eT~program also provides Davidson-like reduced space solvers\cite{DAVIDSON197587,hirao1982generalization} that can be {used} for any eigenvalue or linear equation problem, as long as the linear transformation {is} provided. {These solvers are flexible and only require setting up a transformation object and passing it along to the solver, facilitating future extensions.}

    Particular improvements have been made for the calculation of excited states with the perturbative coupled cluster models (CC2\cite{christiansen1995second} and CC3\cite{koch1997cc3,paul2020new}). For these models, the eigenvalue problem may be recast into a non-linear eigenvalue problem in a reduced parameter space. In \eTone, these equations were solved using a DIIS algorithm\cite{SCUSERIA1986236,hamilton1986direct} that suffers from the occasional appearance of duplicate roots and does not ensure convergence to the lowest-energy solutions. In \eTtwo, a non-linear Davidson solver that avoids both these problems is available. Furthermore, for CC3, the multimodel solvers by Kjønstad \textit{et al.}\cite{kjonstad2020accelerated} can be used for both the ground and excited state equations. 
    {These solvers {improve} convergence, resulting in significant computational savings in CC3 calculations.}
    {Finally, the code also features a {band}-Lanczos solver\cite{skeidsvoll2022} and a damped response solver for linear response.~\cite{schnack2023new}} In general, the solvers in \eT are built using general tools that {make} it easy to implement new solvers.

    The robustness of the program is supported by a broad and comprehensive suite of tests, and we have made several improvements in this direction since the original release. Our tests are optimized for speed, single out selected features for testing, and target near-complete code coverage (monitored using {\href{https://coveralls.io/}{Coveralls}}). 
    Our code repository is connected to a CI/CD setup that automatically checks the validity of the code upon every commit to the repository, both in the development and the release branch. The test set includes a large number of integration tests (using the Runtest library\cite{runtest}) and some unit tests, \citep{rilee2014towards} {{checking} a range of possible configurations of the program (with different compilers, dependencies, compiler flags, etc.)}. The test suite provides confidence in the correctness of the program and aids in refactoring and extending the program with the knowledge that new bugs are likely not being introduced in existing features.


\section{Highlighted program capabilities}
\begin{figure}[b!]
    \centering
    \includegraphics[width=\linewidth]{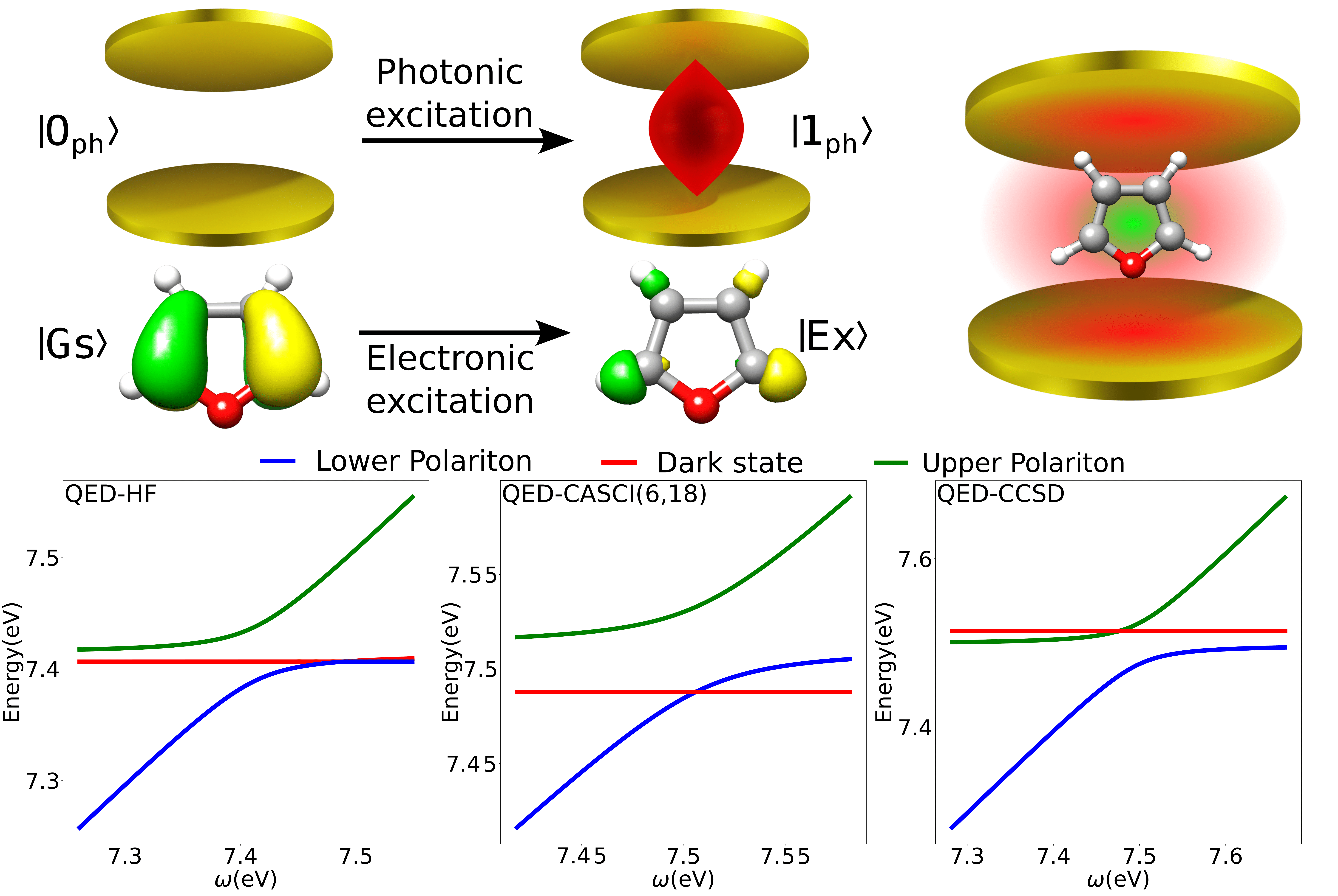}
    \caption{Dispersion of the excitation energies {of furan using the aug-cc-pVDZ basis} in an optical cavity, as a function of the cavity frequency, $\omega$. On the top panel of the image, we display the orbital transition mainly featured in the excitation. On the bottom panel of the figure, we show the formation of the lower and upper polaritons.}
    \label{fig:Rabi}
\end{figure}
\begin{figure}
    \centering
    \includegraphics[width=0.7\linewidth]{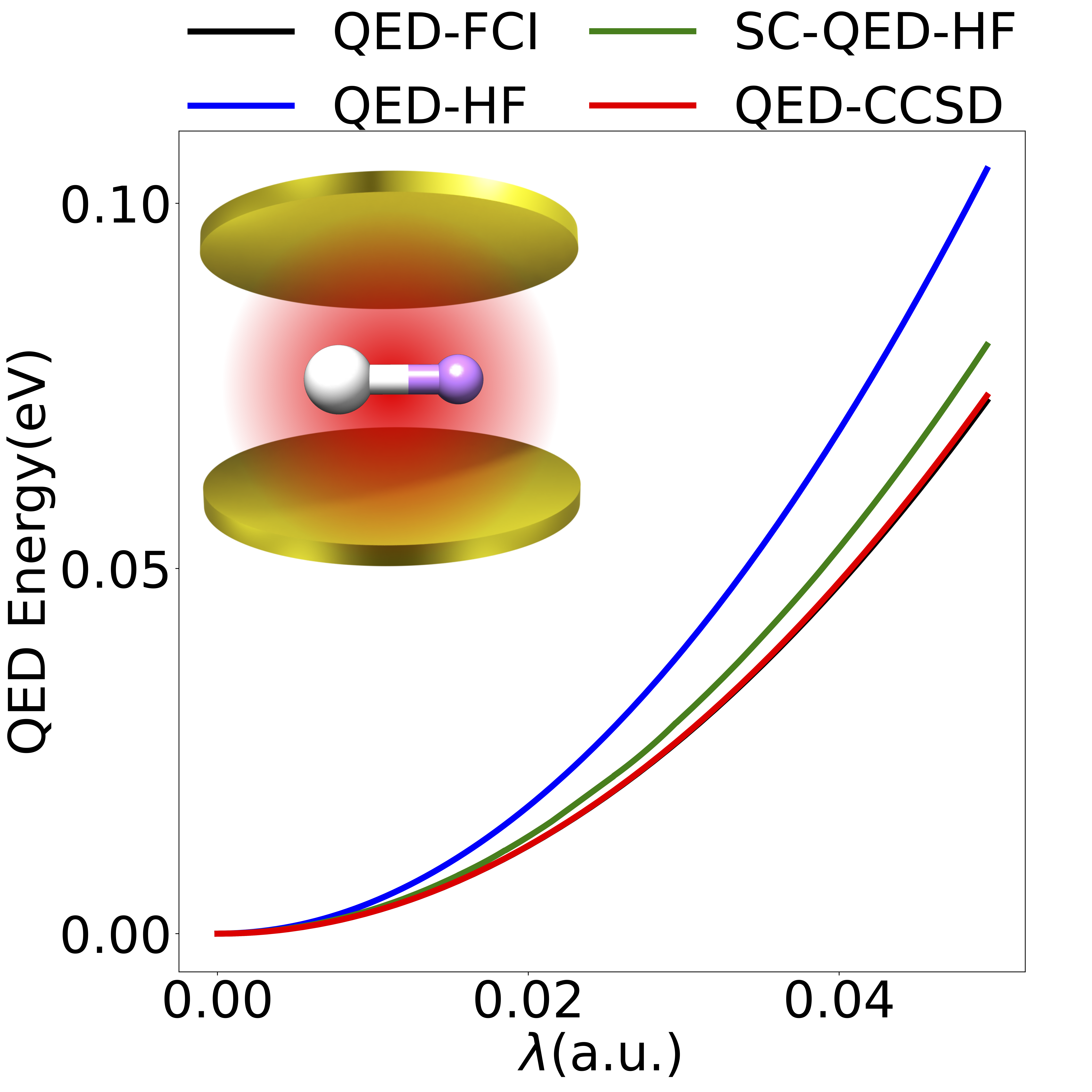}
    \caption{Dispersion of the ground state energy {of LiH using the aug-cc-pVDZ basis} with respect to the light-matter coupling, $\lambda$. {While} QED-CCSD clearly outperforms QED-HF and SC-QED-HF, all the implemented methods capture the qualitative behavior of the function.}
    \label{fig:coup}
\end{figure}
\begin{figure}
    \centering
    \includegraphics[width=0.7\linewidth]{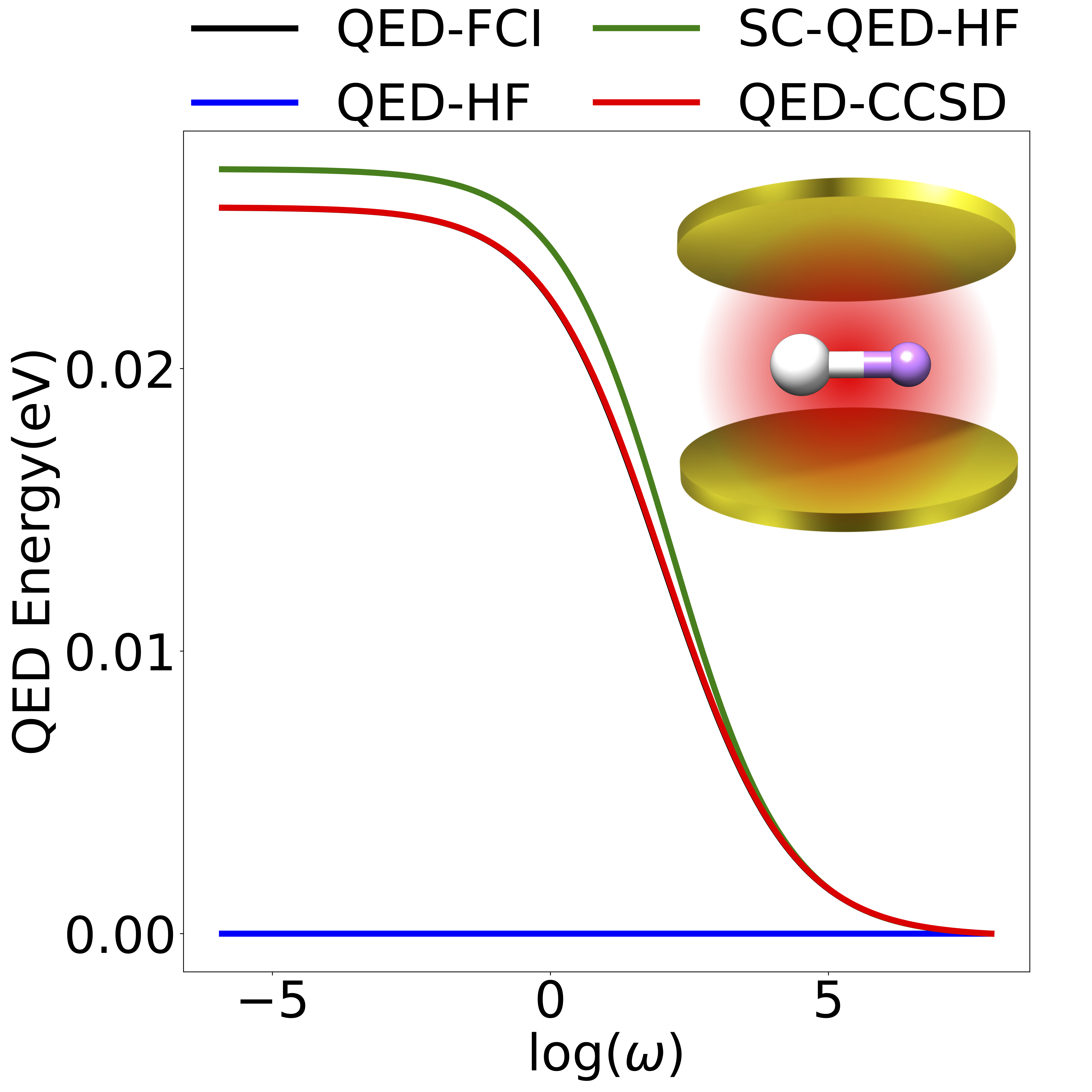}
    \caption{Dispersion of the ground state energy {of LiH using the aug-cc-pVDZ basis} with respect to the cavity frequency{, $\omega$}. The QED-HF energy is completely independent {of} $\omega$ while SC-QED-HF, despite being a mean-field approach, reproduces the QED-FCI behavior {qualitatively}. Excellent agreement is observed between QED-CCSD and QED-FCI results.}
    \label{fig:omega}
\end{figure}
\subsection{Quantum-electrodynamical electronic structure theory}
Strong coupling between electromagnetic fields and molecular systems gives rise to hybrid light-matter states known as polaritons.\cite{bloch2022strongly,fregoni2022theoretical,ruggenthaler2023understanding,feist2018polaritonic} Due to photon-molecule entanglement, polaritons exhibit unique properties that cannot be explained by their uncoupled counterparts.\cite{schafer2024machine,george2023polaritonic,ebbesen2023introduction} Crucially, they can be engineered by tuning the photonic component, enabling non-invasive modulation of matter properties.\cite{campos2019resonant,wellnitz2022disorder,nagarajan2021chemistry}
Optical cavities, composed of mirrors that confine the electromagnetic field within a small quantization volume, provide an optimal platform for achieving strong light-matter coupling.\cite{garcia2021manipulating,castagnola2024strong,sandik2024cavity} 
The interaction strength increases as the quantization volume decreases, driving experimental efforts toward extreme field confinement. Plasmonic nanocavities currently achieve the strongest coupling, with quantization volumes below {a single} $\si{\nano\meter\cubed}$.\cite{chikkaraddy2016single,santhosh2016vacuum}

By engineering cavity mirrors, key field properties--such as frequency, shape, and polarization--can be controlled.\cite{wu2023bottom} Polaritonic effects have been experimentally {observed} in absorption spectra, photochemical reaction rates, and conductivity.\cite{sandik2024cavity,thomas2019tilting,hutchison2012modifying,thomas2016ground,ahn2023modification} Notably, it has been shown that {photonic} coupling to molecular vibrations can catalyze, slow down, or induce selectivity in chemical reactions.\cite{thomas2019tilting,hutchison2012modifying,thomas2016ground} However, reproducing experimental results remains challenging,\cite{imperatore2021reproducibility} highlighting the need for theoretical modeling to understand the complex light-matter interplay. Developing \emph{ab initio} electron-photon methods 
{an important step}
toward understanding cavity effects. Since photons play a fundamental role, they must be treated as quantum particles following {QED} principles. 
{The} first \emph{ab initio} methodology {that adopted standard quantum chemistry methods} for electron-photon systems was quantum electrodynamical density functional theory (QEDFT).\cite{ruggenthaler2014quantum,flick2015kohn,lu2024electron} While computationally efficient, QEDFT inherits functional limitations in describing electron-electron and electron-photon correlations. Subsequently, {we have developed and implemented} QED-Hartree-Fock (QED-HF),\cite{haugland2020coupled,haugland2021intermolecular} 
strong {coupling} QED-HF (SC-QED-HF),\cite{riso2022molecular,el2024toward,elmoutakal2025strong} QED-coupled cluster (QED-CC),\cite{haugland2020coupled,haugland2021intermolecular} QED-FCI and complete active space configuration interaction (QED-FCI/QED-CAS{CI}){, as well as the cavity Born-Oppenheimer (CBO) approximation for all electronic structure methods,\cite{angelico2023coupled} } in the \eT~program. These methods now enable the computation of both ground and excited state properties of molecules within optical cavities.
In Figure \ref{fig:Rabi}, for example, we display how the excitation energies of a furan molecule depend on the cavity frequency $\omega$. The excited state results are computed using QED-HF, QED-CAS and QED-CCSD. For all three cases, we observe that when the photonic energy is resonant with a molecular excitation, an avoided crossing is observed. The accuracy of the excitation energies increases going from HF to CCSD, but the qualitative picture is {the same, independent of} the chosen approach.

In Figures \ref{fig:coup} and \ref{fig:omega} we show the {cavity frequency ($\omega$) and light-matter coupling ($\lambda$)} dispersions for the ground state energy of lithium hydrate as computed using {mean-field} methods (QED-HF, SC-QED-HF) and correlated methods (QED-CCSD, QED-FCI). While all approaches capture the qualitative behavior of the $\lambda$ energy dispersion, the same is not true for the $\omega$ dispersion. In particular, we notice that the QED-HF ground state energy is independent {of} the field frequency. This problem is solved using the SC-QED-HF mean-field approach, which also provides a fully consistent set of molecular orbitals.\cite{riso2022molecular,el2024toward}

\subsection{Real-time time-dependent equation-of-motion coupled cluster theory}
parametrizes {time dependence} {with} the linear coefficient{s} of {the} EOM-CC {states}. This allows {one} to compute the {electronic} time evolution of a system, its {observables,} and the populations of excited states. 
The TD-EOM-CC approach has application in, e.g., the simulations of ultra-fast phenomena {such as stimulated X-ray Raman scattering.}\cite{skeidsvoll2022,balbi2023coupled} 
{The TD-EOM-CC} method can provide greater numerical stability\cite{Skeidsvoll2023Rabi} {compared to} {the} TDCC method which was already present in \eTone. In addition to the already present numerical integrators, {adaptive time-stepping capabilities and} Dormand-Prince {integrators} (with orders 5(4) and 8(5,3))\cite{DORMAND198019,Hairer1993} {have} been added {to the new version of the program}.~\cite{skeidsvoll2022}
\begin{figure}[h!]
    \centering
    \includegraphics[width=0.9\linewidth]{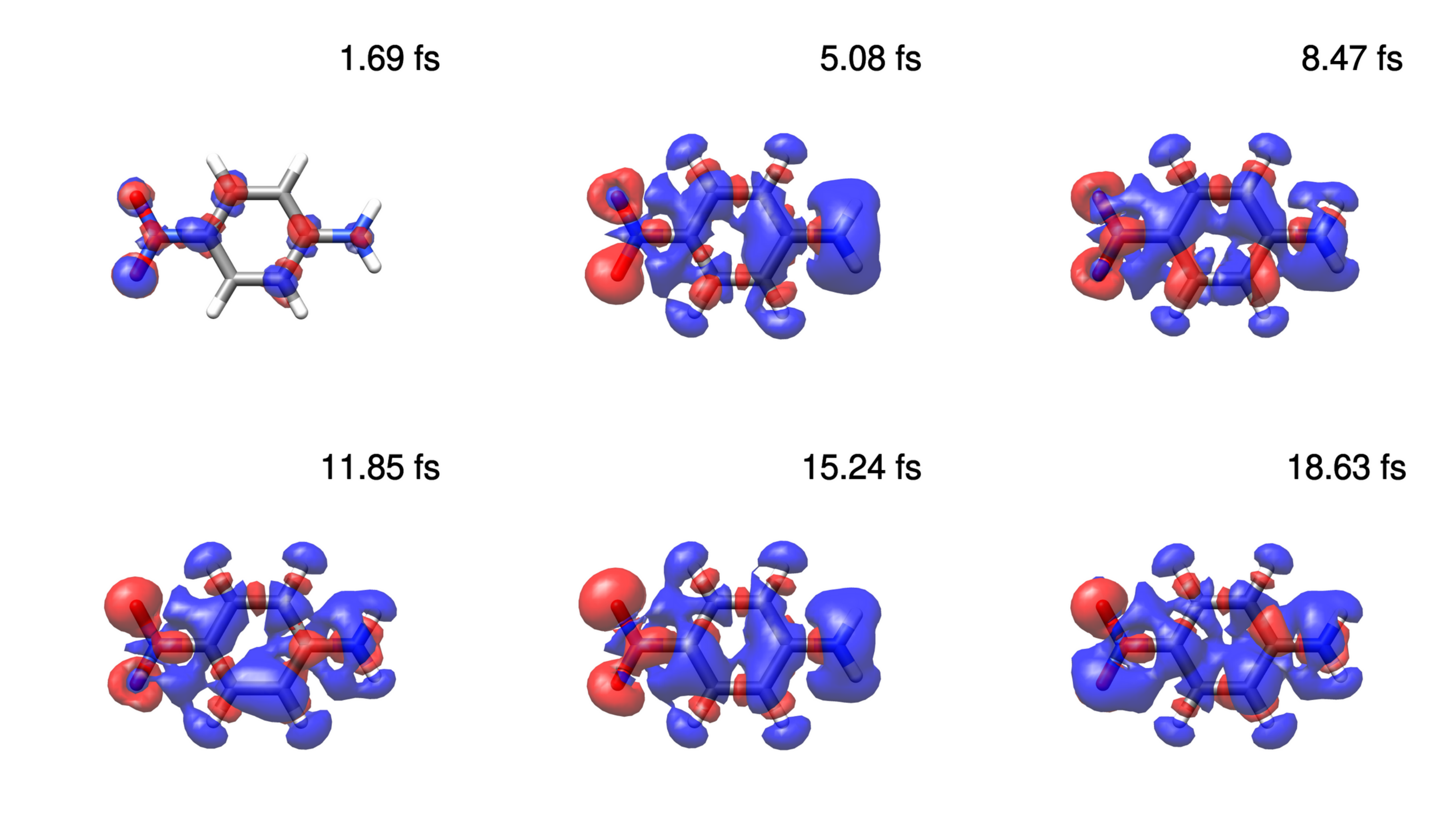}
    \caption{Time evolution of the isodensity surfaces after subtracting the ground state electron density, plotted using UCSF Chimera.\citep{Pettersen2004Chimera} The calculation was performed with {TD-EOM-}CCSD/aug-cc-pVDZ. The system is a molecule of \textit{p}-nitroaniline in {the} presence of an external electric field. The red surface depicts the isosurface at 0.001 contour level, while the blue one refers to the -0.001 contour level. {T}he interaction with the external electric field induces a migration of electrons that leads to an increase of negative charge on the nitro group while the amino group becomes more positively charged with respect to the ground state.}
    \label{fig:tdeomcc_pna}
\end{figure}

{As an illustration of these capabilities, we}
consider the \textit{p}-nitroaniline molecule in {the} presence of an external electric field{. Figure \ref{fig:tdeomcc_pna} shows the density difference relative to the ground state electron density at different time steps.} The {molecular} geometry is provided in the {{supporting} information} of {Ref.~\citenum{haugland2020coupled}}, {and it} was optimized at the DFT/B3LYP level of theory using a 6-31+G* basis set.\cite{Hariharan1973basis} The time{-dependent state} {is} propagated for $800$ a.u. and the chosen electric field has a Gaussian envelope with a width of $5$ a.u. 
The chosen carrier angular frequency is
$0.166152$
a.u.{, which} corresponds to the excitation energy between the ground state and the second excited state. The peak strength {was} set to $1$ a.u., {the envelope has} a central time {set to} $80$ a.u., and {the} polarization {was} oriented in the $[1, 1, 1]$ {(non-normalized)} direction.

\subsection{Ground and excited state molecular geometry optimization, harmonic frequencies, and normal modes}

Identifying stationary structures is useful in various contexts, allowing the user to inspect vibrational modes and frequencies and potentially aid in the interpretation of excited state dynamics, as well as in isolating spectral signatures {by extraction of spectra at stationary points}, e.g.~{for interpreting} time-resolved X-ray spectra. In \eTtwo, we have implemented ground and excited state singlet gradients at the EOM-CCSD level of theory (both for valence and core excited states),~\cite{schnack2022efficient} allowing for easy determination of stationary structures as well as harmonic frequencies, {associated normal} modes, and Wigner samples at {$\SI{0}{\kelvin}$}. {The new version of the program now also features a more robust Broyden-Fletcher-Goldfarb-Shanno (BFGS) optimizer that uses redundant internal coordinates\citep{bakken2002efficient} and rotational coordinates{.\citep{wang2016geometry} {In this way,} optimization of single-molecule systems but also the orientation of fragments as well as cavity-induced orientation effects (in QED-HF){ is made possible}.} In Figure \ref{fig:adenine-s0s1}, we provide an illustration of {these new capabilities} for the ground and excited state minima in 9H-adenine {determined with the EOM-CCSD model}. {Associated harmonic frequencies are given in Tables \ref{tab:adenine-freq} and \ref{tab:adenine-freq-s1}.}

\begin{figure}[h!]
    \centering
    \includegraphics[width=\linewidth]{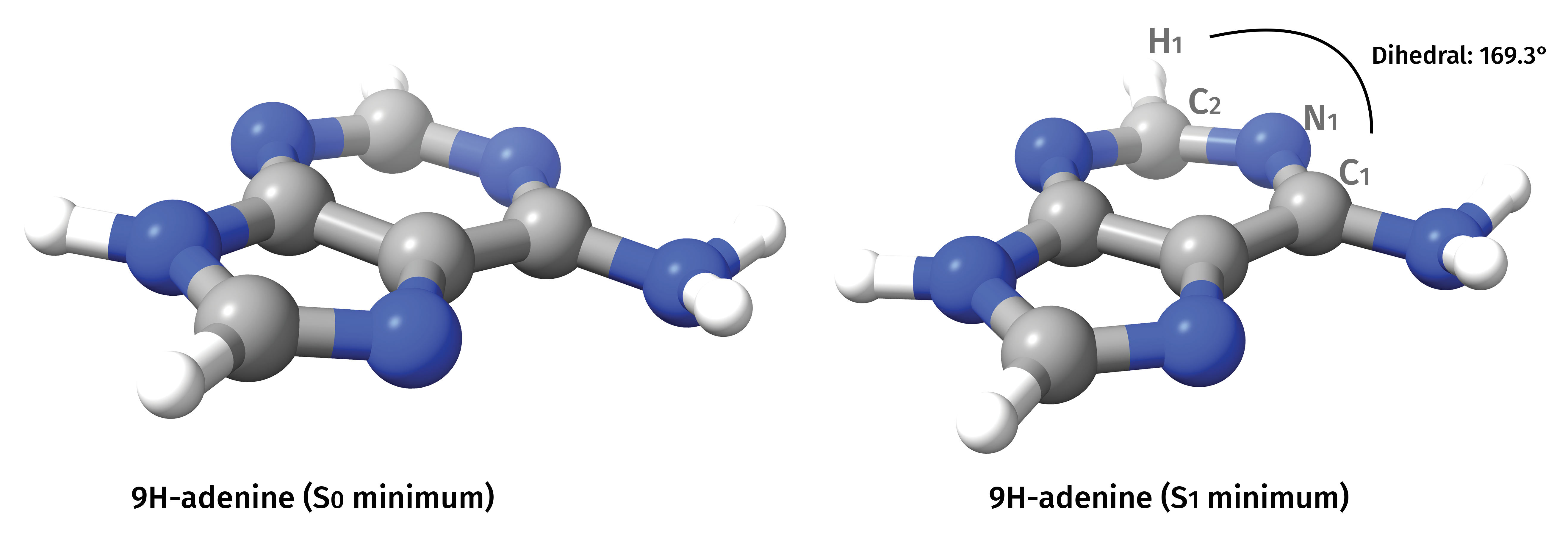}
    \caption{An $S_0$ and an $S_1$ minimum in 9H-adenine determined with EOM-CCSD/cc-pVDZ. In the $S_1$ minimum, we observe a puckering of the {C\scriptsize{2}} carbon atom which is not present in the $S_0$ minimum. }
    \label{fig:adenine-s0s1}
\end{figure}

\begin{table}[h!]
    \centering
    \caption{Harmonic frequencies ($\omega$) for the ground state equilibrium geometry of 9H-adenine determined with EOM-CCSD/cc-pVDZ. }
    \begin{tabular}{cc cc cc}
    \toprule
    Mode & $\omega [\si{\per\centi\meter}]$ & Mode & $\omega [\si{\per\centi\meter}]$ & Mode & $\omega [\si{\per\centi\meter}]$\\
    \cmidrule(lr){1-2} \cmidrule(lr){3-4} \cmidrule(lr){5-6}
    1  & 3736  & 14 & 1380 & 27 & 697 \\
    2  & 3702  & 15 & 1337 & 28 & 673 \\
    3  & 3606  & 16 & 1277 & 29 & 624 \\
    4  & 3285  & 17 & 1273 & 30 & 581 \\
    5  & 3217  & 18 & 1165 & 31 & 541 \\
    6  & 1709  & 19 & 1097 & 32 & 527 \\
    7  & 1694  & 20 & 1059 & 33 & 499 \\
    8  & 1640  & 21 & 1000 & 34 & 476 \\
    9  & 1572  & 22 & 950  & 35 & 465 \\
    10 & 1548  & 23 & 910  & 36 & 302 \\
    11 & 1488  & 24 & 882  & 37 & 275 \\
    12 & 1444  & 25 & 818  & 38 & 220 \\
    13 & 1394  & 26 & 731  & 39 & 170 \\
 \bottomrule
    \end{tabular}
    \label{tab:adenine-freq}
\end{table}

\begin{table}[ht]
\centering
\caption{Harmonic frequencies ($\omega$) for the excited state minimum of 9H-adenine determined with EOM-CCSD/cc-pVDZ.}
\begin{tabular}{cc cc cc}
\toprule
    Mode & $\omega [\si{\per\centi\meter}]$ & Mode & $\omega [\si{\per\centi\meter}]$ & Mode & $\omega [\si{\per\centi\meter}]$\\
    \cmidrule(lr){1-2} \cmidrule(lr){3-4} \cmidrule(lr){5-6}
 1 & 3704 & 14 & 1312 & 27 &  638 \\
 2 & 3696 & 15 & 1273 & 28 &  590 \\
 3 & 3576 & 16 & 1254 & 29 &  572 \\
 4 & 3302 & 17 & 1165 & 30 &  529 \\
 5 & 3273 & 18 & 1097 & 31 &  470 \\
 6 & 1654 & 19 & 1059 & 32 &  458 \\
 7 & 1643 & 20 & 1000 & 33 &  414 \\
 8 & 1602 & 21 &  926 & 34 &  375 \\
 9 & 1564 & 22 &  796 & 35 &  298 \\
10 & 1509 & 23 &  735 & 36 &  256 \\
11 & 1440 & 24 &  714 & 37 &  238 \\
12 & 1376 & 25 &  693 & 38 &  161 \\
13 & 1334 & 26 &  660 & 39 &  147 \\
\bottomrule
\end{tabular}
    \label{tab:adenine-freq-s1}
\end{table}

\begin{figure}
    \centering
    \includegraphics[width=0.5\linewidth]{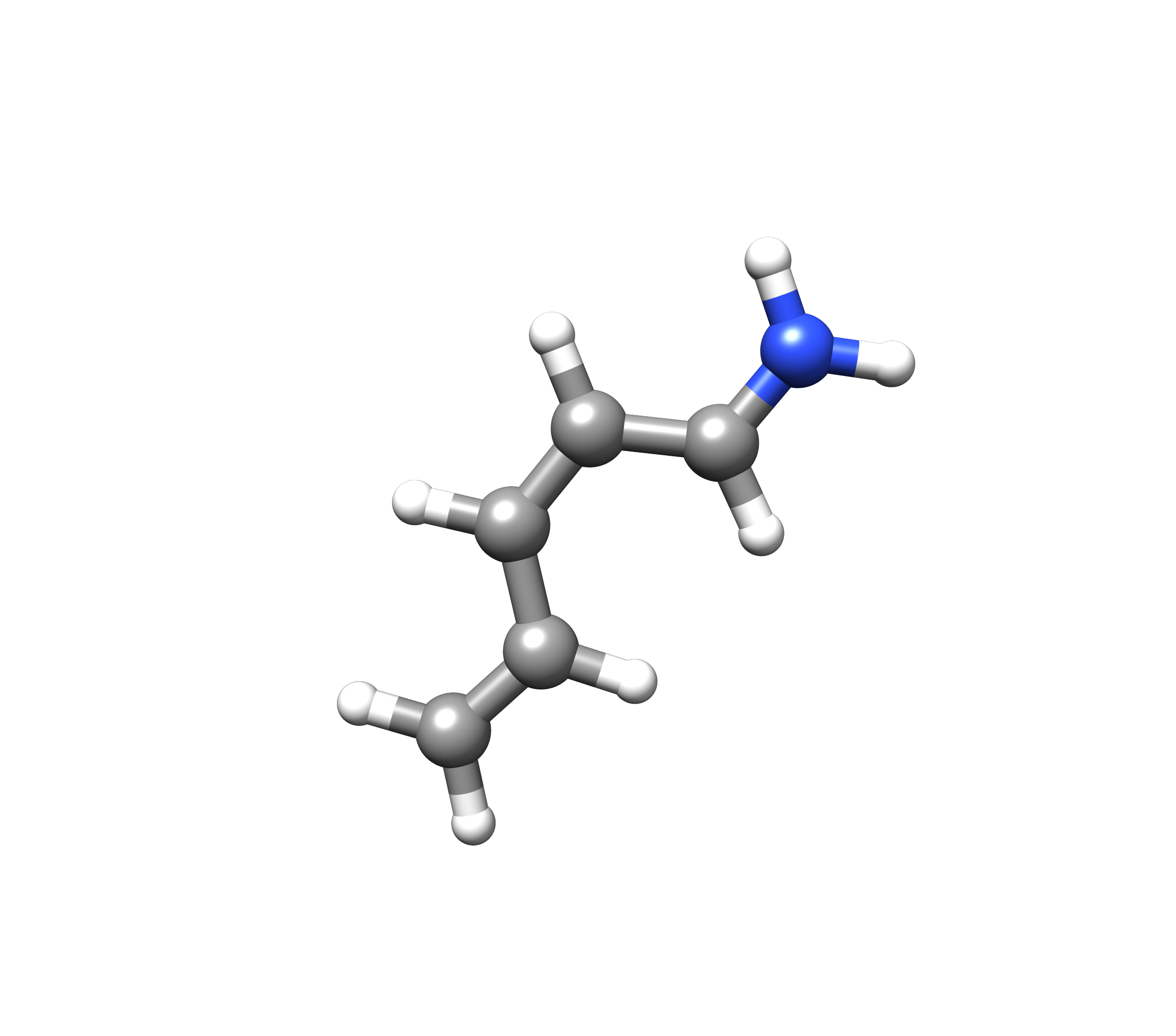}
    \caption{{The penta-2,4-dieniminium cation (PSB3)} {plotted using UCSF Chimera.\citep{Pettersen2004Chimera}}}
    \label{fig:psb3}
\end{figure}
\begin{figure}
    \centering
    \includegraphics[width=0.9\linewidth]{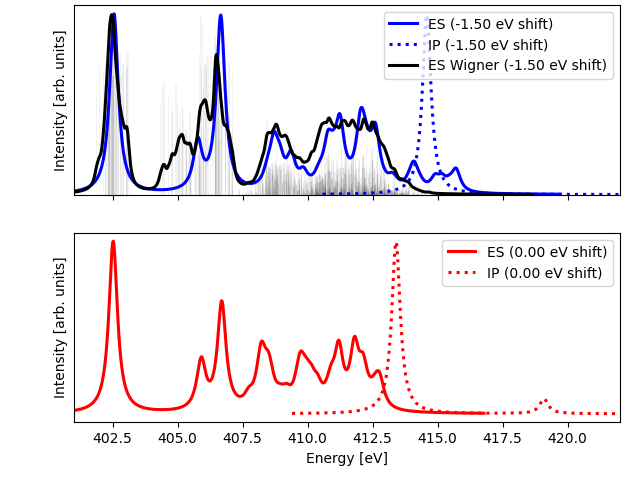}
    \caption{{Nitrogen K-edge excitation energies (ES) and ionization potentials (IP)} for PSB3, calculated at the CCSD/aug-cc-pVDZ (top) and CC3/aug-cc-pVDZ (bottom) levels of theory. The NEXAFS and XPS peaks are re-scaled separately. For the CCSD/aug-cc-pVDZ spectrum, the NEXAFS obtained from a Wigner {sampling} ($n = 40$) is shown . The CCSD spectra are shifted by $-\SI{1.5}{\eV}$.  {A Lorentzian broadening is applied, with $\SI{0.4}{\eV}$ FWHM for the S0-geometry calculations, and $\SI{0.2}{\eV}$ FWHM for the {Wigner-sampling} calculations.}
    }
    \label{fig:PSB3_s0_core}
\end{figure}

\begin{figure}
    \centering
 \includegraphics[width=0.9\linewidth]{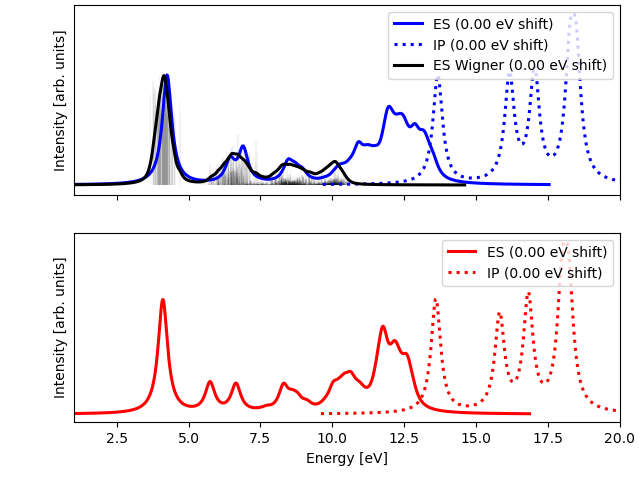}
    \caption{{Excitation energies (ES) and ionization potentials (IP) for PSB3, calculated at the CCSD/aug-cc-pVDZ (top) and CC3/aug-cc-pVDZ (bottom) levels of theory. The absorption and photoelectron spectra} are re-scaled separately. For the CCSD/aug-cc-pVDZ spectrum, the UV spectrum obtained from a Wigner sample ($n = 100$) is shown (with states limited to $< 10$ eV).  {A Lorentzian broadening is applied, with $\SI{0.4}{\eV}$ FWHM for the S0-geometry calculations, and $\SI{0.2}{\eV}$ FWHM for the Wigner sample calculations.}
    }
    \label{fig:PSB3_s0_valence}
\end{figure}

\subsection{Modeling UV-vis and X-ray absorption and photoelectron spectroscopy}

Modeling spectroscopies with coupled cluster theory has been a primary driver for developments {in $e^T$.} 
One-photon absorption and photoelectron spectra can be generated in the UV-visible and X-ray regions of the spectrum. Core excitations are obtained using the core-valence separation (CVS){\citep{Coriani2015,Coriani2016erratum}} approximation and ionization potentials are generated through inclusion of a non-interacting orbital into the excited state calculation{, with intensities obtained through the use of Dyson orbitals.\citep{moitra2022multi}} 

In Figures \ref{fig:PSB3_s0_core} and \ref{fig:PSB3_s0_valence}, we have plotted the core {(X-ray)} and valence {(UV)} absorption and photoelectron spectra of {the penta-2,4-dieniminium cation (PSB3, see Figure \ref{fig:psb3}).} The calculations are performed with the CCSD and CC3 models. The geometry of PSB3 is optimized at the CCSD level of theory, and the CCSD spectra are also generated {using 40 (NEXAFS) and 100 (UV)} geometries obtained from a {$\SI{0}{\kelvin}$ Wigner sample} {(see, e.g., Ref.~\citenum{curchod2018ab})}
of geometries around the {$S_0$} minimum.

\clearpage

\subsection{Multilevel coupled cluster theory}

\begin{figure}[h!]
    \centering
    \includegraphics[width=0.5\linewidth]{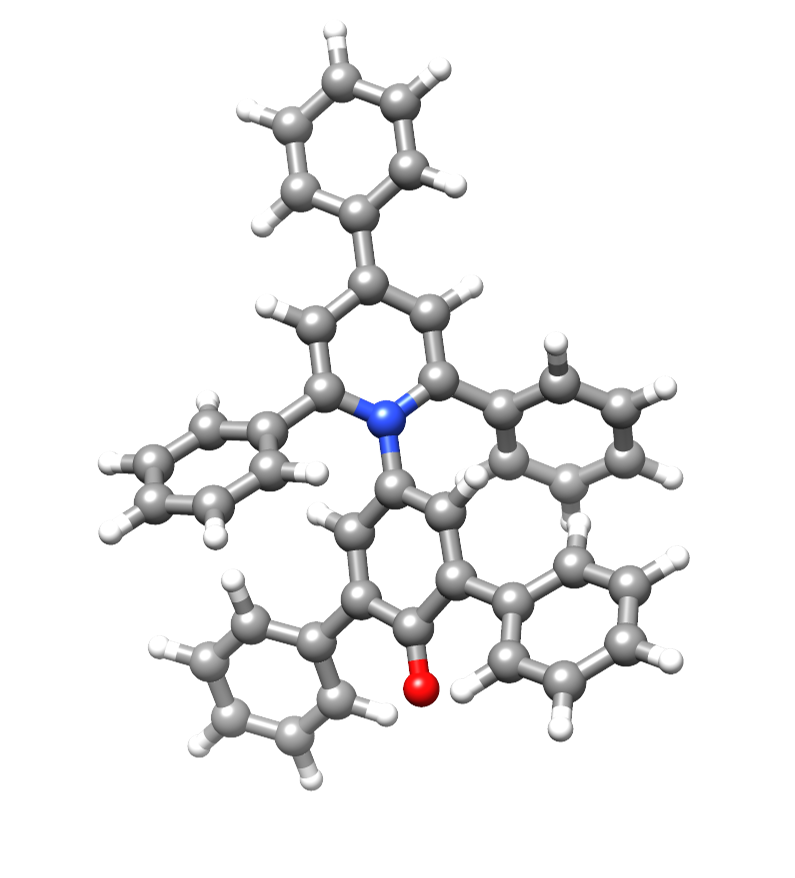}
    \includegraphics[width=0.4\linewidth]{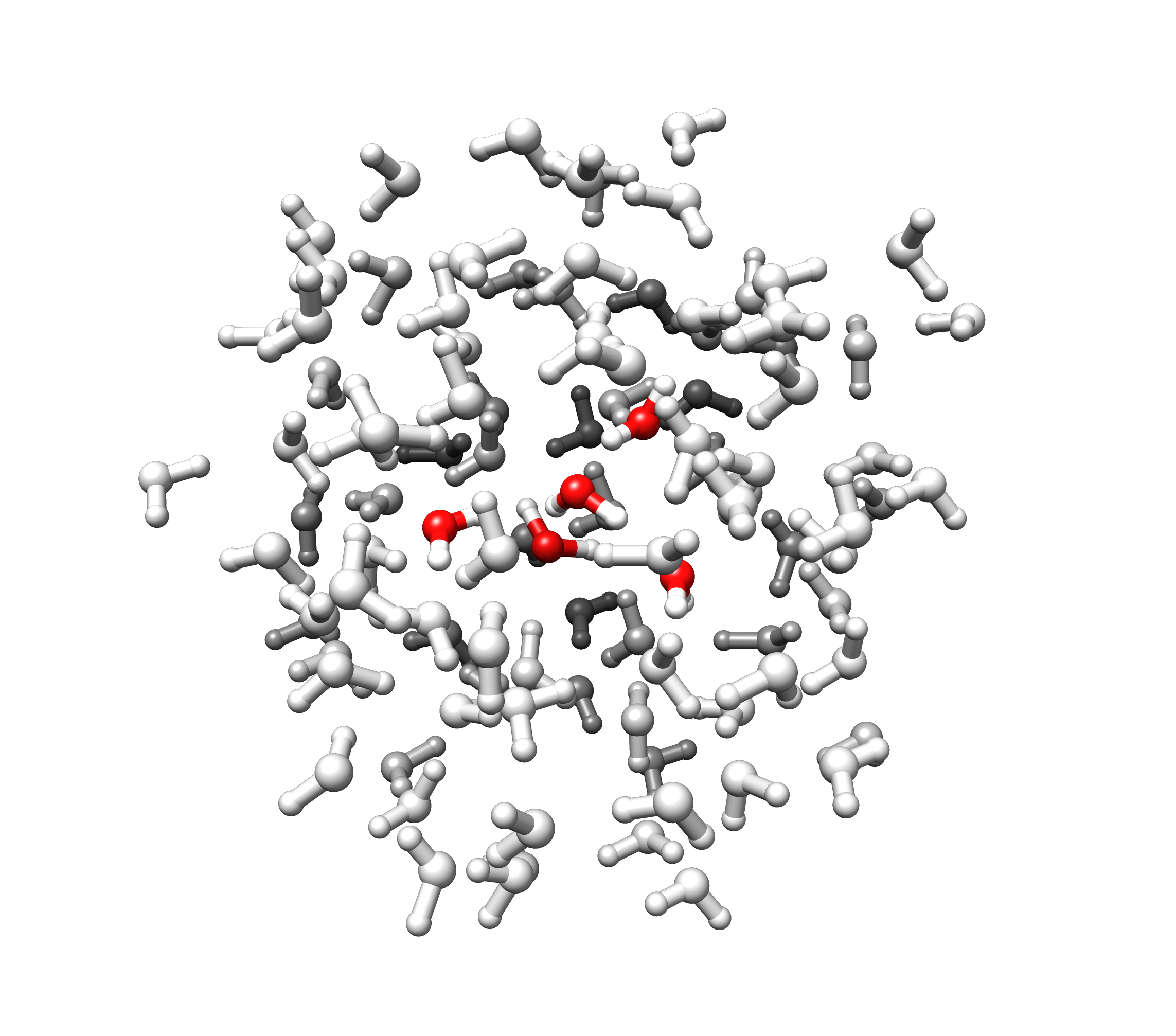}
    \caption{The betaine-30 molecule and a water cluster {plotted using UCSF Chimera.\citep{Pettersen2004Chimera}}}
    \label{fig:mlcc_molecules}
\end{figure}

\begin{figure}[h!]
    \centering
    \includegraphics[width=0.8\linewidth]{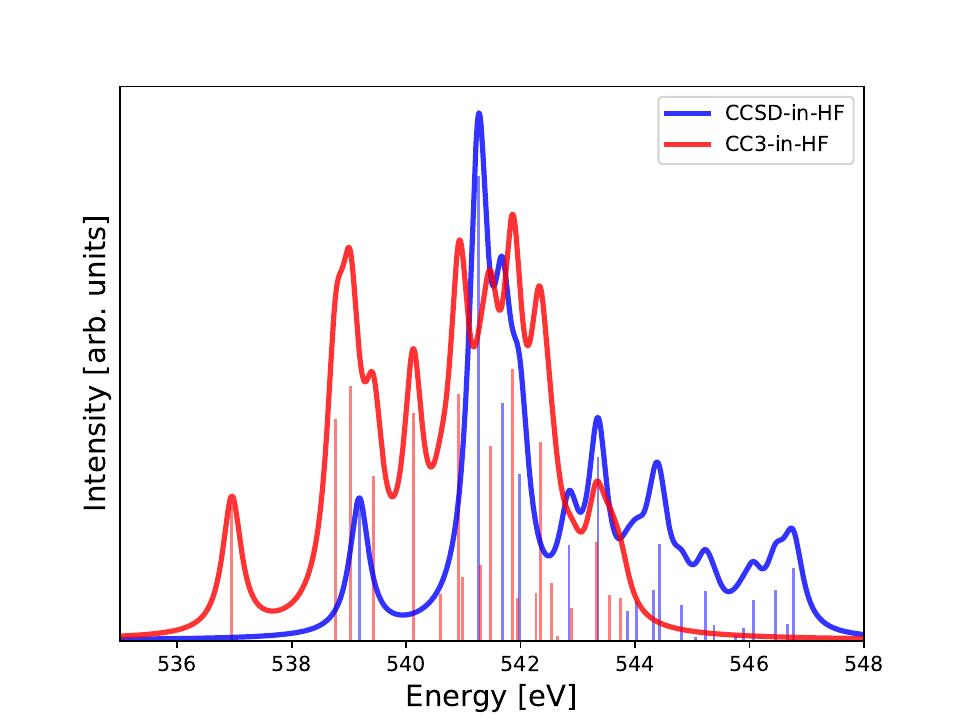}
    \caption{X-ray absorption spectrum of a water molecule at the center of a water cluster, calculated at the CCSD-in-HF and CC3-in-HF levels of theory. The aug-cc-pVDZ basis is used for the five central water molecules, while the cc-pVDZ basis is used for the rest of the water cluster. The CCSD and CC3 spectra are normalized separately, {such that the first peak has the same intensity for both models. A Lorentzian broadening with $\SI{0.4}{\eV}$ FWHM is used.} }
    \label{fig:wiw-plot}
\end{figure}

\begin{table}[h!]
    \centering
    \caption{The lowest excitation energy, $\omega$, of betaine-30 calculated with CC2 and CCSD, MLCC2, and MLCCSD using the aug-cc-pVDZ basis. The active orbital space used in the MLCC2 and MLCCSD calculations consists of 40 occupied and 400 virtual orbitals, and was determined using approximate correlated natural transition orbitals (CNTOs)\cite{hoyvik2017correlated,folkestad2019multilevel}. In total, there are 102 occupied and 1102 virtual orbitals (under the frozen core approximation). The calculations were done on two Intel Xeon Platinum 8380 CPUs ($\SI{2.30}{\giga\hertz}$), using 20 threads and $\SI{2}{\tera\byte}$ memory. {The} Cholesky decomposition threshold {is} $10^{-3}$. The reported wall times are for the full calculation, including determination of the reference state. }
    \begin{tabular}{l c c c}
    \toprule
    Model & $\omega\;[\si{\eV}]$ & Wall time & Peak memory usage\\
    \midrule
    Low memory CC2& $1.2157$ & $\SI{7.1}{\hour}$ & $\SI{2.0}{\tera\byte}$  \\
    MLCC2 & $1.2918$ & $\SI{2.0}{\hour}$ & $\SI{365.3}{\giga\byte}$ \\
    CCSD   & $1.7000$ & $\SI{137.5}{\hour}$ & $\SI{2.0}{\tera\byte}$  \\
    MLCCSD & $1.7064$ & $\SI{3.7}{\hour}$ & \SI{868.1}{\giga\byte} \\
    \bottomrule
    \end{tabular}
    \label{tab:betaine}
\end{table}

For calculation of intensive properties in molecular systems {that} are too large to be considered with standard coupled cluster methods, multilevel or multiscale (embedding) methods can be used.\cite{senn2009qm,tomasi2005quantum,warshel1972calculation, warshel1976theoretical,wesolowski1993frozen} In these approaches, an active region, or a set of active orbitals, are modeled with higher accuracy than the rest of the molecular system. In multiscale models, the environment is treated classically, for instance with molecular mechanics\cite{senn2009qm} or as a polarizable continuum.\cite{tomasi2002molecular,tomasi2005quantum} In multilevel models, different {approximate} levels of quantum mechanics are used for different orbital spaces.

The \eT~program provides a range of multilevel and multiscale features. 
{In particular, \eT~provides an efficient implementation of the polarizable QM/MM method based on the fluctuating charge force field,\cite{cappelli2016integrated, giovannini2020molecular,giovannini2020theory} and the multilevel coupled cluster and Hartree--Fock models, all of which are particularly powerful for spectroscopic and response properties.\cite{folkestad2024understanding, folkestad2024quantum,giovannini2025modeling}
} 

In multilevel coupled cluster theory, the cluster operator is partially (MLCC) or fully (CC-in-HF) restricted to an active orbital space.
For instance, in \eTtwo, excitation energies can be calculated with MLCC2 and MLCCSD. In Table \ref{tab:betaine}, {we present excitation energies of the betaine-30 molecule (see Figure \ref{fig:mlcc_molecules}), calculated with CC2, CCSD, MLCC2 and MLCCSD.} The wall time to calculate a single excitation energy with CCSD/aug-cc-pVDZ is around $\SI{137}{\hour}$, and the full $\SI{2}{\tera\byte}$ of memory allocated for the calculation is utilized. {An} MLCCSD calculation, using approximate correlated natural transition orbitals (CNTOs)\cite{hoyvik2017correlated,folkestad2019multilevel} to select the active orbital space, yields an excitation energy with an error of {less than} $\SI{10}{\milli\eV}$ in less than $\SI{4}{\hour}$, {and the memory usage is more than halved.}
\begin{table}
    \caption{{Cholesky decomposition wall time and the number of Cholesky vectors {($n_J$)} for an X-ray absorption spectra (XAS) calculation of a water molecule in liquid water. Calculations are reported with a standard decomposition procedure and a decomposition targeting accuracy in the electron repulsion integrals in the reduced space MO basis. {The decomposition time ($t_{\mathrm{CD}}$), the number of Cholesky vectors ($n_J$), and the total computational time for the XAS calculation ($t_{\mathrm{total}}$) are given.}}}
    \centering
    \begin{tabular}{l c c c}
    \toprule
    Type & $n_J$ & $t_{\text{CD}}$ & $t_{\text{total}}$ \\
    \midrule
    Standard & $12689$ & $\SI{179}{\second}$ & $\SI{6.4}{\hour}$ \\
    MO-screened & $1195$ &  $\SI{17}{\second}$  & $\SI{4.1}{\hour}$\\
    \bottomrule
    \end{tabular}
    \label{tab:CD_timings}
\end{table}

\begin{figure}
    \centering
    \includegraphics[width=0.8\linewidth]{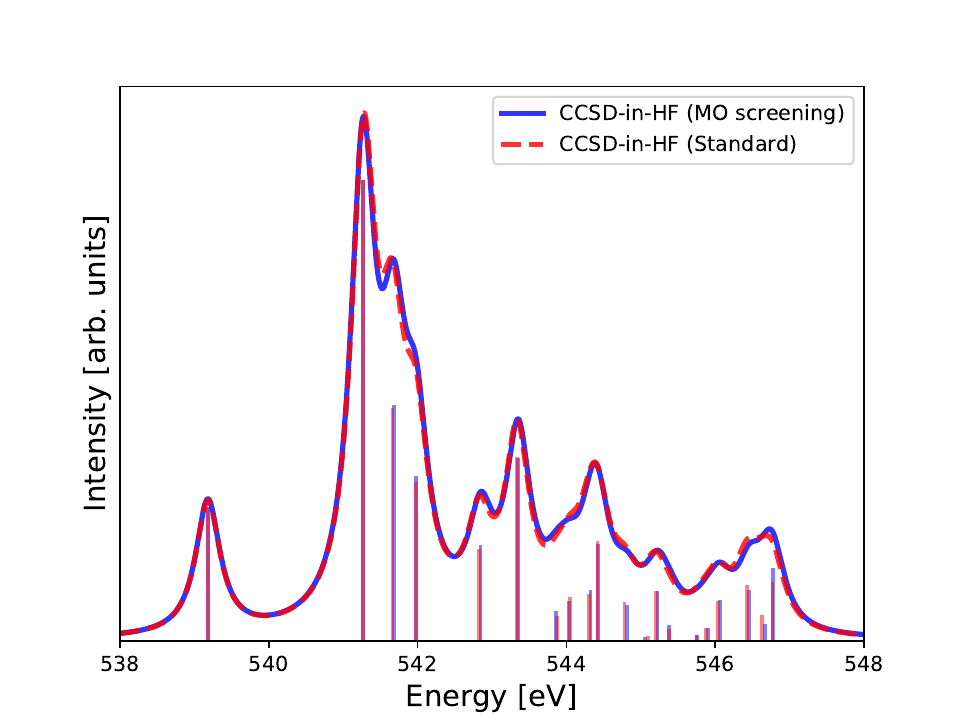}
    \caption{Comparison of the X-ray absorption spectra of a water molecule in liquid water calculated with CCSD-in-HF/aug-cc-pVDZ using the standard and MO-screened Cholesky decompositions. {A Lorentzian broadening with $\SI{0.4}{\eV}$ FWHM is used.} }
    \label{fig:CD_comparison}
\end{figure}

In Figure \ref{fig:wiw-plot}, the X-ray absorption spectrum of a water molecule in liquid water is modeled by use of CCSD-in-HF and CC3-in-HF. The core-excited water molecule and its four closest neighbors are included in the coupled cluster calculation, while the remaining molecules in the water cluster are modeled with a frozen Hartree--Fock density.
{The aug-cc-pVDZ basis is used for the five water molecules included in the coupled cluster calculation, the cc-pVDZ basis is used for the rest.}
The single snapshot in Figure \ref{fig:wiw-plot} echoes the findings of Ref.~\citenum{folkestad2024understanding}, where it was shown that inclusion of triple excitations in the cluster operator is necessary to model the relative intensities between the main- and post-edges of the spectrum. 

{The coupled cluster implementation in \eT~is based on Cholesky-decomposed electron repulsion integrals.
In multilevel calculations of this type, where a large part of the molecular system is described by a frozen Hartree--Fock density, the number of Cholesky vectors can be significantly reduced if one targets accuracy in the reduced active molecular orbital basis during the decomposition. This is done through the method-specific screening procedure described in {Refs.~\citenum{Folkestad2019} and \citenum{folkestad2021multilevel}}{, where accuracy of the integrals in the active (reduced) {molecular orbital (MO)} basis is targeted, rather than in the full orbital space.}
For the water cluster calculations, the standard decomposition yields $n_J = 12689$ Cholesky vectors, 
while with active space screening $n_J = 1195$ Cholesky vectors are obtained, reducing the cost of the 
{subsequent}
coupled cluster calculation. In Table \ref{tab:CD_timings}, we report wall-time comparisons of the two decompositions, and {of} the 
{final}
CCSD-in-HF calculations. In Figure \ref{fig:CD_comparison}, we show that the two X-ray absorption spectra generated with CCSD-in-HF, using different Cholesky decompositions{,} 
{are nearly identical.}
}

\subsection{Efficient solvers for high-level coupled cluster theory\label{sec:solvers}}
In \eTtwo, several new solvers have been added that facilitate new calculations or significantly speed up convergence.
In the calculation of excited states {with} the perturbative coupled cluster models (CC2 and CC3), the eigenvalue problem may be recast into a non-linear eigenvalue problem in a reduced parameter space. In \eTone, these equations were solved using a DIIS algorithm, which suffers from the occasional appearance of duplicate roots, and does not ensure convergence to the lowest-energy solutions. In \eTtwo, a non-linear Davidson solver that avoids both these problems is available. Furthermore, for the CC3 model{,} an efficient multimodel {solver based on the} Olsen {algorithm} can be used.{\cite{olsen1990passing,kjonstad2020accelerated}} 
The multimodel Olsen solver exploits the lower-level CCSD model to accelerate convergence of the CC3 {excited state} equations {with the Olsen algorithm.\citep{olsen1990passing}}Similarly, for CC3 {ground state {and response} equations}, a quasi-Newton solver that uses the CCSD Jacobian, rather than the orbital differences{,} to approximate the CC3 Jacobian{,} {improves convergence.}

In {Table} \ref{tab:nicotine_gs}{,} we compare timings of the CC3/aug-cc-pVDZ calculation of nicotine (C$_{10}$N$_{2}$H$_{14}$, 402 MOs) using \eTone~and \eTtwo.
The overall wall time to determine the ground state amplitudes ($\boldsymbol{t}$ and $\boldsymbol{\bar{t}}$ determined from in eqs. \eqref{eq:omega_standard_cc_omega} and \eqref{eq:multipliers}) is more than halved in \eTtwo.
The {ground state and left amplitude equations have been optimized following the release of \eTone. I.e, in constructing $\boldsymbol{\Omega}$ and performing the linear transformation by $\boldsymbol{A}^T$, see Appendix \ref{sec:cc}.} 
However, the computational savings are primarily a result of {changing the default solver to } the multimodel Olsen solvers,\citep{kjonstad2020accelerated} which approximately halve the number of $\boldsymbol{\Omega}$ constructions and linear transformation{s} by $\boldsymbol{A}^T$, compared to the pure DIIS-accelerated algorithms used in \eTone.

For excited states, 
{CC3 calculations}
can be performed with three different solvers: a DIIS-accelerated steepest-descent algorithm, a non-linear Davidson algorithm, or a multimodel Olsen algorithm. In Table \ref{tab:nicotine_es}, we report timings comparing the different algorithms when we calculate a single excited state with \eTtwo.
The DIIS and Olsen algorithms require fewer transformations than the non-linear Davidson algorithm and therefore significantly reduce the 
{wall time}. 
This is especially pronounced for the right excited state{, for which the initial guesses are less accurate}; the left excited states are calculated  
{with}
the right excited states {as the initial guesses}. 
In addition to optimizations of the $\boldsymbol{\Omega}$ construction and the linear transformation by $\boldsymbol{A}^T$, we have optimized the transformation by $\boldsymbol{A}$ since the release of \eTone. For nicotine, CC3/aug-cc-pVDZ, the average wall time of an $\boldsymbol{A}$ transformation is $\SI{1}{\hour}\;\SI{30}{\minute}$ with \eTone~ and $\SI{1}{\hour}\;\SI{20}{\minute}$ with \eTtwo.

From these calculations, the DIIS solver seems competitive, especially compared to the non-linear Davidson solver.  This picture, however, {changes} when one calculates several excitation energies. In Table \ref{tab:PSB3_es}, we report the calculation of ten excitation energies of the PSB3 molecule.  The DIIS solver is the least efficient, and it also skips a root that is identified with both the non-linear Davidson and the multimodel Olsen algorithms (compare roots 9 and 10 in Tab. \ref{tab:nicotine_es}).  Both for nicotine and PSB3, the multimodel Olsen algorithm proves to be the more efficient algorithm for the CC3 model.
\begin{table}[]
    \small
    \centering
    \caption{CC3/aug-cc-pVDZ ground state calculations for nicotine (C$_{10}$N$_{2}$H$_{14}$, 402 MOs). Reported timings are wall times {per call, and we also report the number of calls}. Calculations were performed using 20 cores with $\SI{2}{\tera\byte}$ memory available.}
    \begin{tabular}{l c c c c}
    \toprule
   & \multicolumn{2}{c}{\eTone} & \multicolumn{2}{c}{\eTtwo} \\
   \cmidrule(lr){2-5}
     Operation &  t  & $n_{\text{calls}}$ & t  & $n_{\text{calls}}$ \\
    \midrule
    $\boldsymbol\Omega$ construction & $\SI{48}{\minute}$ & 10 & $\SI{41}{\minute}$ & 5 \\
    $\boldsymbol{A}^T$ transformation & $\SI{87}{\minute}$ & 11 & $\SI{76}{\minute}$ & 6 \\
    \midrule
    Determine $\boldsymbol{t}$ and $\boldsymbol{\bar{t}}$ amplitudes & $\SI{24.7}{\hour}$ & -- & $\SI{12.4}{\hour}$  & -- \\
    \midrule
    \end{tabular}
    \label{tab:nicotine_gs}
\end{table}

\begin{table}[]
    \small
    \centering
    \caption{CC3/aug-cc-pVDZ excited state calculations for nicotine (C$_{10}$N$_{2}$H$_{14}$, 402 MOs). Calculations were performed using 20 cores with $\SI{2}{\tera\byte}$ memory available.}
    \begin{tabular}{l c c c c}
    \toprule
    & \multicolumn{3}{c}{$n_{\mathrm{calls}}$} \\
    \cmidrule(lr){2-4}
    Operation & DIIS & Non-linear Davidson & Olsen & $t_{\boldsymbol{A}/\boldsymbol{A}^T}$ \\
    \midrule
    $\boldsymbol{A}$ & 8 & 20 & 8 & $\SI{80}{\minute}$\\
    $\boldsymbol{A}^T$ & 8 & 11 & 7 & $\SI{76}{\minute}$\\
    \midrule
    $t_{R}$ &$\SI{11.3}{\hour}$ & $\SI{27.9}{\hour}$ &
     $\SI{12.8}{\hour}$\\
    $t_{L}$ & $\SI{11.0}{\hour}$ & $\SI{15.6}{\hour}$ &
    $\SI{10.8}{\hour}$\\
    \bottomrule
    \end{tabular}
    \label{tab:nicotine_es}
\end{table}

\begin{table}[]
\centering
\caption{Lowest 10 singlet excitation energies of PSB3 calculated at the frozen-core CC3/aug-cc-pVDZ level of theory. Three different solvers have been used: DIIS, non-linear Davidson, and multimodel Olsen. We also report the total time for the calculation of the excitation energies. Convergence threshold for the excited states are $10^{-3}$. All energies are given in $\si{\eV}${.}}
\begin{tabular}{c c c c}
\toprule
State & DIIS & Non-linear Davidson & Multimodel Olsen \\
\midrule
1  & $4.1068$ &  $4.1059$ & $4.1046$ \\
2  & $5.7480$ &  $5.7469$ & $5.7436$ \\
3  & $6.3519$ &  $6.3509$ & $6.3510$ \\
4  & $6.6510$ &  $6.6505$ & $6.6504$ \\
5  & $7.0396$ &  $7.0383$ & $7.0378$ \\
6  & $7.3710$ &  $7.3690$ & $7.3693$ \\
7  & $7.3914$ &  $7.3924$ & $7.3924$ \\
8  & $7.6928$ &  $7.6904$ & $7.6903$ \\
9  & $7.7620$ &  $7.7024$ & $7.7025$ \\
10 & $8.3043$ &  $7.7647$ & $7.7645$ \\
\midrule
Total time & $\SI{5.9}{\hour}$ & $\SI{3.7}{\hour}$ & $\SI{2.4}{\hour}$ \\
\bottomrule
    \end{tabular}
    \label{tab:PSB3_es}
\end{table}

\section{Coupled cluster implementation performance}
\begin{figure*}
    \centering
    \begin{subfigure}[t]{0.2\textwidth}
        \centering
        \includegraphics[width=\linewidth]{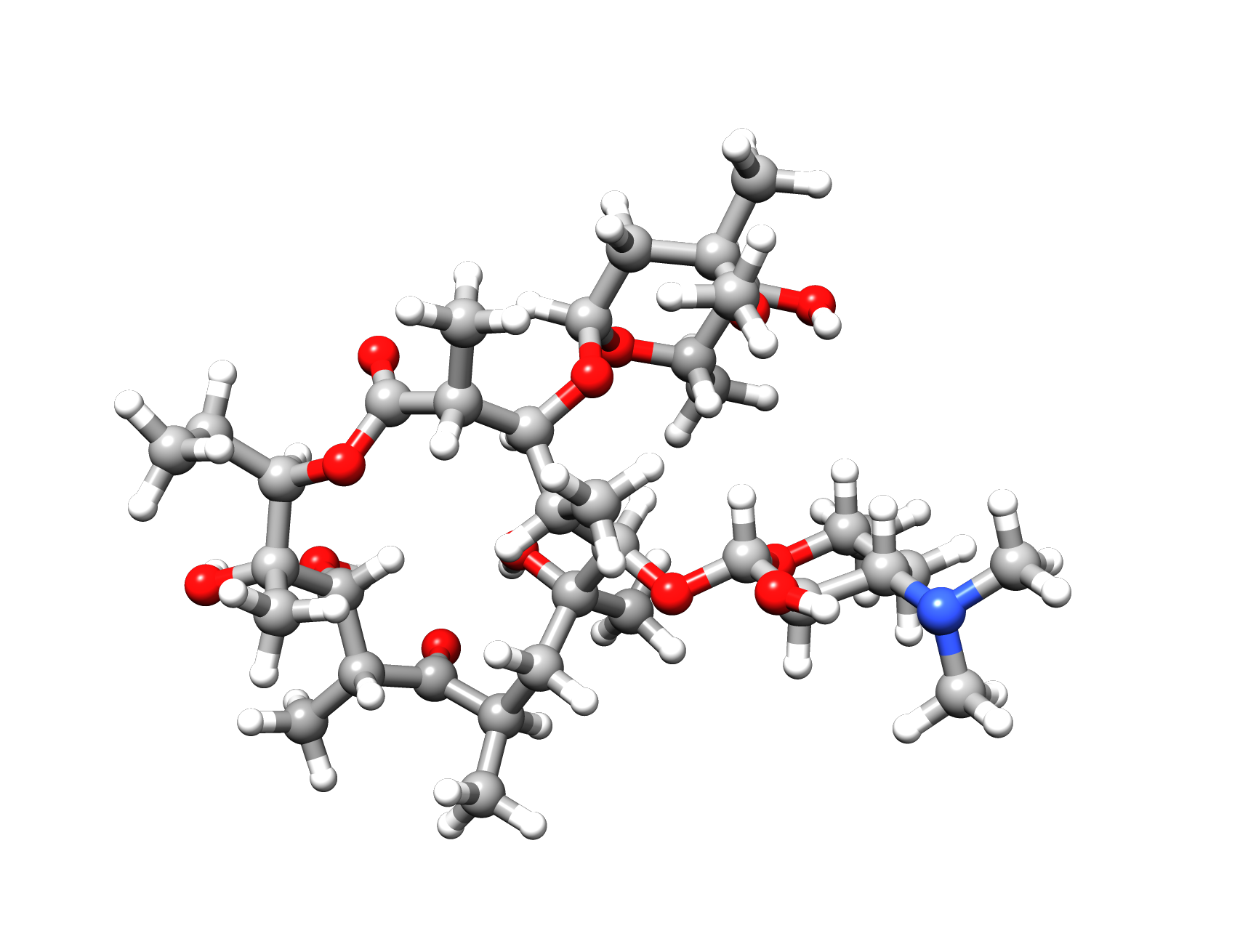}
        \caption{erythromycin\label{fig:erythro}}
    \end{subfigure}
    \begin{subfigure}[t]{0.2\textwidth}
        \centering
        \includegraphics[width=\linewidth]{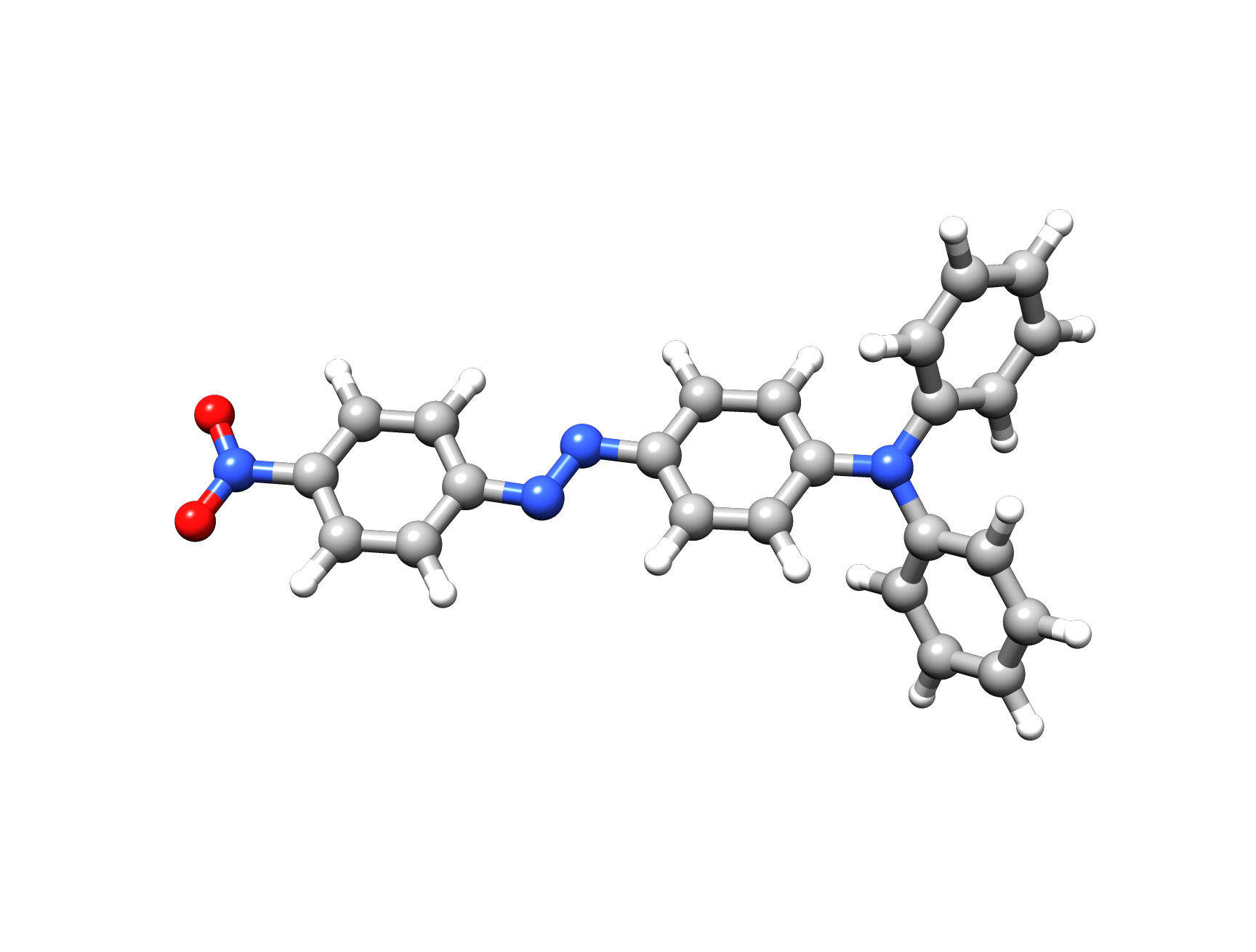}
        \caption{azobenzene dye\label{fig:azo}}
    \end{subfigure}
    \begin{subfigure}[t]{0.2\textwidth}
        \centering
        \includegraphics[width=\linewidth]{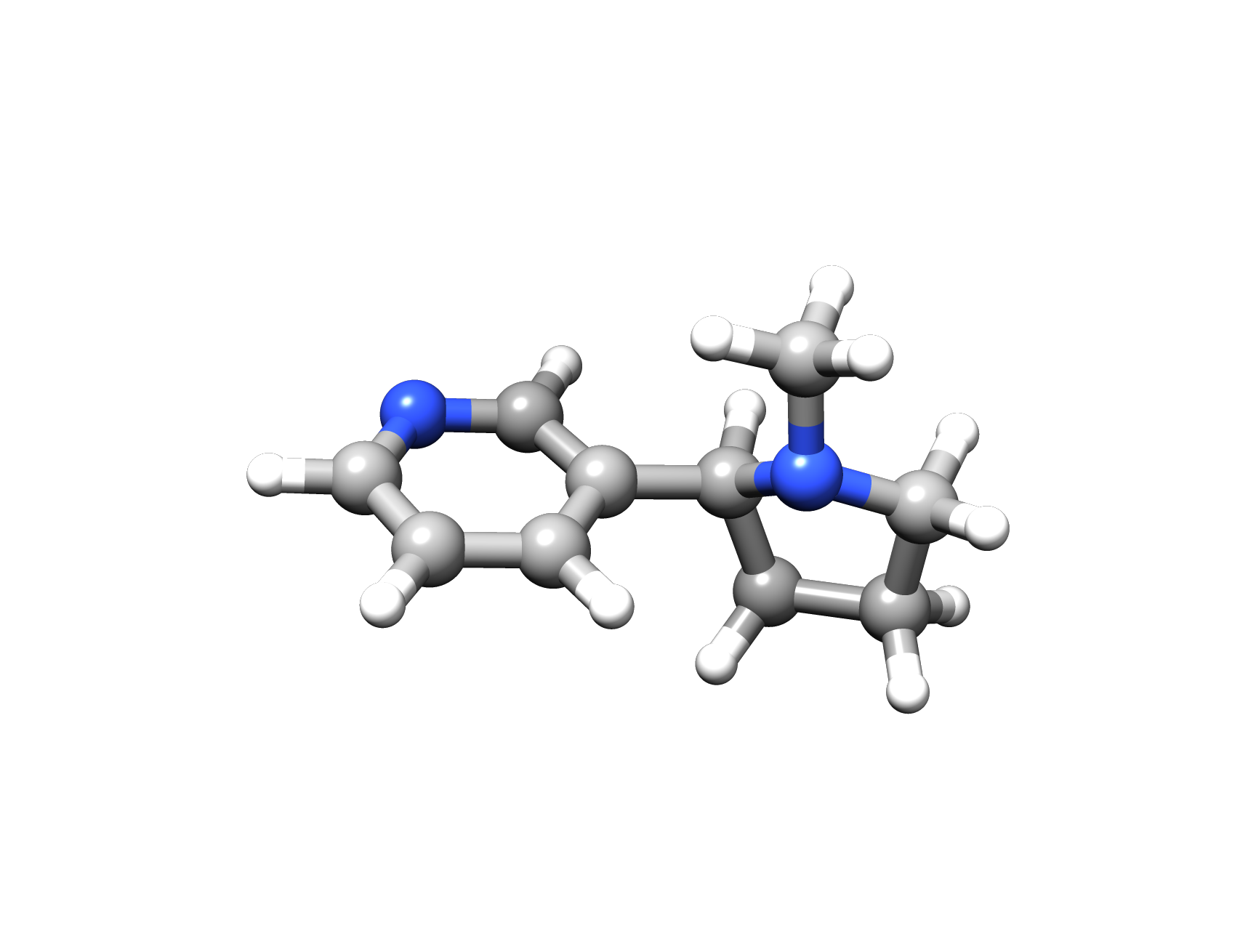}
        \caption{nicotine\label{fig:nico}}
    \end{subfigure}
    \begin{subfigure}[t]{0.2\textwidth}
        \centering
        \includegraphics[width=0.7\linewidth]{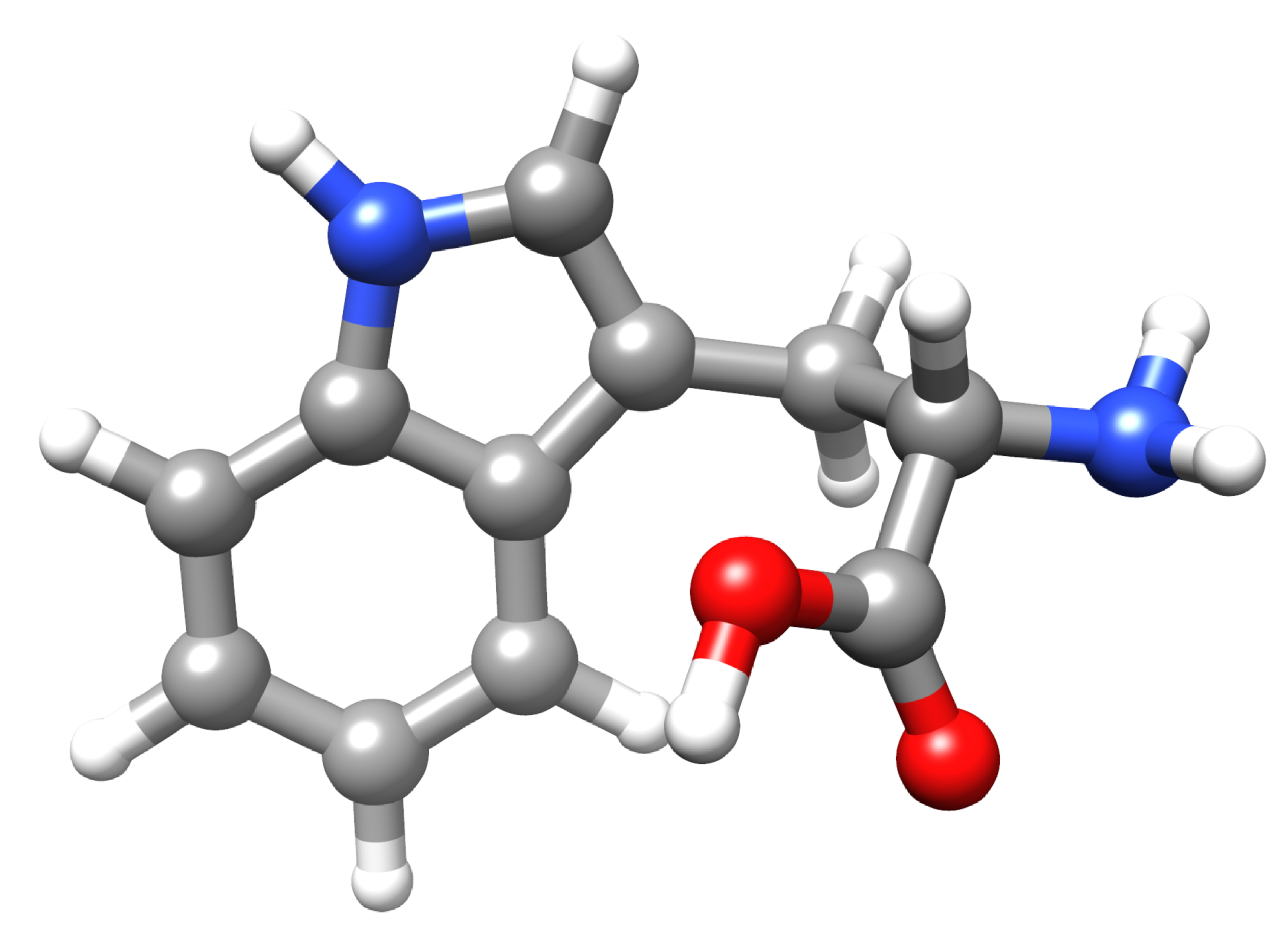}
        \caption{tryptophan\label{fig:trypto}}
    \end{subfigure}
    \caption{Molecules for performance testing of \eT~{plotted using UCSF Chimera.\citep{Pettersen2004Chimera}}}
\end{figure*}

{The \eT program was originally a coupled cluster code, and the ability to perform efficient coupled cluster calculations remains among the most important features of the program. In this section, we demonstrate the capabilities of \eTtwo~for standard coupled cluster calculations. We compare {our current implementations} to \eTone~and other notable electronic structure {software}. The notation and relevant equations are defined in Appendix \ref{sec:cc}. }
We use the default thresholds of \eTtwo, 
{unless stated otherwise:}
the Hartree--Fock equations are solved with a residual threshold of $10^{-7}\;$a.u., Cholesky decomposition of the two-electron integrals is performed with a threshold of $10^{-4}\;$a.u., and the coupled cluster ground and excited state equations are solved to residual thresholds of $10^{-5}\;$a.u. and $10^{-3}\;$a.u., respectively. For calculations with \eTtwo, the default integral library is used (Libcint){. A}ll timed calculations are performed on two Intel Xeon Platinum 8380 CPUs ($\SI{2.30}{\giga\hertz}$) with $\SI{2}{\tera\byte}$ shared memory{, unless otherwise stated}. All calculations are {assigned} 20 threads (or 10 threads per CPU).

\subsection{Comparing \eTtwo~to \eTone}
Since the release of \eTone, significant optimizations have been carried out in the coupled cluster implementations for the CC2, CCSD, and CC3 models. The improvements for {CC3 are} mainly due to the multimodel solvers, as 
{detailed}
in Section \ref{sec:solvers}. {Here, we restrict our attention to CC2 and CCSD.}

\subsubsection*{Calculations with the CC2 model}
There are two implementations of the CC2 model in $e^T$ which differ in their memory requirement. The standard CC2 implementation has the same memory requirements as the CCSD implementation (tensors of dimension $n_o^2n_v^2$ are kept in memory), while the \textit{low memory} CC2 implementation has an $N^2$ memory requirement{, where $N$ denotes the number of orbitals in the system}. While both models have been optimized since \eTone, the most significant optimizations have been implemented for low memory CC2{, where a looping strategy similar to that of the CC3 implementation\cite{myhre2016multilevel,paul2020new} has been introduced.} 

{In Table} \ref{tab:cc2_erythromycin}, we report timings from a calculation of a single excitation energy of the erythromycin molecule (C$_{37}$H$_{67}$O$_{13}$N, $1719$ MOs under the frozen core approximation, see {Figure} \ref{fig:erythro}).
For erythromycin, the time to construct the $\boldsymbol{\Omega}$-vector [eq. \eqref{eq:omega_standard_cc_omega}] and the linear transformation by the $\boldsymbol{A}$-matrix [to solve eq. \eqref{eq:es_rhs}] has been reduced by factors of $\sim6$ and $\sim 4$, respectively. The result is a significant reduction in the total computational times for the ground and excited state. Because different solvers are used (by default) in the two versions, the calculated excitation energies differ. In \eTone, a DIIS solver was used for low memory CC2. This solver does not ensure convergence to the roots {with the lowest energy}, and is also prone to convergence {of} duplicate roots. In \eTtwo, a non-linear Davidson solver is used, ensuring no duplication of roots and convergence to the {roots with lowest energy} (in molecules without symmetry).
\begin{table}[h!]
    \small
    \centering
    \caption{Comparison of \eTone~ and \eTtwo~ CC2 ground and excited state calculations of erythromycin (C$_{37}$H$_{67}$O$_{13}$N, $1719$ MOs under the frozen core approximation). We report the correlation energy ($E_{\mathrm{corr}}$), lowest excitation energy $\omega$, and wall times to calculate the ground and excited state ($t_{\mathrm{gs}}$ and $t_{\mathrm{es}}$), and the average times to construct the $\boldsymbol{\Omega}$  vector ($t_\Omega$) and the linear transformation by the Jacobian matrix (${t_A}$).}
    \begin{tabular}{c c c c c c c}
    \toprule
    \eT & $E_{\mathrm{corr}}$ [$\si{\hartree}$] & $t_{\mathrm{gs}}$ & $t_{\Omega}$ & $\omega$ [$\si{\eV}$]  & $t_{\mathrm{es}}$ & $t_{A}$\\
    \midrule
      v1.0  & $-8.3934$ & $\SI{7.1}{\hour}$  & $\SI{36}{\minute}$ & $5.1858$ & $\SI{53.1}{\hour}$  & $\SI{4.1}{\hour}$ \\
      v2.0  & $-8.3934$ & $\SI{1.8}{\hour}$ & $\SI{6}{\minute}$  & $4.0551$ & $\SI{32.5}{\hour}$ & $\SI{1.3}{\hour}$  \\
    \bottomrule
    \end{tabular}
    \label{tab:cc2_erythromycin}
\end{table}
\subsubsection*{Calculations with the CCSD model}
We test the performance of the CCSD implementation with a calculation of the first four excited states for an azobenzene dye (C$_{24}$H$_{20}$O$_{2}$N$_{4}$, 822 MOs within the frozen core approximation, see Figure \ref{fig:azo}),
and their transition strengths from the ground state; this provides the first features of the dye's ultraviolet spectrum. Here, we are primarily interested in the performance metrics and the comparison with \eTone. Timings are given in Table \ref{tab:ccsd_azobenzene-dye}.

The 
{test consists of}
a ground and an excited state calculation, where we need to determine the 
left and right 
{amplitudes}
of
{both}
the ground state and 
{the}
excited states [eqs. \eqref{eq:multipliers}, \eqref{eq:omega_standard_cc_omega}, \eqref{eq:es_lhs}, \eqref{eq:es_rhs}, and \eqref{eq:os1}], and we present timings for each of these four steps.
For the ground state, we observe {only} a modest performance improvement for the $\boldsymbol{t}$-amplitudes, whereas the $\boldsymbol{\bar{t}}$-amplitudes are obtained almost three times faster than in \eTone. This speedup is primarily due to optimizations of the $\boldsymbol{A}^T$-transformation, which is more than twice as fast {in \eTtwo~compared to \eTone.} 

For the excited states, we 
see 
speedups for both the right and {the} left {amplitudes}. 
The time 
{per $\boldsymbol{A}$-transformation}
is almost halved {compared} to \eTone. 
This is the primary cause of the speedup of the right excited states. For the left states, the speedup is due to both the optimizations of the $\boldsymbol{A}^T$-transformation  
and to an improved starting guess; in \eTtwo, the initial guess {for the left {amplitudes}} is generated from the converged right {amplitudes}. Overall, 
{the optimizations for ground and excited states}
cut the total cost of the calculation by a factor of {$\sim4$} {compared with \eTone}.

\begin{table*}[h!]
    \caption{Comparison of \eTone~ and \eTtwo~ CCSD/aug-cc-pVDZ ground and excited state calculations on the azobenzene dye (C$_{24}$H$_{20}$O$_{2}$N$_{4}$, 822 MOs within the frozen core approximation). Reported timings are wall times. Calculations were performed using 20 cores and $\SI{2}{\tera\byte}$ memory.}
    \centering
    \begin{tabularx}{\linewidth}{p{0.19\linewidth} p{0.19\linewidth} p{0.19\linewidth} p{0.19\linewidth} p{0.19\linewidth}}
        \toprule
         & \multicolumn{2}{c}{\eTone}  & \multicolumn{2}{c}{\eTtwo}\\
         \cmidrule(lr){2-3} \cmidrule(lr){4-5}
         Total time &\multicolumn{2}{c}{$\SI{642.3}{\hour}$} &\multicolumn{2}{c}{$\SI{164.8}{\hour}$} \\
         \midrule    
         & Ground state & Excited state & Ground state & Excited state \\
         \midrule
         Left & $\SI{29.3}{\hour}$& $\SI{336.3}{\hour}$& $\SI{11.8}{\hour}$& $\SI{24.4}{\hour}$ \\
         Right & $\SI{10.8}{\hour}$ & $\SI{265.7}{\hour}$ & $\SI{8.9}{\hour}$& $\SI{119.3}{\hour}$\\
         \bottomrule
    \end{tabularx}
    \label{tab:ccsd_azobenzene-dye}
\end{table*}

\subsection{Comparisons to other electronic structure programs}
We compare the CCSD implementation of \eTtwo~to the electronic structure programs CFOUR,\cite{cfour2020} Q-Chem (6.3),\cite{qchem2021} and Psi4 (1.9.1),\cite{psi42020} all known for extensive coupled cluster functionality. For the latter two, Cholesky factorization of the electron repulsion integrals is used (with decomposition threshold $\tau$). Psi4 is open-source, whereas both CFOUR and Q-Chem are proprietary codes, 
{with Q-Chem being commercial as}
CFOUR\cite{cfour2020} 
being available for academic use upon request.
Calculation input files 
{can} be found in Ref.~\citenum{geometries_and_ouputs}.

\begin{table*}[tb!]
    \scriptsize
    \centering
    \caption{CCSD/aug-cc-pVDZ calculations of the ground state and four excited states of nicotine (C$_{10}$H$_{14}$N$_2$, 402 MOs) using \eT, CFOUR, Q-Chem (6.3) and Psi4 (1.9.1). For \eT, Q-Chem, and Psi4, $\tau$ is the thresholds for the Cholesky decomposition {of the} electron repulsion integrals. {Calculations were performed using 20 cores and the available memory is either $\SI{2}{\tera\byte}$ or $\SI{200}{\giga\byte}$.}\label{tab:nico_ccsd}}
    \begin{tabular}{c c c c c c c c c c c c c c c}
    \toprule
    & \multicolumn{7}{c}{$\SI{2}{\tera\byte}$} & \multicolumn{7}{c}{$\SI{200}{\giga\byte}$} \\
     \cmidrule(lr){2-8}\cmidrule(lr){9-15}
    & Psi4 & \multicolumn{2}{c}{CFOUR} & \multicolumn{2}{c}{\eTtwo} & \multicolumn{2}{c}{Q-Chem} & Psi4 & \multicolumn{2}{c}{CFOUR} & \multicolumn{2}{c}{\eTtwo} & \multicolumn{2}{c}{Q-Chem}\\
    \cmidrule(lr){3-4} \cmidrule(lr){5-6}  \cmidrule(lr){7-8} \cmidrule(lr){10-11} \cmidrule(lr){12-13} \cmidrule(lr){14-15}
    & $\tau = 10^{-4}$ & ECC & NCC & $\tau = 10^{-4}$ & $\tau = 10^{-8}$ & $\tau = 10^{-4}$ & $\tau = 10^{-8}$ &$\tau = 10^{-4}$  & ECC & NCC & $\tau = 10^{-4}$ & $\tau = 10^{-8}$ & $\tau = 10^{-4}$ & $\tau = 10^{-8}$\\
    \midrule
    $t_{\mrm{gs}}$    & $\SI{0.5}{\hour}^\dagger$ & $\SI{0.7}{\hour}$  & $\SI{0.3}{\hour}$ & $\SI{0.3}{\hour}$ & $\SI{0.5}{\hour}$ & $\SI{0.6}{\hour}$ & $\SI{0.8}{\hour}$& $\SI{0.5}{\hour}^\dagger$ &$\SI{0.7}{\hour}$ & $\SI{0.7}{\hour}$  & $\SI{0.2}{\hour}$ & $\SI{0.3}{\hour}$&$\SI{0.6}{\hour}$ & $\SI{0.8}{\hour}$\\
    $t_{\mrm{es}}$ & $\SI{10.7}{\hour}$ & $\SI{7.0}{\hour}$  & $\SI{3.8}{\hour}$ & $\SI{3.0}{\hour}$ & $\SI{3.3}{\hour}$ & $\SI{2.3}{\hour}$& $\SI{2.8}{\hour}$& $\SI{10.8}{\hour}$& $\SI{6.9}{\hour}$  & $\SI{3.1}{\hour}$  & $\SI{3.3}{\hour}$ & $\SI{4.5}{\hour}$&$\SI{2.5}{\hour}$ & $\SI{3.1}{\hour}$ \\
    \midrule
    $t_{\mrm{tot}}$  & $\SI{11.2}{\hour}$ & $\SI{12.3}{\hour}$ & $\SI{8.6}{\hour}$ & $\SI{3.3}{\hour}$ & $\SI{3.9}{\hour}$ & $\SI{3.1}{\hour}$ & $\SI{3.8}{\hour}$ & $\SI{11.8}{\hour}$ & $\SI{12.1}{\hour}$ & $\SI{8.3}{\hour}$ & $\SI{3.5}{\hour}$ & $\SI{4.8}{\hour}$& $\SI{3.1}{\hour}$ & $\SI{3.9}{\hour}$ \\
    \bottomrule
    \end{tabular}
    
    {\raggedright\scriptsize $^\dagger$Ground state equations are solved to ($10^{-3}$).}
    \label{tab:nico-comparison}
\end{table*}
\begin{table}[h!]
    \caption{{CCSD/aug-cc-pVDZ ground and excited state calculation of t}ryptophan (C$_{11}$H$_{12}$N$_2$O$_2$, 438 MOs under the frozen core approximation). Cholesky decomposed electron repulsion integrals are used, and the decomposition threshold is {$\tau = 10^{-4}$.} }
    \centering
    \begin{tabular}{l c c c c c c c c}
        \toprule
         & $t_{\mrm{gs}}$
         & $t_{\mrm{es}}$
         & $t_{\mrm{total}}$
         & $n_{\mrm{calls}}^{\mrm{gs, R}}$
         & $n_{\mrm{calls}}^{\mrm{gs, L}}$
         & $n_{\mrm{calls}}^{\mrm{R}}$
         & $n_{\mrm{calls}}^{\mrm{L}}$ \\
        \midrule
        \eTone~   & $\SI{1.0}{\hour}$ & $\SI{12.3}{\hour}$& $\SI{13.3}{\hour}$ & 12 & 13 & 79 &  81\\
        \eTtwo~   & $\SI{0.6}{\hour}$ & $\SI{2.8}{\hour}$ & $\SI{3.5}{\hour} $ & 12 & 11 & 79 & 29 \\
        Q-Chem    & $\SI{1.8}{\hour}$ & $\SI{7.9}{\hour}$ & $\SI{10.0}{\hour}$ & 13 & 8 & 83 &  24\\
        \bottomrule
    \end{tabular}
    \label{tab:tryptophan_timings}
\end{table}
\begin{table*}[tbh!]
    \centering
    \caption{CC3/aug-cc-pVDZ calculations of the ground state and a single excited state of {nicotine} (C$_{10}$H$_{14}$N$_2$, 402 MOs) using \eT~and  CFOUR. For \eT, we report timings with different Cholesky decomposition thresholds $\tau$. {The available memory is either $\SI{2}{\tera\byte}$, $\SI{400}{\giga\byte}$, or $\SI{200}{\giga\byte}$.}}
    \begin{tabular}{ l c c c c c c c c}
    \toprule
    Available memory & \multicolumn{2}{c}{$\SI{200}{\giga\byte}$}&\multicolumn{3}{c}{$\SI{400}{\giga\byte}$} & \multicolumn{3}{c}{$\SI{2}{\tera\byte}$} \\
    \cmidrule(lr){2-3}\cmidrule(lr){4-6}\cmidrule(lr){7-9} 
    Program & \multicolumn{2}{c}{\eTtwo} & {CFOUR} & \multicolumn{2}{c}{\eTtwo}& {CFOUR} & \multicolumn{2}{c}{\eTtwo}  \\
    \cmidrule(lr){2-3}\cmidrule(lr){4-6}\cmidrule(lr){8-9} 
     & $\tau=10^{-4}$ &$\tau=10^{-8}$ & & $\tau=10^{-4}$ & $\tau=10^{-8}$ & & $\tau=10^{-4}$ & $\tau=10^{-8}$\\
    \midrule
    Wall time & $\SI{208}{\hour}$ & $\SI{210}{\hour}$ &$\SI{226}{\hour}$ & $\SI{209}{\hour}$ & $\SI{214}{\hour}$ & $\SI{198}{\hour}$ & $\SI{210}{\hour}$& $\SI{210}{\hour}$ \\
    \bottomrule
    \end{tabular}
    \label{tab:nico-comparison-cc3}
\end{table*}
In Table \ref{tab:nico_ccsd}, {we present wall times for calculations} of the first four excited {states} of nicotine using the aug-cc-pVDZ basis (C$_{10}$H$_{14}$N$_2$, 402 MOs, see {Figure} \ref{fig:nico}). The wall times to solve the ground and excited state equations [eqs \eqref{eq:omega_standard_cc_omega}-- \eqref{eq:es_rhs}] are given.
The calculations are performed with either $\SI{200}{\giga\byte}$ or $\SI{2}{\tera\byte}$ of memory available; with $\SI{2}{\tera\byte}${,} all electron repulsion integrals can be stored in memory.
{B}oth \eTtwo~and Q-Chem {obtain} four excited states of nicotine in 3--4 hours, depending on the Cholesky decomposition threshold. While \eT~solves the ground states more efficiently, Q-Chem is more efficient for the excited state equations of this molecule.

{In Table \ref{tab:nico_ccsd},} we also compare to the two coupled cluster programs 
{included in}
the public release of CFOUR (ECC and NCC). The NCC program solves the excited state equations in less than 4 hours, and is approximately twice as fast as the ECC program. However, the total wall time in all CFOUR calculations {is} high, due to the integral transformations following the mean-field calculation; this bottleneck is eliminated in \eT~and Q-Chem with the Cholesky decomposition of the integrals.

As the {size of the molecular system} increases, there is a crossover point at which \eTtwo~becomes more efficient than the Q-Chem software. In Table \ref{tab:tryptophan_timings}, we compare Q-Chem to \eTone~and \eTtwo~for the first four excitations of tryptophan (C$_{11}$H$_{12}$N$_2$O$_2$, aug-cc-pVDZ{,} 438 MOs under the frozen core approximation, see {Figure} \ref{fig:trypto}). In these calculations we calculate the energies and the transition strengths, i.e.~{we solve} eqs. \eqref{eq:omega_standard_cc_omega}, \eqref{eq:es_rhs}, \eqref{eq:es_lhs}, \eqref{eq:multipliers}, and \eqref{eq:os1}. From these calculations, {we see} that \eTtwo~{provides} a significant improvement on the first release, {and that the first four excitation energies and transition strengths of tryptophan are obtained} almost three times faster than {with} Q-Chem.

Finally, we compare the CC3 implementation of \eTtwo~to the CC3 implementation of the NCC program in CFOUR.
Previously, the {CC3} implementation has been compared to the implementations in the Dalton program,\cite{dalton2014} in the ECC program of CFOUR,\cite{cfour2020} and in Psi4.\citep{psi42020}\cite{paul2020new} It was shown that \eT~significantly outperformed the other implementations.
From Table \ref{tab:nico-comparison-cc3}, where we present the calculation of a single excited state of nicotine, {it is evident} that both \eT and the NCC program provide efficient implementations of the CC3 model. With a Cholesky decomposition threshold of $\tau=10^{-8}$, \eTtwo~performs better under low-memory conditions and CFOUR performs better under high-memory conditions{, although the differences in performance are relatively minor}. Note that the minimal memory requirement for CFOUR is $\SI{245}{\giga\byte}$, whereas {\eTtwo~is able to run with less memory due to its batching algorithm\citep{paul2020new} and} {never uses more than $\SI{427}{\giga\byte}$ in these calculations on nicotine.}

\section{Future developments}
The \eT~program is a product of, and a framework for, significant and ongoing research activities within quantum chemistry. New functionality will be introduced in the program that reflect these activities. {Below, we list some of the major ongoing developments that are underway.} 

{Several recent developments in the code have {targeted} ground and excited state molecular dynamics, with recent developments of efficient analytical nuclear gradients and coupling elements, at the CCSD\citep{schnack2022efficient,kjonstad2023communication} and similarity constrained CCSD (SCCSD)\citep{kjonstad2017resolving,kjonstad2019orbital,lexander2024analytical,kjonstad2024coupled} levels, which recently enabled nonadiabatic excited state dynamics simulations with coupled cluster theory.\citep{hait2024prediction,kjonstad2024photoinduced}}
{Very recent work has also made progress on the description of conical intersections at the CC2 level (SCC2),\citep{stoll2025similarity} determination of ground {and excited} state minimum-energy intersection geometries {at the {coupled cluster} level},\citep{angelico2025determining} as well as a correct description of 
{ground state}
intersections
{using} the generalized coupled cluster (GCC)\citep{rossi2025generalized} and convex Hartree-Fock (CVX-HF){\citep{rossi2025convex}} {methods}.}
{These}
developments are planned to be included in the next release, enabling the adoption of the developed methods for dynamics {simulations.} 

Within multicomponent chemistry, additional functionality for conditions with strong light-matter coupling, such as triplet polaritonic states at the CCSD level of theory,\citep{smedsrud2024coupled} response theory for SC-QED-HF,\cite{castagnola2025, castagnola2024polaritonic} real-time time-dependent QED-{CCSD,}\cite{castagnola2024strong} 
as well as QED-HF harmonic frequencies ({molecular gradient} and Hessian)\citep{barlini2025}
and magnetic properties\cite{barilini2024} are underway. {Recently, the QED implementations have been extended to treat chiral cavities\cite{Riso2023,Riso_2025} that enable the study of enantiomeric selectivity in ground and excited states. A model for collective strong coupling using QED-CCSD will be included in the next release.~\cite{castagnola2025} }
Finally, additional wave function models, such as CASSCF{,} QED--CASSCF\cite{Alessandro2025} and SC-QED-MP2,\cite{elmoutakal2025strong}, and, 
in the context of electronically open molecules, the particle-breaking Hartree--Fock model\cite{Matveeva:2023aa,Paul:2024aa} and its time-dependent version\cite{jacobsen:2025} has been developed within the \eT~community. 

{To study electronic excitations, time-dependent DFT (TDDFT) will soon be released, \cite{marques2004time} in addition to new multilevel methods based on DFT.\cite{marrazzini2021multilevel,giovannini2023integrated, giovannini2024time} This will allow for the study of large, realistic condensed phase systems. To calculate local response properties, such as polarizabilities and electronic excitations, energy-based orbital localizations in specific fragment regions at the HF and DFT levels of theory\cite{giovannini2020energy,giovannini2022fragment,goletto2022linear,goletto2021combining, giovannini2024kohn} will allow for novel energy decomposition analysis.\cite{giovannini2022fragment,giovannini2024kohn}} 
{Regarding the calculation of spectroscopic properties at the coupled cluster levels of theory, several upcoming features are in store. Resonant inelastic X-ray scattering (RIXS) spectroscopy\cite{schnack2023new} and X-ray emission spectroscopy (XES), implemented using a damped response solver (available in \eTtwo), will soon be available. Transition moments will be available with the MLCC2 and MLCCSD models, as well as triplet excitation energies.}

Another strand of developments in the code moves away from manual implementations of equations. Indeed, several upcoming developments in the program has been aided with the equation and $e^T$ code generation tool SpinAdaptedSecondQuantization (SASQ).\citep{lexander2025spinadaptedsecondquantization}

A significant emerging challenge for established electronic structure software,  and also for the \eT~program, is to adapt to a further shift in HPC facilities towards heterogeneous computer architectures. Support for GPU acceleration{, and/or parallelization over multiple {CPU/GPU nodes,} could be a necessary step to ensure that the \eT~program remains competitive.

\section{Concluding remarks}
\eTtwo~is an electronic structure program with 
{emphasis on efficiency and modularity, featuring}
efficient coupled cluster {implementations} and unique multicomponent and multilevel features.
The program is open-source {and} available to {everyone} {free of cost,}
both for use and modification {within the GPLv3.0 license conditions}. {The development cycle of the program is also open, with public {code} review and continuous improvements added to the development version hosted on GitLab.} Rigorous testing and a continuous integration/continuous delivery (CI/CD) setup ensures that modifications are correct and that new additions follow style guidelines. Since the release of \eTone, the program has undergone significant {improvements}. The coupled cluster implementation{s} {have} been made more efficient with new solvers, improved starting guesses, and refinements in both memory handling and how terms are evaluated.

{Since its first major release, the program has also evolved in new directions.}
Additional {established} mean-field and correlated models have been added to the program---such as ROHF, DFT, FCI, and CASCI---and \eT~has also been extended to treat new {chemical and physical phenomena}, such as strong light-matter coupling considered {with the CBO approximation and} at the mean-field, and correlated levels of theory.
{With this} major release, the \eT~program is no longer {only} a coupled cluster code,{ even {though} it continues to prominently feature the functionality that gives the program its name.}

\section{Data availability statement}
Molecular geometries and calculation output files can be found in Ref.~\citenum{geometries_and_ouputs}.

\section{Acknowledgments}
This work was supported by the European Research Council (ERC) under the European Union's Horizon 2020 Research and Innovation Program (grant agreement No. 101020016), Marie Sk{\l}odowska-Curie European Training Network ``COSINE - COmputational Spectroscopy In Natural sciences and Engineering'' (grant agreement No. 765739) and Research Council of Norway through FRINATEK (projects 263110 and 275506). {The work was also supported by advanced User Support provided by Sigma2 - the National Infrastructure for High-Performance Computing and Data Storage in Norway}
J.P. acknowledges financial support from the Technical University of Denmark within the Ph.D. Alliance Programme. 
S.C. acknowledges financial support from the
Independent Research Fund Denmark-Natural Sciences,
Research Project 2 grant 
no. 7014-00258B.
{T.G acknowledges financial support from the European Union – Next Generation EU in the framework of the PRIN 2022 PNRR project POSEIDON – Code P2022J9C3R.}
{R. A., G. T. and E. R. acknowledge financial support from the European Research Council (ERC) under the European Union's Horizon 2020 Research and Innovation Program (grant agreement No. 101040197). }

\subsection*{Conflict of interest}
The authors have no conflicts to disclose.

\appendix

\section{Solving the coupled cluster equations \label{sec:cc}}

In coupled cluster theory, the wave function is given by~\citep{helgaker2014molecular}
\begin{align}
   \ket{\mathrm{CC}} = \exp(T)\ket{\text{R}},\; T = \sum_{\mu}t_\mu\tau_\mu
\end{align}
where $\ket{\text{HF}}$ is the Hartree--Fock determinant, $T$
is the cluster operator which generates excitations of the reference determinant through excitation operators $\tau_\mu$. 
{The ground state coupled cluster equations are obtained by projecting the Schrödinger equation onto the Hartree--Fock {state} and excited {configurations}:
\begin{align}
    E_0 &= \Tbraket{\mathrm{R}}{\simH}{\mathrm{R}} \label{eq:omega_standard_cc_E}\\
    \Omega_{\mu} &= \Tbraket{\mu}{\simH}{\mathrm{R}} = 0.\label{eq:omega_standard_cc_omega}
\end{align}
Here, we have introduced the similarity-transformed Hamiltonian ${\simH} = \exp(-T)H\exp(T)$. The $\boldsymbol{\Omega}$-equations [eq. \eqref{eq:omega_standard_cc_omega}] must be solved to determine the cluster amplitudes ($t_\mu$). }

{Coupled cluster excitation energies can be calculated with linear response theory or the equation-of-motion (EOM) approach. In EOM coupled cluster theory, the states $\ket{k}$ are defined by the expansion
\begin{align}
    \ket{k}=\sum_{\mu\geq0}\exp({T})R^k_{\mu}\ket{\mu},\label{eq:es_rhs}
\end{align}
where $\boldsymbol{R^k}$ are the right eigenvectors of the similarity-transformed Hamiltonian:
\begin{align}
   \boldsymbol{\simH} \boldsymbol{R}^k = E_k \boldsymbol{R}^k.
\end{align}
The similarity-transformed Hamiltonian has the form
\begin{align}
    \boldsymbol{\simH} =
    \begin{pmatrix}
     E_0 & \boldsymbol{\eta}^\mathrm{T} \\
     \boldsymbol\Omega &\boldsymbol{A} + E_0\boldsymbol{I} \\
    \end{pmatrix}=
        \begin{pmatrix}
     E_0 & \boldsymbol{\eta}^\mathrm{T} \\
     \boldsymbol 0 & \boldsymbol{A} + E_0\boldsymbol{I} \\
    \end{pmatrix},\label{eq:shape_of_simH}
\end{align}
where $\boldsymbol{A}$ is the Jacobian matrix with elements
\begin{align}
    A_{\mu\nu}=\Tbraket{\mu}{[\simH,\tau_{\nu}]}{\mathrm{R}},
\end{align}
and
$ \eta_{\nu} = \Tbraket{\mathrm{R}}{[\simH, \tau_{\nu}]}{\mathrm{R}}$.
The eigenvalues of $\boldsymbol{\simH}$ are the energies of the electronic states in EOM coupled cluster theory, and the excitation energies $\omega_k$ are the eigenvalues of $\boldsymbol{A}$.}

{The similarity-transformed Hamiltonian is non-Hermitian {and} its left and right eigenvectors differ. We may express the left EOM coupled cluster states as
\begin{align}
   \bra{k} = \sum_{\mu\geq0}L^k_{\mu}\bra{\mu}\exp(-T),
\end{align}
where
\begin{align}
   \boldsymbol{\simH}^T {\boldsymbol{L}^{k}} = E_k {\boldsymbol{L}^{k}},\label{eq:es_lhs}
\end{align}
and we require that the left and right states form a biorthonormal set. With this biorthonormalization condition, the right vectors are given by
\begin{align}
    \boldsymbol R^0 =
    \begin{pmatrix}
    1\\
    0\\
    \end{pmatrix}, \; \;
    \boldsymbol R^k =
    \begin{pmatrix}
    \omega_k^{-1}\boldsymbol{\eta}^\mathrm{T}\boldsymbol{r}_k \\
    \boldsymbol{r}_k\\
    \end{pmatrix}\; \text{for}\; k>0, \label{eq:right_std}
\end{align}
where $\boldsymbol{r}_k$ are the right eigenvectors of $\boldsymbol{A}$, corresponding to the eigenvalue $\omega_k$.
The left vectors are given by
\begin{align}
    \boldsymbol L^0 =
    \begin{pmatrix}
    1 \\
    \boldsymbol{\bar{t}}\\
    \end{pmatrix}, \; \;
    \boldsymbol L^k =
    \begin{pmatrix}
    0 \\
    \boldsymbol{l}_k\\
    \end{pmatrix}\;\text{for}\; k>0,\label{eq:left_std}
\end{align}
where $\boldsymbol{\bar{t}}$ are the left ground state amplitudes, determined by solving
\begin{align}
    \boldsymbol{A}^\mathrm{T}\boldsymbol{\bar{t}} = -\boldsymbol{\eta},\label{eq:multipliers}
\end{align}
and $\boldsymbol{l}_k$ is a left eigenvector of $\boldsymbol{A}$, corresponding to the eigenvalue $\omega_k$.
The eigenvector equations for the Jacobian matrix, and eq. \eqref{eq:multipliers} are solved by using the linear transformation by $\boldsymbol{A}$ or its transpose. This transformation usually dominates the computational cost to solve these equations.
}

{Molecular response properties and oscillator strengths can also be obtained from linear response theory or within the EOM framework. For instance, the EOM oscillator strengths (for excitations {from} the ground state) are obtained from the expression
\begin{align}
    f_k = \frac{2}{3}\omega_k\Tbraket{0}{\boldsymbol{\mu}}{k}\Tbraket{k}{\boldsymbol{\mu}}{0},\label{eq:os1}
\end{align}
{which require} the calculation of left and right transition dipole moments.}

\bibliography{eTbibs.bib}

\providecommand{\noopsort}[1]{}\providecommand{\singleletter}[1]{#1}%
\begin{thebibliography}{149}%
\makeatletter
\providecommand \@ifxundefined [1]{%
 \@ifx{#1\undefined}
}%
\providecommand \@ifnum [1]{%
 \ifnum #1\expandafter \@firstoftwo
 \else \expandafter \@secondoftwo
 \fi
}%
\providecommand \@ifx [1]{%
 \ifx #1\expandafter \@firstoftwo
 \else \expandafter \@secondoftwo
 \fi
}%
\providecommand \natexlab [1]{#1}%
\providecommand \enquote  [1]{``#1''}%
\providecommand \bibnamefont  [1]{#1}%
\providecommand \bibfnamefont [1]{#1}%
\providecommand \citenamefont [1]{#1}%
\providecommand \href@noop [0]{\@secondoftwo}%
\providecommand \href [0]{\begingroup \@sanitize@url \@href}%
\providecommand \@href[1]{\@@startlink{#1}\@@href}%
\providecommand \@@href[1]{\endgroup#1\@@endlink}%
\providecommand \@sanitize@url [0]{\catcode `\\12\catcode `\$12\catcode
  `\&12\catcode `\#12\catcode `\^12\catcode `\_12\catcode `\%12\relax}%
\providecommand \@@startlink[1]{}%
\providecommand \@@endlink[0]{}%
\providecommand \url  [0]{\begingroup\@sanitize@url \@url }%
\providecommand \@url [1]{\endgroup\@href {#1}{\urlprefix }}%
\providecommand \urlprefix  [0]{URL }%
\providecommand \Eprint [0]{\href }%
\providecommand \doibase [0]{http://dx.doi.org/}%
\providecommand \selectlanguage [0]{\@gobble}%
\providecommand \bibinfo  [0]{\@secondoftwo}%
\providecommand \bibfield  [0]{\@secondoftwo}%
\providecommand \translation [1]{[#1]}%
\providecommand \BibitemOpen [0]{}%
\providecommand \bibitemStop [0]{}%
\providecommand \bibitemNoStop [0]{.\EOS\space}%
\providecommand \EOS [0]{\spacefactor3000\relax}%
\providecommand \BibitemShut  [1]{\csname bibitem#1\endcsname}%
\let\auto@bib@innerbib\@empty
\bibitem [{\citenamefont {Frisch}\ \emph {et~al.}(2016)\citenamefont {Frisch},
  \citenamefont {Trucks}, \citenamefont {Schlegel}, \citenamefont {Scuseria},
  \citenamefont {Robb}, \citenamefont {Cheeseman}, \citenamefont {Scalmani},
  \citenamefont {Barone}, \citenamefont {Petersson}, \citenamefont {Nakatsuji},
  \citenamefont {Li}, \citenamefont {Caricato}, \citenamefont {Marenich},
  \citenamefont {Bloino}, \citenamefont {Janesko}, \citenamefont {Gomperts},
  \citenamefont {Mennucci}, \citenamefont {Hratchian}, \citenamefont {Ortiz},
  \citenamefont {Izmaylov}, \citenamefont {Sonnenberg}, \citenamefont
  {Williams-Young}, \citenamefont {Ding}, \citenamefont {Lipparini},
  \citenamefont {Egidi}, \citenamefont {Goings}, \citenamefont {Peng},
  \citenamefont {Petrone}, \citenamefont {Henderson}, \citenamefont
  {Ranasinghe}, \citenamefont {Zakrzewski}, \citenamefont {Gao}, \citenamefont
  {Rega}, \citenamefont {Zheng}, \citenamefont {Liang}, \citenamefont {Hada},
  \citenamefont {Ehara}, \citenamefont {Toyota}, \citenamefont {Fukuda},
  \citenamefont {Hasegawa}, \citenamefont {Ishida}, \citenamefont {Nakajima},
  \citenamefont {Honda}, \citenamefont {Kitao}, \citenamefont {Nakai},
  \citenamefont {Vreven}, \citenamefont {Throssell}, \citenamefont
  {Montgomery}, \citenamefont {Peralta}, \citenamefont {Ogliaro}, \citenamefont
  {Bearpark}, \citenamefont {Heyd}, \citenamefont {Brothers}, \citenamefont
  {Kudin}, \citenamefont {Staroverov}, \citenamefont {Keith}, \citenamefont
  {Kobayashi}, \citenamefont {Normand}, \citenamefont {Raghavachari},
  \citenamefont {Rendell}, \citenamefont {Burant}, \citenamefont {Iyengar},
  \citenamefont {Tomasi}, \citenamefont {Cossi}, \citenamefont {Millam},
  \citenamefont {Klene}, \citenamefont {Adamo}, \citenamefont {Cammi},
  \citenamefont {Ochterski}, \citenamefont {Martin}, \citenamefont {Morokuma},
  \citenamefont {Farkas}, \citenamefont {Foresman},\ and\ \citenamefont
  {Fox}}]{g16}%
  \BibitemOpen
  \bibfield  {author} {\bibinfo {author} {\bibfnamefont {M.~J.}\ \bibnamefont
  {Frisch}}, \bibinfo {author} {\bibfnamefont {G.~W.}\ \bibnamefont {Trucks}},
  \bibinfo {author} {\bibfnamefont {H.~B.}\ \bibnamefont {Schlegel}}, \bibinfo
  {author} {\bibfnamefont {G.~E.}\ \bibnamefont {Scuseria}}, \bibinfo {author}
  {\bibfnamefont {M.~A.}\ \bibnamefont {Robb}}, \bibinfo {author}
  {\bibfnamefont {J.~R.}\ \bibnamefont {Cheeseman}}, \bibinfo {author}
  {\bibfnamefont {G.}~\bibnamefont {Scalmani}}, \bibinfo {author}
  {\bibfnamefont {V.}~\bibnamefont {Barone}}, \bibinfo {author} {\bibfnamefont
  {G.~A.}\ \bibnamefont {Petersson}}, \bibinfo {author} {\bibfnamefont
  {H.}~\bibnamefont {Nakatsuji}}, \bibinfo {author} {\bibfnamefont
  {X.}~\bibnamefont {Li}}, \bibinfo {author} {\bibfnamefont {M.}~\bibnamefont
  {Caricato}}, \bibinfo {author} {\bibfnamefont {A.~V.}\ \bibnamefont
  {Marenich}}, \bibinfo {author} {\bibfnamefont {J.}~\bibnamefont {Bloino}},
  \bibinfo {author} {\bibfnamefont {B.~G.}\ \bibnamefont {Janesko}}, \bibinfo
  {author} {\bibfnamefont {R.}~\bibnamefont {Gomperts}}, \bibinfo {author}
  {\bibfnamefont {B.}~\bibnamefont {Mennucci}}, \bibinfo {author}
  {\bibfnamefont {H.~P.}\ \bibnamefont {Hratchian}}, \bibinfo {author}
  {\bibfnamefont {J.~V.}\ \bibnamefont {Ortiz}}, \bibinfo {author}
  {\bibfnamefont {A.~F.}\ \bibnamefont {Izmaylov}}, \bibinfo {author}
  {\bibfnamefont {J.~L.}\ \bibnamefont {Sonnenberg}}, \bibinfo {author}
  {\bibfnamefont {D.}~\bibnamefont {Williams-Young}}, \bibinfo {author}
  {\bibfnamefont {F.}~\bibnamefont {Ding}}, \bibinfo {author} {\bibfnamefont
  {F.}~\bibnamefont {Lipparini}}, \bibinfo {author} {\bibfnamefont
  {F.}~\bibnamefont {Egidi}}, \bibinfo {author} {\bibfnamefont
  {J.}~\bibnamefont {Goings}}, \bibinfo {author} {\bibfnamefont
  {B.}~\bibnamefont {Peng}}, \bibinfo {author} {\bibfnamefont {A.}~\bibnamefont
  {Petrone}}, \bibinfo {author} {\bibfnamefont {T.}~\bibnamefont {Henderson}},
  \bibinfo {author} {\bibfnamefont {D.}~\bibnamefont {Ranasinghe}}, \bibinfo
  {author} {\bibfnamefont {V.~G.}\ \bibnamefont {Zakrzewski}}, \bibinfo
  {author} {\bibfnamefont {J.}~\bibnamefont {Gao}}, \bibinfo {author}
  {\bibfnamefont {N.}~\bibnamefont {Rega}}, \bibinfo {author} {\bibfnamefont
  {G.}~\bibnamefont {Zheng}}, \bibinfo {author} {\bibfnamefont
  {W.}~\bibnamefont {Liang}}, \bibinfo {author} {\bibfnamefont
  {M.}~\bibnamefont {Hada}}, \bibinfo {author} {\bibfnamefont {M.}~\bibnamefont
  {Ehara}}, \bibinfo {author} {\bibfnamefont {K.}~\bibnamefont {Toyota}},
  \bibinfo {author} {\bibfnamefont {R.}~\bibnamefont {Fukuda}}, \bibinfo
  {author} {\bibfnamefont {J.}~\bibnamefont {Hasegawa}}, \bibinfo {author}
  {\bibfnamefont {M.}~\bibnamefont {Ishida}}, \bibinfo {author} {\bibfnamefont
  {T.}~\bibnamefont {Nakajima}}, \bibinfo {author} {\bibfnamefont
  {Y.}~\bibnamefont {Honda}}, \bibinfo {author} {\bibfnamefont
  {O.}~\bibnamefont {Kitao}}, \bibinfo {author} {\bibfnamefont
  {H.}~\bibnamefont {Nakai}}, \bibinfo {author} {\bibfnamefont
  {T.}~\bibnamefont {Vreven}}, \bibinfo {author} {\bibfnamefont
  {K.}~\bibnamefont {Throssell}}, \bibinfo {author} {\bibfnamefont {J.~A.}\
  \bibnamefont {Montgomery}, \bibfnamefont {{Jr.}}}, \bibinfo {author}
  {\bibfnamefont {J.~E.}\ \bibnamefont {Peralta}}, \bibinfo {author}
  {\bibfnamefont {F.}~\bibnamefont {Ogliaro}}, \bibinfo {author} {\bibfnamefont
  {M.~J.}\ \bibnamefont {Bearpark}}, \bibinfo {author} {\bibfnamefont {J.~J.}\
  \bibnamefont {Heyd}}, \bibinfo {author} {\bibfnamefont {E.~N.}\ \bibnamefont
  {Brothers}}, \bibinfo {author} {\bibfnamefont {K.~N.}\ \bibnamefont {Kudin}},
  \bibinfo {author} {\bibfnamefont {V.~N.}\ \bibnamefont {Staroverov}},
  \bibinfo {author} {\bibfnamefont {T.~A.}\ \bibnamefont {Keith}}, \bibinfo
  {author} {\bibfnamefont {R.}~\bibnamefont {Kobayashi}}, \bibinfo {author}
  {\bibfnamefont {J.}~\bibnamefont {Normand}}, \bibinfo {author} {\bibfnamefont
  {K.}~\bibnamefont {Raghavachari}}, \bibinfo {author} {\bibfnamefont {A.~P.}\
  \bibnamefont {Rendell}}, \bibinfo {author} {\bibfnamefont {J.~C.}\
  \bibnamefont {Burant}}, \bibinfo {author} {\bibfnamefont {S.~S.}\
  \bibnamefont {Iyengar}}, \bibinfo {author} {\bibfnamefont {J.}~\bibnamefont
  {Tomasi}}, \bibinfo {author} {\bibfnamefont {M.}~\bibnamefont {Cossi}},
  \bibinfo {author} {\bibfnamefont {J.~M.}\ \bibnamefont {Millam}}, \bibinfo
  {author} {\bibfnamefont {M.}~\bibnamefont {Klene}}, \bibinfo {author}
  {\bibfnamefont {C.}~\bibnamefont {Adamo}}, \bibinfo {author} {\bibfnamefont
  {R.}~\bibnamefont {Cammi}}, \bibinfo {author} {\bibfnamefont {J.~W.}\
  \bibnamefont {Ochterski}}, \bibinfo {author} {\bibfnamefont {R.~L.}\
  \bibnamefont {Martin}}, \bibinfo {author} {\bibfnamefont {K.}~\bibnamefont
  {Morokuma}}, \bibinfo {author} {\bibfnamefont {O.}~\bibnamefont {Farkas}},
  \bibinfo {author} {\bibfnamefont {J.~B.}\ \bibnamefont {Foresman}}, \ and\
  \bibinfo {author} {\bibfnamefont {D.~J.}\ \bibnamefont {Fox}},\ }\href@noop
  {} {\enquote {\bibinfo {title} {Gaussian~16 {R}evision {C}.01},}\ } (\bibinfo
  {year} {2016}),\ \bibinfo {note} {{Gaussian Inc. Wallingford CT}}\BibitemShut
  {NoStop}%
\bibitem [{\citenamefont {Matthews}\ \emph {et~al.}(2020)\citenamefont
  {Matthews}, \citenamefont {Cheng}, \citenamefont {Harding}, \citenamefont
  {Lipparini}, \citenamefont {Stopkowicz}, \citenamefont {Jagau}, \citenamefont
  {Szalay}, \citenamefont {Gauss},\ and\ \citenamefont {Stanton}}]{cfour2020}%
  \BibitemOpen
  \bibfield  {author} {\bibinfo {author} {\bibfnamefont {D.~A.}\ \bibnamefont
  {Matthews}}, \bibinfo {author} {\bibfnamefont {L.}~\bibnamefont {Cheng}},
  \bibinfo {author} {\bibfnamefont {M.~E.}\ \bibnamefont {Harding}}, \bibinfo
  {author} {\bibfnamefont {F.}~\bibnamefont {Lipparini}}, \bibinfo {author}
  {\bibfnamefont {S.}~\bibnamefont {Stopkowicz}}, \bibinfo {author}
  {\bibfnamefont {T.-C.}\ \bibnamefont {Jagau}}, \bibinfo {author}
  {\bibfnamefont {P.~G.}\ \bibnamefont {Szalay}}, \bibinfo {author}
  {\bibfnamefont {J.}~\bibnamefont {Gauss}}, \ and\ \bibinfo {author}
  {\bibfnamefont {J.~F.}\ \bibnamefont {Stanton}},\ }\bibfield  {title}
  {\enquote {\bibinfo {title} {Coupled-cluster techniques for computational
  chemistry: The {CFOUR} program package},}\ }\href@noop {} {\bibfield
  {journal} {\bibinfo  {journal} {J. Chem. Phys.}\ }\textbf {\bibinfo {volume}
  {152}},\ \bibinfo {pages} {214108} (\bibinfo {year} {2020})}\BibitemShut
  {NoStop}%
\bibitem [{\citenamefont {Sun}\ \emph {et~al.}(2020)\citenamefont {Sun},
  \citenamefont {Zhang}, \citenamefont {Banerjee}, \citenamefont {Bao},
  \citenamefont {Barbry}, \citenamefont {Blunt}, \citenamefont {Bogdanov},
  \citenamefont {Booth}, \citenamefont {Chen}, \citenamefont {Cui} \emph
  {et~al.}}]{pyscf2020}%
  \BibitemOpen
  \bibfield  {author} {\bibinfo {author} {\bibfnamefont {Q.}~\bibnamefont
  {Sun}}, \bibinfo {author} {\bibfnamefont {X.}~\bibnamefont {Zhang}}, \bibinfo
  {author} {\bibfnamefont {S.}~\bibnamefont {Banerjee}}, \bibinfo {author}
  {\bibfnamefont {P.}~\bibnamefont {Bao}}, \bibinfo {author} {\bibfnamefont
  {M.}~\bibnamefont {Barbry}}, \bibinfo {author} {\bibfnamefont {N.~S.}\
  \bibnamefont {Blunt}}, \bibinfo {author} {\bibfnamefont {N.~A.}\ \bibnamefont
  {Bogdanov}}, \bibinfo {author} {\bibfnamefont {G.~H.}\ \bibnamefont {Booth}},
  \bibinfo {author} {\bibfnamefont {J.}~\bibnamefont {Chen}}, \bibinfo {author}
  {\bibfnamefont {Z.-H.}\ \bibnamefont {Cui}},  \emph {et~al.},\ }\bibfield
  {title} {\enquote {\bibinfo {title} {Recent developments in the {PySCF}
  program package},}\ }\href@noop {} {\bibfield  {journal} {\bibinfo  {journal}
  {J. Chem. Phys.}\ }\textbf {\bibinfo {volume} {153}},\ \bibinfo {pages}
  {024109} (\bibinfo {year} {2020})}\BibitemShut {NoStop}%
\bibitem [{\citenamefont {Aidas}\ \emph {et~al.}(2014)\citenamefont {Aidas},
  \citenamefont {Angeli}, \citenamefont {Bak}, \citenamefont {Bakken},
  \citenamefont {Bast}, \citenamefont {Boman}, \citenamefont {Christiansen},
  \citenamefont {Cimiraglia}, \citenamefont {Coriani}, \citenamefont {Dahle}
  \emph {et~al.}}]{dalton2014}%
  \BibitemOpen
  \bibfield  {author} {\bibinfo {author} {\bibfnamefont {K.}~\bibnamefont
  {Aidas}}, \bibinfo {author} {\bibfnamefont {C.}~\bibnamefont {Angeli}},
  \bibinfo {author} {\bibfnamefont {K.~L.}\ \bibnamefont {Bak}}, \bibinfo
  {author} {\bibfnamefont {V.}~\bibnamefont {Bakken}}, \bibinfo {author}
  {\bibfnamefont {R.}~\bibnamefont {Bast}}, \bibinfo {author} {\bibfnamefont
  {L.}~\bibnamefont {Boman}}, \bibinfo {author} {\bibfnamefont
  {O.}~\bibnamefont {Christiansen}}, \bibinfo {author} {\bibfnamefont
  {R.}~\bibnamefont {Cimiraglia}}, \bibinfo {author} {\bibfnamefont
  {S.}~\bibnamefont {Coriani}}, \bibinfo {author} {\bibfnamefont
  {P.}~\bibnamefont {Dahle}},  \emph {et~al.},\ }\bibfield  {title} {\enquote
  {\bibinfo {title} {{The Dalton quantum chemistry program system}},}\
  }\href@noop {} {\bibfield  {journal} {\bibinfo  {journal} {{WIREs} Comput.
  Mol. Sci.}\ }\textbf {\bibinfo {volume} {4}},\ \bibinfo {pages} {269--284}
  (\bibinfo {year} {2014})}\BibitemShut {NoStop}%
\bibitem [{\citenamefont {Saue}\ \emph {et~al.}(2020)\citenamefont {Saue},
  \citenamefont {Bast}, \citenamefont {Gomes}, \citenamefont {Jensen},
  \citenamefont {Visscher}, \citenamefont {Aucar}, \citenamefont {Di~Remigio},
  \citenamefont {Dyall}, \citenamefont {Eliav}, \citenamefont {Fasshauer} \emph
  {et~al.}}]{dirac2020}%
  \BibitemOpen
  \bibfield  {author} {\bibinfo {author} {\bibfnamefont {T.}~\bibnamefont
  {Saue}}, \bibinfo {author} {\bibfnamefont {R.}~\bibnamefont {Bast}}, \bibinfo
  {author} {\bibfnamefont {A.~S.~P.}\ \bibnamefont {Gomes}}, \bibinfo {author}
  {\bibfnamefont {H.~J.~A.}\ \bibnamefont {Jensen}}, \bibinfo {author}
  {\bibfnamefont {L.}~\bibnamefont {Visscher}}, \bibinfo {author}
  {\bibfnamefont {I.~A.}\ \bibnamefont {Aucar}}, \bibinfo {author}
  {\bibfnamefont {R.}~\bibnamefont {Di~Remigio}}, \bibinfo {author}
  {\bibfnamefont {K.~G.}\ \bibnamefont {Dyall}}, \bibinfo {author}
  {\bibfnamefont {E.}~\bibnamefont {Eliav}}, \bibinfo {author} {\bibfnamefont
  {E.}~\bibnamefont {Fasshauer}},  \emph {et~al.},\ }\bibfield  {title}
  {\enquote {\bibinfo {title} {{The DIRAC code for relativistic molecular
  calculations}},}\ }\href@noop {} {\bibfield  {journal} {\bibinfo  {journal}
  {J. Chem. Phys.}\ }\textbf {\bibinfo {volume} {152}} (\bibinfo {year}
  {2020})}\BibitemShut {NoStop}%
\bibitem [{\citenamefont {Niemeyer}\ \emph {et~al.}(2023)\citenamefont
  {Niemeyer}, \citenamefont {Eschenbach}, \citenamefont {Bensberg},
  \citenamefont {T{\"o}lle}, \citenamefont {Hellmann}, \citenamefont {Lampe},
  \citenamefont {Massolle}, \citenamefont {Rikus}, \citenamefont {Schnieders},
  \citenamefont {Unsleber} \emph {et~al.}}]{serenity2023}%
  \BibitemOpen
  \bibfield  {author} {\bibinfo {author} {\bibfnamefont {N.}~\bibnamefont
  {Niemeyer}}, \bibinfo {author} {\bibfnamefont {P.}~\bibnamefont
  {Eschenbach}}, \bibinfo {author} {\bibfnamefont {M.}~\bibnamefont
  {Bensberg}}, \bibinfo {author} {\bibfnamefont {J.}~\bibnamefont {T{\"o}lle}},
  \bibinfo {author} {\bibfnamefont {L.}~\bibnamefont {Hellmann}}, \bibinfo
  {author} {\bibfnamefont {L.}~\bibnamefont {Lampe}}, \bibinfo {author}
  {\bibfnamefont {A.}~\bibnamefont {Massolle}}, \bibinfo {author}
  {\bibfnamefont {A.}~\bibnamefont {Rikus}}, \bibinfo {author} {\bibfnamefont
  {D.}~\bibnamefont {Schnieders}}, \bibinfo {author} {\bibfnamefont {J.~P.}\
  \bibnamefont {Unsleber}},  \emph {et~al.},\ }\bibfield  {title} {\enquote
  {\bibinfo {title} {The subsystem quantum chemistry program {Serenity}},}\
  }\href@noop {} {\bibfield  {journal} {\bibinfo  {journal} {WIREs Comput. Mol.
  Sci.}\ }\textbf {\bibinfo {volume} {13}},\ \bibinfo {pages} {e1647} (\bibinfo
  {year} {2023})}\BibitemShut {NoStop}%
\bibitem [{\citenamefont {Zahariev}\ \emph {et~al.}(2023)\citenamefont
  {Zahariev}, \citenamefont {Xu}, \citenamefont {Westheimer}, \citenamefont
  {Webb}, \citenamefont {Galvez~Vallejo}, \citenamefont {Tiwari}, \citenamefont
  {Sundriyal}, \citenamefont {Sosonkina}, \citenamefont {Shen}, \citenamefont
  {Schoendorff} \emph {et~al.}}]{gamess2023}%
  \BibitemOpen
  \bibfield  {author} {\bibinfo {author} {\bibfnamefont {F.}~\bibnamefont
  {Zahariev}}, \bibinfo {author} {\bibfnamefont {P.}~\bibnamefont {Xu}},
  \bibinfo {author} {\bibfnamefont {B.~M.}\ \bibnamefont {Westheimer}},
  \bibinfo {author} {\bibfnamefont {S.}~\bibnamefont {Webb}}, \bibinfo {author}
  {\bibfnamefont {J.}~\bibnamefont {Galvez~Vallejo}}, \bibinfo {author}
  {\bibfnamefont {A.}~\bibnamefont {Tiwari}}, \bibinfo {author} {\bibfnamefont
  {V.}~\bibnamefont {Sundriyal}}, \bibinfo {author} {\bibfnamefont
  {M.}~\bibnamefont {Sosonkina}}, \bibinfo {author} {\bibfnamefont
  {J.}~\bibnamefont {Shen}}, \bibinfo {author} {\bibfnamefont {G.}~\bibnamefont
  {Schoendorff}},  \emph {et~al.},\ }\bibfield  {title} {\enquote {\bibinfo
  {title} {{The General Atomic and Molecular Electronic Structure System
  (GAMESS): Novel Methods on Novel Architectures}},}\ }\href@noop {} {\bibfield
   {journal} {\bibinfo  {journal} {J. Chem. Theory Comput.}\ }\textbf {\bibinfo
  {volume} {19}},\ \bibinfo {pages} {7031--7055} (\bibinfo {year}
  {2023})}\BibitemShut {NoStop}%
\bibitem [{\citenamefont {Balasubramani}\ \emph {et~al.}(2020)\citenamefont
  {Balasubramani}, \citenamefont {Chen}, \citenamefont {Coriani}, \citenamefont
  {Diedenhofen}, \citenamefont {Frank}, \citenamefont {Franzke}, \citenamefont
  {Furche}, \citenamefont {Grotjahn}, \citenamefont {Harding}, \citenamefont
  {H{\"a}ttig} \emph {et~al.}}]{turbomole2020}%
  \BibitemOpen
  \bibfield  {author} {\bibinfo {author} {\bibfnamefont {S.~G.}\ \bibnamefont
  {Balasubramani}}, \bibinfo {author} {\bibfnamefont {G.~P.}\ \bibnamefont
  {Chen}}, \bibinfo {author} {\bibfnamefont {S.}~\bibnamefont {Coriani}},
  \bibinfo {author} {\bibfnamefont {M.}~\bibnamefont {Diedenhofen}}, \bibinfo
  {author} {\bibfnamefont {M.~S.}\ \bibnamefont {Frank}}, \bibinfo {author}
  {\bibfnamefont {Y.~J.}\ \bibnamefont {Franzke}}, \bibinfo {author}
  {\bibfnamefont {F.}~\bibnamefont {Furche}}, \bibinfo {author} {\bibfnamefont
  {R.}~\bibnamefont {Grotjahn}}, \bibinfo {author} {\bibfnamefont {M.~E.}\
  \bibnamefont {Harding}}, \bibinfo {author} {\bibfnamefont {C.}~\bibnamefont
  {H{\"a}ttig}},  \emph {et~al.},\ }\bibfield  {title} {\enquote {\bibinfo
  {title} {{TURBOMOLE}: {Modular} program suite for ab initio quantum-chemical
  and condensed-matter simulations},}\ }\href@noop {} {\bibfield  {journal}
  {\bibinfo  {journal} {J. Chem. Phys.}\ }\textbf {\bibinfo {volume} {152}},\
  \bibinfo {pages} {184107} (\bibinfo {year} {2020})}\BibitemShut {NoStop}%
\bibitem [{\citenamefont {Franzke}\ \emph {et~al.}(2023)\citenamefont
  {Franzke}, \citenamefont {Holzer}, \citenamefont {Andersen}, \citenamefont
  {Begušić}, \citenamefont {Bruder}, \citenamefont {Coriani}, \citenamefont
  {Della~Sala}, \citenamefont {Fabiano}, \citenamefont {Fedotov}, \citenamefont
  {F{\"u}rst}, \citenamefont {Gillhuber}, \citenamefont {Grotjahn},
  \citenamefont {Kaupp}, \citenamefont {Kehry}, \citenamefont {Krstić},
  \citenamefont {Mack}, \citenamefont {Majumdar}, \citenamefont {Nguyen},
  \citenamefont {Parker}, \citenamefont {Pauly}, \citenamefont {Pausch},
  \citenamefont {Perlt}, \citenamefont {Phun}, \citenamefont {Rajabi},
  \citenamefont {Rappoport}, \citenamefont {Samal}, \citenamefont {Schrader},
  \citenamefont {Sharma}, \citenamefont {Tapavicza}, \citenamefont {Tre\ss},
  \citenamefont {Voora}, \citenamefont {Wodyński}, \citenamefont {Yu},
  \citenamefont {Zerulla}, \citenamefont {Furche}, \citenamefont {H{\"a}ttig},
  \citenamefont {Sierka}, \citenamefont {Tew},\ and\ \citenamefont
  {Weigend}}]{turbomole2023}%
  \BibitemOpen
  \bibfield  {author} {\bibinfo {author} {\bibfnamefont {Y.~J.}\ \bibnamefont
  {Franzke}}, \bibinfo {author} {\bibfnamefont {C.}~\bibnamefont {Holzer}},
  \bibinfo {author} {\bibfnamefont {J.~H.}\ \bibnamefont {Andersen}}, \bibinfo
  {author} {\bibfnamefont {T.}~\bibnamefont {Begušić}}, \bibinfo {author}
  {\bibfnamefont {F.}~\bibnamefont {Bruder}}, \bibinfo {author} {\bibfnamefont
  {S.}~\bibnamefont {Coriani}}, \bibinfo {author} {\bibfnamefont
  {F.}~\bibnamefont {Della~Sala}}, \bibinfo {author} {\bibfnamefont
  {E.}~\bibnamefont {Fabiano}}, \bibinfo {author} {\bibfnamefont {D.~A.}\
  \bibnamefont {Fedotov}}, \bibinfo {author} {\bibfnamefont {S.}~\bibnamefont
  {F{\"u}rst}}, \bibinfo {author} {\bibfnamefont {S.}~\bibnamefont
  {Gillhuber}}, \bibinfo {author} {\bibfnamefont {R.}~\bibnamefont {Grotjahn}},
  \bibinfo {author} {\bibfnamefont {M.}~\bibnamefont {Kaupp}}, \bibinfo
  {author} {\bibfnamefont {M.}~\bibnamefont {Kehry}}, \bibinfo {author}
  {\bibfnamefont {M.}~\bibnamefont {Krstić}}, \bibinfo {author} {\bibfnamefont
  {F.}~\bibnamefont {Mack}}, \bibinfo {author} {\bibfnamefont {S.}~\bibnamefont
  {Majumdar}}, \bibinfo {author} {\bibfnamefont {B.~D.}\ \bibnamefont
  {Nguyen}}, \bibinfo {author} {\bibfnamefont {S.~M.}\ \bibnamefont {Parker}},
  \bibinfo {author} {\bibfnamefont {F.}~\bibnamefont {Pauly}}, \bibinfo
  {author} {\bibfnamefont {A.}~\bibnamefont {Pausch}}, \bibinfo {author}
  {\bibfnamefont {E.}~\bibnamefont {Perlt}}, \bibinfo {author} {\bibfnamefont
  {G.~S.}\ \bibnamefont {Phun}}, \bibinfo {author} {\bibfnamefont
  {A.}~\bibnamefont {Rajabi}}, \bibinfo {author} {\bibfnamefont
  {D.}~\bibnamefont {Rappoport}}, \bibinfo {author} {\bibfnamefont
  {B.}~\bibnamefont {Samal}}, \bibinfo {author} {\bibfnamefont
  {T.}~\bibnamefont {Schrader}}, \bibinfo {author} {\bibfnamefont
  {M.}~\bibnamefont {Sharma}}, \bibinfo {author} {\bibfnamefont
  {E.}~\bibnamefont {Tapavicza}}, \bibinfo {author} {\bibfnamefont {R.~S.}\
  \bibnamefont {Tre\ss}}, \bibinfo {author} {\bibfnamefont {V.}~\bibnamefont
  {Voora}}, \bibinfo {author} {\bibfnamefont {A.}~\bibnamefont {Wodyński}},
  \bibinfo {author} {\bibfnamefont {J.~M.}\ \bibnamefont {Yu}}, \bibinfo
  {author} {\bibfnamefont {B.}~\bibnamefont {Zerulla}}, \bibinfo {author}
  {\bibfnamefont {F.}~\bibnamefont {Furche}}, \bibinfo {author} {\bibfnamefont
  {C.}~\bibnamefont {H{\"a}ttig}}, \bibinfo {author} {\bibfnamefont
  {M.}~\bibnamefont {Sierka}}, \bibinfo {author} {\bibfnamefont {D.~P.}\
  \bibnamefont {Tew}}, \ and\ \bibinfo {author} {\bibfnamefont
  {F.}~\bibnamefont {Weigend}},\ }\bibfield  {title} {\enquote {\bibinfo
  {title} {{TURBOMOLE: Today and Tomorrow}},}\ }\href {\doibase
  10.1021/acs.jctc.3c00347} {\bibfield  {journal} {\bibinfo  {journal} {J.
  Chem. Theory Comput.}\ }\textbf {\bibinfo {volume} {19}},\ \bibinfo {pages}
  {6859--6890} (\bibinfo {year} {2023})}\BibitemShut {NoStop}%
\bibitem [{\citenamefont {Seritan}\ \emph {et~al.}(2020)\citenamefont
  {Seritan}, \citenamefont {Bannwarth}, \citenamefont {Fales}, \citenamefont
  {Hohenstein}, \citenamefont {Kokkila-Schumacher}, \citenamefont {Luehr},
  \citenamefont {Snyder}, \citenamefont {Song}, \citenamefont {Titov},
  \citenamefont {Ufimtsev} \emph {et~al.}}]{terachem2020}%
  \BibitemOpen
  \bibfield  {author} {\bibinfo {author} {\bibfnamefont {S.}~\bibnamefont
  {Seritan}}, \bibinfo {author} {\bibfnamefont {C.}~\bibnamefont {Bannwarth}},
  \bibinfo {author} {\bibfnamefont {B.~S.}\ \bibnamefont {Fales}}, \bibinfo
  {author} {\bibfnamefont {E.~G.}\ \bibnamefont {Hohenstein}}, \bibinfo
  {author} {\bibfnamefont {S.~I.}\ \bibnamefont {Kokkila-Schumacher}}, \bibinfo
  {author} {\bibfnamefont {N.}~\bibnamefont {Luehr}}, \bibinfo {author}
  {\bibfnamefont {J.~W.}\ \bibnamefont {Snyder}}, \bibinfo {author}
  {\bibfnamefont {C.}~\bibnamefont {Song}}, \bibinfo {author} {\bibfnamefont
  {A.~V.}\ \bibnamefont {Titov}}, \bibinfo {author} {\bibfnamefont {I.~S.}\
  \bibnamefont {Ufimtsev}},  \emph {et~al.},\ }\bibfield  {title} {\enquote
  {\bibinfo {title} {{TeraChem}: Accelerating electronic structure and ab
  initio molecular dynamics with graphical processing units},}\ }\href@noop {}
  {\bibfield  {journal} {\bibinfo  {journal} {J. Chem. Phys.}\ }\textbf
  {\bibinfo {volume} {152}},\ \bibinfo {pages} {224110} (\bibinfo {year}
  {2020})}\BibitemShut {NoStop}%
\bibitem [{\citenamefont {Tancogne-Dejean}\ \emph {et~al.}(2020)\citenamefont
  {Tancogne-Dejean}, \citenamefont {Oliveira}, \citenamefont {Andrade},
  \citenamefont {Appel}, \citenamefont {Borca}, \citenamefont {Le~Breton},
  \citenamefont {Buchholz}, \citenamefont {Castro}, \citenamefont {Corni},
  \citenamefont {Correa} \emph {et~al.}}]{octopus}%
  \BibitemOpen
  \bibfield  {author} {\bibinfo {author} {\bibfnamefont {N.}~\bibnamefont
  {Tancogne-Dejean}}, \bibinfo {author} {\bibfnamefont {M.~J.}\ \bibnamefont
  {Oliveira}}, \bibinfo {author} {\bibfnamefont {X.}~\bibnamefont {Andrade}},
  \bibinfo {author} {\bibfnamefont {H.}~\bibnamefont {Appel}}, \bibinfo
  {author} {\bibfnamefont {C.~H.}\ \bibnamefont {Borca}}, \bibinfo {author}
  {\bibfnamefont {G.}~\bibnamefont {Le~Breton}}, \bibinfo {author}
  {\bibfnamefont {F.}~\bibnamefont {Buchholz}}, \bibinfo {author}
  {\bibfnamefont {A.}~\bibnamefont {Castro}}, \bibinfo {author} {\bibfnamefont
  {S.}~\bibnamefont {Corni}}, \bibinfo {author} {\bibfnamefont {A.~A.}\
  \bibnamefont {Correa}},  \emph {et~al.},\ }\bibfield  {title} {\enquote
  {\bibinfo {title} {Octopus, a computational framework for exploring
  light-driven phenomena and quantum dynamics in extended and finite
  systems},}\ }\href@noop {} {\bibfield  {journal} {\bibinfo  {journal} {J.
  Chem. Phys.}\ }\textbf {\bibinfo {volume} {152}} (\bibinfo {year}
  {2020})}\BibitemShut {NoStop}%
\bibitem [{\citenamefont {Neese}(2025)}]{orca2025}%
  \BibitemOpen
  \bibfield  {author} {\bibinfo {author} {\bibfnamefont {F.}~\bibnamefont
  {Neese}},\ }\bibfield  {title} {\enquote {\bibinfo {title} {{Software Update:
  The ORCA Program System—Version 6.0}},}\ }\href {\doibase
  https://doi.org/10.1002/wcms.70019} {\bibfield  {journal} {\bibinfo
  {journal} {WIREs Comput. Mol. Sci.}\ }\textbf {\bibinfo {volume} {15}},\
  \bibinfo {pages} {e70019} (\bibinfo {year} {2025})},\ \bibinfo {note} {e70019
  CMS-1186.R1}\BibitemShut {NoStop}%
\bibitem [{\citenamefont {Epifanovsky}\ \emph {et~al.}(2021)\citenamefont
  {Epifanovsky}, \citenamefont {Gilbert}, \citenamefont {Feng}, \citenamefont
  {Lee}, \citenamefont {Mao}, \citenamefont {Mardirossian}, \citenamefont
  {Pokhilko}, \citenamefont {White}, \citenamefont {Coons}, \citenamefont
  {Dempwolff} \emph {et~al.}}]{qchem2021}%
  \BibitemOpen
  \bibfield  {author} {\bibinfo {author} {\bibfnamefont {E.}~\bibnamefont
  {Epifanovsky}}, \bibinfo {author} {\bibfnamefont {A.~T.}\ \bibnamefont
  {Gilbert}}, \bibinfo {author} {\bibfnamefont {X.}~\bibnamefont {Feng}},
  \bibinfo {author} {\bibfnamefont {J.}~\bibnamefont {Lee}}, \bibinfo {author}
  {\bibfnamefont {Y.}~\bibnamefont {Mao}}, \bibinfo {author} {\bibfnamefont
  {N.}~\bibnamefont {Mardirossian}}, \bibinfo {author} {\bibfnamefont
  {P.}~\bibnamefont {Pokhilko}}, \bibinfo {author} {\bibfnamefont {A.~F.}\
  \bibnamefont {White}}, \bibinfo {author} {\bibfnamefont {M.~P.}\ \bibnamefont
  {Coons}}, \bibinfo {author} {\bibfnamefont {A.~L.}\ \bibnamefont
  {Dempwolff}},  \emph {et~al.},\ }\bibfield  {title} {\enquote {\bibinfo
  {title} {Software for the frontiers of quantum chemistry: {A}n overview of
  developments in the {Q-Chem} 5 package},}\ }\href@noop {} {\bibfield
  {journal} {\bibinfo  {journal} {J. Chem. Phys.}\ }\textbf {\bibinfo {volume}
  {155}},\ \bibinfo {pages} {084801} (\bibinfo {year} {2021})}\BibitemShut
  {NoStop}%
\bibitem [{\citenamefont {Kowalski}\ \emph {et~al.}(2021)\citenamefont
  {Kowalski}, \citenamefont {Bair}, \citenamefont {Bauman}, \citenamefont
  {Boschen}, \citenamefont {Bylaska}, \citenamefont {Daily}, \citenamefont
  {De~Jong}, \citenamefont {Dunning~Jr}, \citenamefont {Govind}, \citenamefont
  {Harrison} \emph {et~al.}}]{nwchem2021}%
  \BibitemOpen
  \bibfield  {author} {\bibinfo {author} {\bibfnamefont {K.}~\bibnamefont
  {Kowalski}}, \bibinfo {author} {\bibfnamefont {R.}~\bibnamefont {Bair}},
  \bibinfo {author} {\bibfnamefont {N.~P.}\ \bibnamefont {Bauman}}, \bibinfo
  {author} {\bibfnamefont {J.~S.}\ \bibnamefont {Boschen}}, \bibinfo {author}
  {\bibfnamefont {E.~J.}\ \bibnamefont {Bylaska}}, \bibinfo {author}
  {\bibfnamefont {J.}~\bibnamefont {Daily}}, \bibinfo {author} {\bibfnamefont
  {W.~A.}\ \bibnamefont {De~Jong}}, \bibinfo {author} {\bibfnamefont
  {T.}~\bibnamefont {Dunning~Jr}}, \bibinfo {author} {\bibfnamefont
  {N.}~\bibnamefont {Govind}}, \bibinfo {author} {\bibfnamefont {R.~J.}\
  \bibnamefont {Harrison}},  \emph {et~al.},\ }\bibfield  {title} {\enquote
  {\bibinfo {title} {{From NWChem to NWChemEx: Evolving with the computational
  chemistry landscape}},}\ }\href@noop {} {\bibfield  {journal} {\bibinfo
  {journal} {Chem. Rev.}\ }\textbf {\bibinfo {volume} {121}},\ \bibinfo {pages}
  {4962--4998} (\bibinfo {year} {2021})}\BibitemShut {NoStop}%
\bibitem [{\citenamefont {Smith}\ \emph {et~al.}(2020)\citenamefont {Smith},
  \citenamefont {Burns}, \citenamefont {Simmonett}, \citenamefont {Parrish},
  \citenamefont {Schieber}, \citenamefont {Galvelis}, \citenamefont {Kraus},
  \citenamefont {Kruse}, \citenamefont {Di~Remigio}, \citenamefont {Alenaizan}
  \emph {et~al.}}]{psi42020}%
  \BibitemOpen
  \bibfield  {author} {\bibinfo {author} {\bibfnamefont {D.~G.~A.}\
  \bibnamefont {Smith}}, \bibinfo {author} {\bibfnamefont {L.~A.}\ \bibnamefont
  {Burns}}, \bibinfo {author} {\bibfnamefont {A.~C.}\ \bibnamefont
  {Simmonett}}, \bibinfo {author} {\bibfnamefont {R.~M.}\ \bibnamefont
  {Parrish}}, \bibinfo {author} {\bibfnamefont {M.~C.}\ \bibnamefont
  {Schieber}}, \bibinfo {author} {\bibfnamefont {R.}~\bibnamefont {Galvelis}},
  \bibinfo {author} {\bibfnamefont {P.}~\bibnamefont {Kraus}}, \bibinfo
  {author} {\bibfnamefont {H.}~\bibnamefont {Kruse}}, \bibinfo {author}
  {\bibfnamefont {R.}~\bibnamefont {Di~Remigio}}, \bibinfo {author}
  {\bibfnamefont {A.}~\bibnamefont {Alenaizan}},  \emph {et~al.},\ }\bibfield
  {title} {\enquote {\bibinfo {title} {{PSI4 1.4: Open-source software for
  high-throughput quantum chemistry}},}\ }\href@noop {} {\bibfield  {journal}
  {\bibinfo  {journal} {J. Chem. Phys.}\ }\textbf {\bibinfo {volume} {152}},\
  \bibinfo {pages} {184108} (\bibinfo {year} {2020})}\BibitemShut {NoStop}%
\bibitem [{\citenamefont {Werner}\ \emph {et~al.}(2020)\citenamefont {Werner},
  \citenamefont {Knowles}, \citenamefont {Manby}, \citenamefont {Black},
  \citenamefont {Doll}, \citenamefont {He{\ss}elmann}, \citenamefont {Kats},
  \citenamefont {K{\"o}hn}, \citenamefont {Korona}, \citenamefont {Kreplin}
  \emph {et~al.}}]{molpro2020}%
  \BibitemOpen
  \bibfield  {author} {\bibinfo {author} {\bibfnamefont {H.-J.}\ \bibnamefont
  {Werner}}, \bibinfo {author} {\bibfnamefont {P.~J.}\ \bibnamefont {Knowles}},
  \bibinfo {author} {\bibfnamefont {F.~R.}\ \bibnamefont {Manby}}, \bibinfo
  {author} {\bibfnamefont {J.~A.}\ \bibnamefont {Black}}, \bibinfo {author}
  {\bibfnamefont {K.}~\bibnamefont {Doll}}, \bibinfo {author} {\bibfnamefont
  {A.}~\bibnamefont {He{\ss}elmann}}, \bibinfo {author} {\bibfnamefont
  {D.}~\bibnamefont {Kats}}, \bibinfo {author} {\bibfnamefont {A.}~\bibnamefont
  {K{\"o}hn}}, \bibinfo {author} {\bibfnamefont {T.}~\bibnamefont {Korona}},
  \bibinfo {author} {\bibfnamefont {D.~A.}\ \bibnamefont {Kreplin}},  \emph
  {et~al.},\ }\bibfield  {title} {\enquote {\bibinfo {title} {The {Molpro}
  quantum chemistry package},}\ }\href@noop {} {\bibfield  {journal} {\bibinfo
  {journal} {J. Chem. Phys.}\ }\textbf {\bibinfo {volume} {152}},\ \bibinfo
  {pages} {144107} (\bibinfo {year} {2020})}\BibitemShut {NoStop}%
\bibitem [{\citenamefont {K{\'a}llay}\ \emph {et~al.}(2020)\citenamefont
  {K{\'a}llay}, \citenamefont {Nagy}, \citenamefont {Mester}, \citenamefont
  {Rolik}, \citenamefont {Samu}, \citenamefont {Csontos}, \citenamefont
  {Cs{\'o}ka}, \citenamefont {Szab{\'o}}, \citenamefont {Gyevi-Nagy},
  \citenamefont {H{\'e}gely} \emph {et~al.}}]{mrcc2020}%
  \BibitemOpen
  \bibfield  {author} {\bibinfo {author} {\bibfnamefont {M.}~\bibnamefont
  {K{\'a}llay}}, \bibinfo {author} {\bibfnamefont {P.~R.}\ \bibnamefont
  {Nagy}}, \bibinfo {author} {\bibfnamefont {D.}~\bibnamefont {Mester}},
  \bibinfo {author} {\bibfnamefont {Z.}~\bibnamefont {Rolik}}, \bibinfo
  {author} {\bibfnamefont {G.}~\bibnamefont {Samu}}, \bibinfo {author}
  {\bibfnamefont {J.}~\bibnamefont {Csontos}}, \bibinfo {author} {\bibfnamefont
  {J.}~\bibnamefont {Cs{\'o}ka}}, \bibinfo {author} {\bibfnamefont {P.~B.}\
  \bibnamefont {Szab{\'o}}}, \bibinfo {author} {\bibfnamefont {L.}~\bibnamefont
  {Gyevi-Nagy}}, \bibinfo {author} {\bibfnamefont {B.}~\bibnamefont
  {H{\'e}gely}},  \emph {et~al.},\ }\bibfield  {title} {\enquote {\bibinfo
  {title} {The {MRCC} program system: Accurate quantum chemistry from water to
  proteins},}\ }\href@noop {} {\bibfield  {journal} {\bibinfo  {journal} {J.
  Chem. Phys.}\ }\textbf {\bibinfo {volume} {152}},\ \bibinfo {pages} {074107}
  (\bibinfo {year} {2020})}\BibitemShut {NoStop}%
\bibitem [{\citenamefont {Aquilante}\ \emph {et~al.}(2020)\citenamefont
  {Aquilante}, \citenamefont {Autschbach}, \citenamefont {Baiardi},
  \citenamefont {Battaglia}, \citenamefont {Borin}, \citenamefont {Chibotaru},
  \citenamefont {Conti}, \citenamefont {De~Vico}, \citenamefont {Delcey},
  \citenamefont {Ferr{\'e}} \emph {et~al.}}]{molcas2020}%
  \BibitemOpen
  \bibfield  {author} {\bibinfo {author} {\bibfnamefont {F.}~\bibnamefont
  {Aquilante}}, \bibinfo {author} {\bibfnamefont {J.}~\bibnamefont
  {Autschbach}}, \bibinfo {author} {\bibfnamefont {A.}~\bibnamefont {Baiardi}},
  \bibinfo {author} {\bibfnamefont {S.}~\bibnamefont {Battaglia}}, \bibinfo
  {author} {\bibfnamefont {V.~A.}\ \bibnamefont {Borin}}, \bibinfo {author}
  {\bibfnamefont {L.~F.}\ \bibnamefont {Chibotaru}}, \bibinfo {author}
  {\bibfnamefont {I.}~\bibnamefont {Conti}}, \bibinfo {author} {\bibfnamefont
  {L.}~\bibnamefont {De~Vico}}, \bibinfo {author} {\bibfnamefont
  {M.}~\bibnamefont {Delcey}}, \bibinfo {author} {\bibfnamefont
  {N.}~\bibnamefont {Ferr{\'e}}},  \emph {et~al.},\ }\bibfield  {title}
  {\enquote {\bibinfo {title} {Modern quantum chemistry with [{O}pen]
  {M}olcas},}\ }\href@noop {} {\bibfield  {journal} {\bibinfo  {journal} {J.
  Chem. Phys.}\ }\textbf {\bibinfo {volume} {152}},\ \bibinfo {pages} {214117}
  (\bibinfo {year} {2020})}\BibitemShut {NoStop}%
\bibitem [{\citenamefont {Li~Manni}\ \emph {et~al.}(2023)\citenamefont
  {Li~Manni}, \citenamefont {Fdez.~Galván}, \citenamefont {Alavi},
  \citenamefont {Aleotti}, \citenamefont {Aquilante}, \citenamefont
  {Autschbach}, \citenamefont {Avagliano}, \citenamefont {Baiardi},
  \citenamefont {Bao}, \citenamefont {Battaglia}, \citenamefont {Birnoschi},
  \citenamefont {Blanco-González}, \citenamefont {Bokarev}, \citenamefont
  {Broer}, \citenamefont {Cacciari}, \citenamefont {Calio}, \citenamefont
  {Carlson}, \citenamefont {Carvalho~Couto}, \citenamefont {Cerdán},
  \citenamefont {Chibotaru}, \citenamefont {Chilton}, \citenamefont {Church},
  \citenamefont {Conti}, \citenamefont {Coriani}, \citenamefont
  {Cu{\'e}llar-Zuquin}, \citenamefont {Daoud}, \citenamefont {Dattani},
  \citenamefont {Decleva}, \citenamefont {de~Graaf}, \citenamefont {Delcey},
  \citenamefont {De~Vico}, \citenamefont {Dobrautz}, \citenamefont {Dong},
  \citenamefont {Feng}, \citenamefont {Ferr{\'e}}, \citenamefont
  {Filatov(Gulak)}, \citenamefont {Gagliardi}, \citenamefont {Garavelli},
  \citenamefont {González}, \citenamefont {Guan}, \citenamefont {Guo},
  \citenamefont {Hennefarth}, \citenamefont {Hermes}, \citenamefont {Hoyer},
  \citenamefont {Huix-Rotllant}, \citenamefont {Jaiswal}, \citenamefont
  {Kaiser}, \citenamefont {Kaliakin}, \citenamefont {Khamesian}, \citenamefont
  {King}, \citenamefont {Kochetov}, \citenamefont {Krośnicki}, \citenamefont
  {Kumaar}, \citenamefont {Larsson}, \citenamefont {Lehtola}, \citenamefont
  {Lepetit}, \citenamefont {Lischka}, \citenamefont {López~Ríos},
  \citenamefont {Lundberg}, \citenamefont {Ma}, \citenamefont {Mai},
  \citenamefont {Marquetand}, \citenamefont {Merritt}, \citenamefont
  {Montorsi}, \citenamefont {M{\"o}rchen}, \citenamefont {Nenov}, \citenamefont
  {Nguyen}, \citenamefont {Nishimoto}, \citenamefont {Oakley}, \citenamefont
  {Olivucci}, \citenamefont {Oppel}, \citenamefont {Padula}, \citenamefont
  {Pandharkar}, \citenamefont {Phung}, \citenamefont {Plasser}, \citenamefont
  {Raggi}, \citenamefont {Rebolini}, \citenamefont {Reiher}, \citenamefont
  {Rivalta}, \citenamefont {Roca-Sanjuán}, \citenamefont {Romig},
  \citenamefont {Safari}, \citenamefont {Sánchez-Mansilla}, \citenamefont
  {Sand}, \citenamefont {Schapiro}, \citenamefont {Scott}, \citenamefont
  {Segarra-Martí}, \citenamefont {Segatta}, \citenamefont {Sergentu},
  \citenamefont {Sharma}, \citenamefont {Shepard}, \citenamefont {Shu},
  \citenamefont {Staab}, \citenamefont {Straatsma}, \citenamefont {Sørensen},
  \citenamefont {Tenorio}, \citenamefont {Truhlar}, \citenamefont {Ungur},
  \citenamefont {Vacher}, \citenamefont {Veryazov}, \citenamefont {Voß},
  \citenamefont {Weser}, \citenamefont {Wu}, \citenamefont {Yang},
  \citenamefont {Yarkony}, \citenamefont {Zhou}, \citenamefont {Zobel},\ and\
  \citenamefont {Lindh}}]{OpenMolcas2023}%
  \BibitemOpen
  \bibfield  {author} {\bibinfo {author} {\bibfnamefont {G.}~\bibnamefont
  {Li~Manni}}, \bibinfo {author} {\bibfnamefont {I.}~\bibnamefont
  {Fdez.~Galván}}, \bibinfo {author} {\bibfnamefont {A.}~\bibnamefont
  {Alavi}}, \bibinfo {author} {\bibfnamefont {F.}~\bibnamefont {Aleotti}},
  \bibinfo {author} {\bibfnamefont {F.}~\bibnamefont {Aquilante}}, \bibinfo
  {author} {\bibfnamefont {J.}~\bibnamefont {Autschbach}}, \bibinfo {author}
  {\bibfnamefont {D.}~\bibnamefont {Avagliano}}, \bibinfo {author}
  {\bibfnamefont {A.}~\bibnamefont {Baiardi}}, \bibinfo {author} {\bibfnamefont
  {J.~J.}\ \bibnamefont {Bao}}, \bibinfo {author} {\bibfnamefont
  {S.}~\bibnamefont {Battaglia}}, \bibinfo {author} {\bibfnamefont
  {L.}~\bibnamefont {Birnoschi}}, \bibinfo {author} {\bibfnamefont
  {A.}~\bibnamefont {Blanco-González}}, \bibinfo {author} {\bibfnamefont
  {S.~I.}\ \bibnamefont {Bokarev}}, \bibinfo {author} {\bibfnamefont
  {R.}~\bibnamefont {Broer}}, \bibinfo {author} {\bibfnamefont
  {R.}~\bibnamefont {Cacciari}}, \bibinfo {author} {\bibfnamefont {P.~B.}\
  \bibnamefont {Calio}}, \bibinfo {author} {\bibfnamefont {R.~K.}\ \bibnamefont
  {Carlson}}, \bibinfo {author} {\bibfnamefont {R.}~\bibnamefont
  {Carvalho~Couto}}, \bibinfo {author} {\bibfnamefont {L.}~\bibnamefont
  {Cerdán}}, \bibinfo {author} {\bibfnamefont {L.~F.}\ \bibnamefont
  {Chibotaru}}, \bibinfo {author} {\bibfnamefont {N.~F.}\ \bibnamefont
  {Chilton}}, \bibinfo {author} {\bibfnamefont {J.~R.}\ \bibnamefont {Church}},
  \bibinfo {author} {\bibfnamefont {I.}~\bibnamefont {Conti}}, \bibinfo
  {author} {\bibfnamefont {S.}~\bibnamefont {Coriani}}, \bibinfo {author}
  {\bibfnamefont {J.}~\bibnamefont {Cu{\'e}llar-Zuquin}}, \bibinfo {author}
  {\bibfnamefont {R.~E.}\ \bibnamefont {Daoud}}, \bibinfo {author}
  {\bibfnamefont {N.}~\bibnamefont {Dattani}}, \bibinfo {author} {\bibfnamefont
  {P.}~\bibnamefont {Decleva}}, \bibinfo {author} {\bibfnamefont
  {C.}~\bibnamefont {de~Graaf}}, \bibinfo {author} {\bibfnamefont {M.~G.}\
  \bibnamefont {Delcey}}, \bibinfo {author} {\bibfnamefont {L.}~\bibnamefont
  {De~Vico}}, \bibinfo {author} {\bibfnamefont {W.}~\bibnamefont {Dobrautz}},
  \bibinfo {author} {\bibfnamefont {S.~S.}\ \bibnamefont {Dong}}, \bibinfo
  {author} {\bibfnamefont {R.}~\bibnamefont {Feng}}, \bibinfo {author}
  {\bibfnamefont {N.}~\bibnamefont {Ferr{\'e}}}, \bibinfo {author}
  {\bibfnamefont {M.}~\bibnamefont {Filatov(Gulak)}}, \bibinfo {author}
  {\bibfnamefont {L.}~\bibnamefont {Gagliardi}}, \bibinfo {author}
  {\bibfnamefont {M.}~\bibnamefont {Garavelli}}, \bibinfo {author}
  {\bibfnamefont {L.}~\bibnamefont {González}}, \bibinfo {author}
  {\bibfnamefont {Y.}~\bibnamefont {Guan}}, \bibinfo {author} {\bibfnamefont
  {M.}~\bibnamefont {Guo}}, \bibinfo {author} {\bibfnamefont {M.~R.}\
  \bibnamefont {Hennefarth}}, \bibinfo {author} {\bibfnamefont {M.~R.}\
  \bibnamefont {Hermes}}, \bibinfo {author} {\bibfnamefont {C.~E.}\
  \bibnamefont {Hoyer}}, \bibinfo {author} {\bibfnamefont {M.}~\bibnamefont
  {Huix-Rotllant}}, \bibinfo {author} {\bibfnamefont {V.~K.}\ \bibnamefont
  {Jaiswal}}, \bibinfo {author} {\bibfnamefont {A.}~\bibnamefont {Kaiser}},
  \bibinfo {author} {\bibfnamefont {D.~S.}\ \bibnamefont {Kaliakin}}, \bibinfo
  {author} {\bibfnamefont {M.}~\bibnamefont {Khamesian}}, \bibinfo {author}
  {\bibfnamefont {D.~S.}\ \bibnamefont {King}}, \bibinfo {author}
  {\bibfnamefont {V.}~\bibnamefont {Kochetov}}, \bibinfo {author}
  {\bibfnamefont {M.}~\bibnamefont {Krośnicki}}, \bibinfo {author}
  {\bibfnamefont {A.~A.}\ \bibnamefont {Kumaar}}, \bibinfo {author}
  {\bibfnamefont {E.~D.}\ \bibnamefont {Larsson}}, \bibinfo {author}
  {\bibfnamefont {S.}~\bibnamefont {Lehtola}}, \bibinfo {author} {\bibfnamefont
  {M.-B.}\ \bibnamefont {Lepetit}}, \bibinfo {author} {\bibfnamefont
  {H.}~\bibnamefont {Lischka}}, \bibinfo {author} {\bibfnamefont
  {P.}~\bibnamefont {López~Ríos}}, \bibinfo {author} {\bibfnamefont
  {M.}~\bibnamefont {Lundberg}}, \bibinfo {author} {\bibfnamefont
  {D.}~\bibnamefont {Ma}}, \bibinfo {author} {\bibfnamefont {S.}~\bibnamefont
  {Mai}}, \bibinfo {author} {\bibfnamefont {P.}~\bibnamefont {Marquetand}},
  \bibinfo {author} {\bibfnamefont {I.~C.~D.}\ \bibnamefont {Merritt}},
  \bibinfo {author} {\bibfnamefont {F.}~\bibnamefont {Montorsi}}, \bibinfo
  {author} {\bibfnamefont {M.}~\bibnamefont {M{\"o}rchen}}, \bibinfo {author}
  {\bibfnamefont {A.}~\bibnamefont {Nenov}}, \bibinfo {author} {\bibfnamefont
  {V.~H.~A.}\ \bibnamefont {Nguyen}}, \bibinfo {author} {\bibfnamefont
  {Y.}~\bibnamefont {Nishimoto}}, \bibinfo {author} {\bibfnamefont {M.~S.}\
  \bibnamefont {Oakley}}, \bibinfo {author} {\bibfnamefont {M.}~\bibnamefont
  {Olivucci}}, \bibinfo {author} {\bibfnamefont {M.}~\bibnamefont {Oppel}},
  \bibinfo {author} {\bibfnamefont {D.}~\bibnamefont {Padula}}, \bibinfo
  {author} {\bibfnamefont {R.}~\bibnamefont {Pandharkar}}, \bibinfo {author}
  {\bibfnamefont {Q.~M.}\ \bibnamefont {Phung}}, \bibinfo {author}
  {\bibfnamefont {F.}~\bibnamefont {Plasser}}, \bibinfo {author} {\bibfnamefont
  {G.}~\bibnamefont {Raggi}}, \bibinfo {author} {\bibfnamefont
  {E.}~\bibnamefont {Rebolini}}, \bibinfo {author} {\bibfnamefont
  {M.}~\bibnamefont {Reiher}}, \bibinfo {author} {\bibfnamefont
  {I.}~\bibnamefont {Rivalta}}, \bibinfo {author} {\bibfnamefont
  {D.}~\bibnamefont {Roca-Sanjuán}}, \bibinfo {author} {\bibfnamefont
  {T.}~\bibnamefont {Romig}}, \bibinfo {author} {\bibfnamefont {A.~A.}\
  \bibnamefont {Safari}}, \bibinfo {author} {\bibfnamefont {A.}~\bibnamefont
  {Sánchez-Mansilla}}, \bibinfo {author} {\bibfnamefont {A.~M.}\ \bibnamefont
  {Sand}}, \bibinfo {author} {\bibfnamefont {I.}~\bibnamefont {Schapiro}},
  \bibinfo {author} {\bibfnamefont {T.~R.}\ \bibnamefont {Scott}}, \bibinfo
  {author} {\bibfnamefont {J.}~\bibnamefont {Segarra-Martí}}, \bibinfo
  {author} {\bibfnamefont {F.}~\bibnamefont {Segatta}}, \bibinfo {author}
  {\bibfnamefont {D.-C.}\ \bibnamefont {Sergentu}}, \bibinfo {author}
  {\bibfnamefont {P.}~\bibnamefont {Sharma}}, \bibinfo {author} {\bibfnamefont
  {R.}~\bibnamefont {Shepard}}, \bibinfo {author} {\bibfnamefont
  {Y.}~\bibnamefont {Shu}}, \bibinfo {author} {\bibfnamefont {J.~K.}\
  \bibnamefont {Staab}}, \bibinfo {author} {\bibfnamefont {T.~P.}\ \bibnamefont
  {Straatsma}}, \bibinfo {author} {\bibfnamefont {L.~K.}\ \bibnamefont
  {Sørensen}}, \bibinfo {author} {\bibfnamefont {B.~N.~C.}\ \bibnamefont
  {Tenorio}}, \bibinfo {author} {\bibfnamefont {D.~G.}\ \bibnamefont
  {Truhlar}}, \bibinfo {author} {\bibfnamefont {L.}~\bibnamefont {Ungur}},
  \bibinfo {author} {\bibfnamefont {M.}~\bibnamefont {Vacher}}, \bibinfo
  {author} {\bibfnamefont {V.}~\bibnamefont {Veryazov}}, \bibinfo {author}
  {\bibfnamefont {T.~A.}\ \bibnamefont {Voß}}, \bibinfo {author}
  {\bibfnamefont {O.}~\bibnamefont {Weser}}, \bibinfo {author} {\bibfnamefont
  {D.}~\bibnamefont {Wu}}, \bibinfo {author} {\bibfnamefont {X.}~\bibnamefont
  {Yang}}, \bibinfo {author} {\bibfnamefont {D.}~\bibnamefont {Yarkony}},
  \bibinfo {author} {\bibfnamefont {C.}~\bibnamefont {Zhou}}, \bibinfo {author}
  {\bibfnamefont {J.~P.}\ \bibnamefont {Zobel}}, \ and\ \bibinfo {author}
  {\bibfnamefont {R.}~\bibnamefont {Lindh}},\ }\bibfield  {title} {\enquote
  {\bibinfo {title} {{The OpenMolcas Web: A Community-Driven Approach to
  Advancing Computational Chemistry}},}\ }\href {\doibase
  10.1021/acs.jctc.3c00182} {\bibfield  {journal} {\bibinfo  {journal} {J.
  Chem. Theory Comput.}\ }\textbf {\bibinfo {volume} {19}},\ \bibinfo {pages}
  {6933--6991} (\bibinfo {year} {2023})}\BibitemShut {NoStop}%
\bibitem [{\citenamefont {Rinkevicius}\ \emph {et~al.}(2020)\citenamefont
  {Rinkevicius}, \citenamefont {Li}, \citenamefont {Vahtras}, \citenamefont
  {Ahmadzadeh}, \citenamefont {Brand}, \citenamefont {Ringholm}, \citenamefont
  {List}, \citenamefont {Scheurer}, \citenamefont {Scott}, \citenamefont
  {Dreuw} \emph {et~al.}}]{veloxchem2020}%
  \BibitemOpen
  \bibfield  {author} {\bibinfo {author} {\bibfnamefont {Z.}~\bibnamefont
  {Rinkevicius}}, \bibinfo {author} {\bibfnamefont {X.}~\bibnamefont {Li}},
  \bibinfo {author} {\bibfnamefont {O.}~\bibnamefont {Vahtras}}, \bibinfo
  {author} {\bibfnamefont {K.}~\bibnamefont {Ahmadzadeh}}, \bibinfo {author}
  {\bibfnamefont {M.}~\bibnamefont {Brand}}, \bibinfo {author} {\bibfnamefont
  {M.}~\bibnamefont {Ringholm}}, \bibinfo {author} {\bibfnamefont {N.~H.}\
  \bibnamefont {List}}, \bibinfo {author} {\bibfnamefont {M.}~\bibnamefont
  {Scheurer}}, \bibinfo {author} {\bibfnamefont {M.}~\bibnamefont {Scott}},
  \bibinfo {author} {\bibfnamefont {A.}~\bibnamefont {Dreuw}},  \emph
  {et~al.},\ }\bibfield  {title} {\enquote {\bibinfo {title} {{VeloxChem}: A
  {P}ython-driven density-functional theory program for spectroscopy
  simulations in high-performance computing environments},}\ }\href@noop {}
  {\bibfield  {journal} {\bibinfo  {journal} {WIREs Comput. Mol. Sci.}\
  }\textbf {\bibinfo {volume} {10}},\ \bibinfo {pages} {e1457} (\bibinfo {year}
  {2020})}\BibitemShut {NoStop}%
\bibitem [{\citenamefont {Boguslawski}\ \emph {et~al.}(2024)\citenamefont
  {Boguslawski}, \citenamefont {Brz{\k{e}}k}, \citenamefont {Chakraborty},
  \citenamefont {Cie{\'s}lak}, \citenamefont {Jahani}, \citenamefont
  {Leszczyk}, \citenamefont {Nowak}, \citenamefont {Sujkowski}, \citenamefont
  {{\'S}wierczy{\'n}ski}, \citenamefont {Ahmadkhani} \emph
  {et~al.}}]{pybest2024}%
  \BibitemOpen
  \bibfield  {author} {\bibinfo {author} {\bibfnamefont {K.}~\bibnamefont
  {Boguslawski}}, \bibinfo {author} {\bibfnamefont {F.}~\bibnamefont
  {Brz{\k{e}}k}}, \bibinfo {author} {\bibfnamefont {R.}~\bibnamefont
  {Chakraborty}}, \bibinfo {author} {\bibfnamefont {K.}~\bibnamefont
  {Cie{\'s}lak}}, \bibinfo {author} {\bibfnamefont {S.}~\bibnamefont {Jahani}},
  \bibinfo {author} {\bibfnamefont {A.}~\bibnamefont {Leszczyk}}, \bibinfo
  {author} {\bibfnamefont {A.}~\bibnamefont {Nowak}}, \bibinfo {author}
  {\bibfnamefont {E.}~\bibnamefont {Sujkowski}}, \bibinfo {author}
  {\bibfnamefont {J.}~\bibnamefont {{\'S}wierczy{\'n}ski}}, \bibinfo {author}
  {\bibfnamefont {S.}~\bibnamefont {Ahmadkhani}},  \emph {et~al.},\ }\bibfield
  {title} {\enquote {\bibinfo {title} {{PyBEST}: Improved functionality and
  enhanced performance},}\ }\href@noop {} {\bibfield  {journal} {\bibinfo
  {journal} {Comput. Phys. Commun.}\ }\textbf {\bibinfo {volume} {297}},\
  \bibinfo {pages} {109049} (\bibinfo {year} {2024})}\BibitemShut {NoStop}%
\bibitem [{\citenamefont {Di~Felice}\ \emph {et~al.}(2023)\citenamefont
  {Di~Felice}, \citenamefont {Mayes}, \citenamefont {Richard}, \citenamefont
  {Williams-Young}, \citenamefont {Chan}, \citenamefont {de~Jong},
  \citenamefont {Govind}, \citenamefont {Head-Gordon}, \citenamefont {Hermes},
  \citenamefont {Kowalski} \emph {et~al.}}]{di2023perspective}%
  \BibitemOpen
  \bibfield  {author} {\bibinfo {author} {\bibfnamefont {R.}~\bibnamefont
  {Di~Felice}}, \bibinfo {author} {\bibfnamefont {M.~L.}\ \bibnamefont
  {Mayes}}, \bibinfo {author} {\bibfnamefont {R.~M.}\ \bibnamefont {Richard}},
  \bibinfo {author} {\bibfnamefont {D.~B.}\ \bibnamefont {Williams-Young}},
  \bibinfo {author} {\bibfnamefont {G.~K.-L.}\ \bibnamefont {Chan}}, \bibinfo
  {author} {\bibfnamefont {W.~A.}\ \bibnamefont {de~Jong}}, \bibinfo {author}
  {\bibfnamefont {N.}~\bibnamefont {Govind}}, \bibinfo {author} {\bibfnamefont
  {M.}~\bibnamefont {Head-Gordon}}, \bibinfo {author} {\bibfnamefont {M.~R.}\
  \bibnamefont {Hermes}}, \bibinfo {author} {\bibfnamefont {K.}~\bibnamefont
  {Kowalski}},  \emph {et~al.},\ }\bibfield  {title} {\enquote {\bibinfo
  {title} {A perspective on sustainable computational chemistry software
  development and integration},}\ }\href@noop {} {\bibfield  {journal}
  {\bibinfo  {journal} {J. Chem. Theory Comput.}\ }\textbf {\bibinfo {volume}
  {19}},\ \bibinfo {pages} {7056--7076} (\bibinfo {year} {2023})}\BibitemShut
  {NoStop}%
\bibitem [{\citenamefont {Krylov}\ \emph {et~al.}(2015)\citenamefont {Krylov},
  \citenamefont {Herbert}, \citenamefont {Furche}, \citenamefont {Head-Gordon},
  \citenamefont {Knowles}, \citenamefont {Lindh}, \citenamefont {Manby},
  \citenamefont {Pulay}, \citenamefont {Skylaris},\ and\ \citenamefont
  {Werner}}]{Krylov2015}%
  \BibitemOpen
  \bibfield  {author} {\bibinfo {author} {\bibfnamefont {A.~I.}\ \bibnamefont
  {Krylov}}, \bibinfo {author} {\bibfnamefont {J.~M.}\ \bibnamefont {Herbert}},
  \bibinfo {author} {\bibfnamefont {F.}~\bibnamefont {Furche}}, \bibinfo
  {author} {\bibfnamefont {M.}~\bibnamefont {Head-Gordon}}, \bibinfo {author}
  {\bibfnamefont {P.~J.}\ \bibnamefont {Knowles}}, \bibinfo {author}
  {\bibfnamefont {R.}~\bibnamefont {Lindh}}, \bibinfo {author} {\bibfnamefont
  {F.~R.}\ \bibnamefont {Manby}}, \bibinfo {author} {\bibfnamefont
  {P.}~\bibnamefont {Pulay}}, \bibinfo {author} {\bibfnamefont {C.-K.}\
  \bibnamefont {Skylaris}}, \ and\ \bibinfo {author} {\bibfnamefont {H.-J.}\
  \bibnamefont {Werner}},\ }\bibfield  {title} {\enquote {\bibinfo {title}
  {What is the price of open-source software?}}\ }\href {\doibase
  10.1021/acs.jpclett.5b01258} {\bibfield  {journal} {\bibinfo  {journal} {J.
  Phys. Chem. Lett.}\ }\textbf {\bibinfo {volume} {6}},\ \bibinfo {pages}
  {2751--2754} (\bibinfo {year} {2015})}\BibitemShut {NoStop}%
\bibitem [{\citenamefont {Lehtola}\ and\ \citenamefont
  {Karttunen}(2022)}]{lehtola2022free}%
  \BibitemOpen
  \bibfield  {author} {\bibinfo {author} {\bibfnamefont {S.}~\bibnamefont
  {Lehtola}}\ and\ \bibinfo {author} {\bibfnamefont {A.~J.}\ \bibnamefont
  {Karttunen}},\ }\bibfield  {title} {\enquote {\bibinfo {title} {Free and open
  source software for computational chemistry education},}\ }\href@noop {}
  {\bibfield  {journal} {\bibinfo  {journal} {WIREs Comput. Mol. Sci.}\
  }\textbf {\bibinfo {volume} {12}},\ \bibinfo {pages} {e1610} (\bibinfo {year}
  {2022})}\BibitemShut {NoStop}%
\bibitem [{\citenamefont {Kj{\o}nstad}\ \emph {et~al.}(2024)\citenamefont
  {Kj{\o}nstad}, \citenamefont {Fajen}, \citenamefont {Paul}, \citenamefont
  {Angelico}, \citenamefont {Mayer}, \citenamefont {G{\"u}hr}, \citenamefont
  {Wolf}, \citenamefont {Mart{\'\i}nez},\ and\ \citenamefont
  {Koch}}]{kjonstad2024photoinduced}%
  \BibitemOpen
  \bibfield  {author} {\bibinfo {author} {\bibfnamefont {E.~F.}\ \bibnamefont
  {Kj{\o}nstad}}, \bibinfo {author} {\bibfnamefont {O.~J.}\ \bibnamefont
  {Fajen}}, \bibinfo {author} {\bibfnamefont {A.~C.}\ \bibnamefont {Paul}},
  \bibinfo {author} {\bibfnamefont {S.}~\bibnamefont {Angelico}}, \bibinfo
  {author} {\bibfnamefont {D.}~\bibnamefont {Mayer}}, \bibinfo {author}
  {\bibfnamefont {M.}~\bibnamefont {G{\"u}hr}}, \bibinfo {author}
  {\bibfnamefont {T.~J.}\ \bibnamefont {Wolf}}, \bibinfo {author}
  {\bibfnamefont {T.~J.}\ \bibnamefont {Mart{\'\i}nez}}, \ and\ \bibinfo
  {author} {\bibfnamefont {H.}~\bibnamefont {Koch}},\ }\bibfield  {title}
  {\enquote {\bibinfo {title} {Photoinduced hydrogen dissociation in thymine
  predicted by coupled cluster theory},}\ }\href@noop {} {\bibfield  {journal}
  {\bibinfo  {journal} {Nat. Commun.}\ }\textbf {\bibinfo {volume} {15}},\
  \bibinfo {pages} {10128} (\bibinfo {year} {2024})}\BibitemShut {NoStop}%
\bibitem [{\citenamefont {Folkestad}\ \emph {et~al.}(2020)\citenamefont
  {Folkestad}, \citenamefont {Kjønstad}, \citenamefont {Myhre}, \citenamefont
  {Andersen}, \citenamefont {Balbi}, \citenamefont {Coriani}, \citenamefont
  {Giovannini}, \citenamefont {Goletto}, \citenamefont {Haugland},
  \citenamefont {Hutcheson}, \citenamefont {Høyvik}, \citenamefont {Moitra},
  \citenamefont {Paul}, \citenamefont {Scavino}, \citenamefont {Skeidsvoll},
  \citenamefont {Tveten},\ and\ \citenamefont {Koch}}]{eTprog}%
  \BibitemOpen
  \bibfield  {author} {\bibinfo {author} {\bibfnamefont {S.~D.}\ \bibnamefont
  {Folkestad}}, \bibinfo {author} {\bibfnamefont {E.~F.}\ \bibnamefont
  {Kjønstad}}, \bibinfo {author} {\bibfnamefont {R.~H.}\ \bibnamefont
  {Myhre}}, \bibinfo {author} {\bibfnamefont {J.~H.}\ \bibnamefont {Andersen}},
  \bibinfo {author} {\bibfnamefont {A.}~\bibnamefont {Balbi}}, \bibinfo
  {author} {\bibfnamefont {S.}~\bibnamefont {Coriani}}, \bibinfo {author}
  {\bibfnamefont {T.}~\bibnamefont {Giovannini}}, \bibinfo {author}
  {\bibfnamefont {L.}~\bibnamefont {Goletto}}, \bibinfo {author} {\bibfnamefont
  {T.~S.}\ \bibnamefont {Haugland}}, \bibinfo {author} {\bibfnamefont
  {A.}~\bibnamefont {Hutcheson}}, \bibinfo {author} {\bibfnamefont {I.-M.}\
  \bibnamefont {Høyvik}}, \bibinfo {author} {\bibfnamefont {T.}~\bibnamefont
  {Moitra}}, \bibinfo {author} {\bibfnamefont {A.~C.}\ \bibnamefont {Paul}},
  \bibinfo {author} {\bibfnamefont {M.}~\bibnamefont {Scavino}}, \bibinfo
  {author} {\bibfnamefont {A.~S.}\ \bibnamefont {Skeidsvoll}}, \bibinfo
  {author} {\bibfnamefont {{\AA}.~H.}\ \bibnamefont {Tveten}}, \ and\ \bibinfo
  {author} {\bibfnamefont {H.}~\bibnamefont {Koch}},\ }\bibfield  {title}
  {\enquote {\bibinfo {title} {{eT 1.0: An open source electronic structure
  program with emphasis on coupled cluster and multilevel methods}},}\
  }\href@noop {} {\bibfield  {journal} {\bibinfo  {journal} {J. Chem. Phys.}\
  }\textbf {\bibinfo {volume} {152}},\ \bibinfo {pages} {184103} (\bibinfo
  {year} {2020})}\BibitemShut {NoStop}%
\bibitem [{\citenamefont {Haugland}\ \emph {et~al.}(2020)\citenamefont
  {Haugland}, \citenamefont {Ronca}, \citenamefont {Kj{\o}nstad}, \citenamefont
  {Rubio},\ and\ \citenamefont {Koch}}]{haugland2020coupled}%
  \BibitemOpen
  \bibfield  {author} {\bibinfo {author} {\bibfnamefont {T.~S.}\ \bibnamefont
  {Haugland}}, \bibinfo {author} {\bibfnamefont {E.}~\bibnamefont {Ronca}},
  \bibinfo {author} {\bibfnamefont {E.~F.}\ \bibnamefont {Kj{\o}nstad}},
  \bibinfo {author} {\bibfnamefont {A.}~\bibnamefont {Rubio}}, \ and\ \bibinfo
  {author} {\bibfnamefont {H.}~\bibnamefont {Koch}},\ }\bibfield  {title}
  {\enquote {\bibinfo {title} {Coupled cluster theory for molecular polaritons:
  {C}hanging ground and excited states},}\ }\href@noop {} {\bibfield  {journal}
  {\bibinfo  {journal} {Phys. Rev. X}\ }\textbf {\bibinfo {volume} {10}},\
  \bibinfo {pages} {041043} (\bibinfo {year} {2020})}\BibitemShut {NoStop}%
\bibitem [{\citenamefont {Haugland}\ \emph {et~al.}(2021)\citenamefont
  {Haugland}, \citenamefont {Sch{\"a}fer}, \citenamefont {Ronca}, \citenamefont
  {Rubio},\ and\ \citenamefont {Koch}}]{haugland2021intermolecular}%
  \BibitemOpen
  \bibfield  {author} {\bibinfo {author} {\bibfnamefont {T.~S.}\ \bibnamefont
  {Haugland}}, \bibinfo {author} {\bibfnamefont {C.}~\bibnamefont
  {Sch{\"a}fer}}, \bibinfo {author} {\bibfnamefont {E.}~\bibnamefont {Ronca}},
  \bibinfo {author} {\bibfnamefont {A.}~\bibnamefont {Rubio}}, \ and\ \bibinfo
  {author} {\bibfnamefont {H.}~\bibnamefont {Koch}},\ }\bibfield  {title}
  {\enquote {\bibinfo {title} {Intermolecular interactions in optical cavities:
  An ab initio {QED} study},}\ }\href@noop {} {\bibfield  {journal} {\bibinfo
  {journal} {J. Chem. Phys.}\ }\textbf {\bibinfo {volume} {154}} (\bibinfo
  {year} {2021})}\BibitemShut {NoStop}%
\bibitem [{\citenamefont {Riso}\ \emph {et~al.}(2022)\citenamefont {Riso},
  \citenamefont {Haugland}, \citenamefont {Ronca},\ and\ \citenamefont
  {Koch}}]{riso2022molecular}%
  \BibitemOpen
  \bibfield  {author} {\bibinfo {author} {\bibfnamefont {R.~R.}\ \bibnamefont
  {Riso}}, \bibinfo {author} {\bibfnamefont {T.~S.}\ \bibnamefont {Haugland}},
  \bibinfo {author} {\bibfnamefont {E.}~\bibnamefont {Ronca}}, \ and\ \bibinfo
  {author} {\bibfnamefont {H.}~\bibnamefont {Koch}},\ }\bibfield  {title}
  {\enquote {\bibinfo {title} {Molecular orbital theory in cavity {QED}
  environments},}\ }\href@noop {} {\bibfield  {journal} {\bibinfo  {journal}
  {Nat. Commun.}\ }\textbf {\bibinfo {volume} {13}},\ \bibinfo {pages} {1368}
  (\bibinfo {year} {2022})}\BibitemShut {NoStop}%
\bibitem [{\citenamefont {El~Moutaoukal}\ \emph {et~al.}(2024)\citenamefont
  {El~Moutaoukal}, \citenamefont {Riso}, \citenamefont {Castagnola},\ and\
  \citenamefont {Koch}}]{el2024toward}%
  \BibitemOpen
  \bibfield  {author} {\bibinfo {author} {\bibfnamefont {Y.}~\bibnamefont
  {El~Moutaoukal}}, \bibinfo {author} {\bibfnamefont {R.~R.}\ \bibnamefont
  {Riso}}, \bibinfo {author} {\bibfnamefont {M.}~\bibnamefont {Castagnola}}, \
  and\ \bibinfo {author} {\bibfnamefont {H.}~\bibnamefont {Koch}},\ }\bibfield
  {title} {\enquote {\bibinfo {title} {Toward polaritonic molecular orbitals
  for large molecular systems},}\ }\href@noop {} {\bibfield  {journal}
  {\bibinfo  {journal} {J. Chem. Theory Comput.}\ }\textbf {\bibinfo {volume}
  {20}},\ \bibinfo {pages} {8911--8920} (\bibinfo {year} {2024})}\BibitemShut
  {NoStop}%
\bibitem [{\citenamefont {El~Moutaoukal}\ \emph {et~al.}(2025)\citenamefont
  {El~Moutaoukal}, \citenamefont {Riso}, \citenamefont {Castagnola},
  \citenamefont {Ronca},\ and\ \citenamefont {Koch}}]{elmoutakal2025strong}%
  \BibitemOpen
  \bibfield  {author} {\bibinfo {author} {\bibfnamefont {Y.}~\bibnamefont
  {El~Moutaoukal}}, \bibinfo {author} {\bibfnamefont {R.~R.}\ \bibnamefont
  {Riso}}, \bibinfo {author} {\bibfnamefont {M.}~\bibnamefont {Castagnola}},
  \bibinfo {author} {\bibfnamefont {E.}~\bibnamefont {Ronca}}, \ and\ \bibinfo
  {author} {\bibfnamefont {H.}~\bibnamefont {Koch}},\ }\bibfield  {title}
  {\enquote {\bibinfo {title} {{Strong Coupling M{\o}ller-Plesset Perturbation
  Theory}},}\ }\href {\doibase 10.1021/acs.jctc.5c00055} {\bibfield  {journal}
  {\bibinfo  {journal} {J. Chem. Theory Comput.}\ }\textbf {\bibinfo {volume}
  {21}},\ \bibinfo {pages} {3981--3992} (\bibinfo {year} {2025})}\BibitemShut
  {NoStop}%
\bibitem [{\citenamefont {Angelico}\ \emph {et~al.}(2023)\citenamefont
  {Angelico}, \citenamefont {Haugland}, \citenamefont {Ronca},\ and\
  \citenamefont {Koch}}]{angelico2023coupled}%
  \BibitemOpen
  \bibfield  {author} {\bibinfo {author} {\bibfnamefont {S.}~\bibnamefont
  {Angelico}}, \bibinfo {author} {\bibfnamefont {T.~S.}\ \bibnamefont
  {Haugland}}, \bibinfo {author} {\bibfnamefont {E.}~\bibnamefont {Ronca}}, \
  and\ \bibinfo {author} {\bibfnamefont {H.}~\bibnamefont {Koch}},\ }\bibfield
  {title} {\enquote {\bibinfo {title} {{Coupled cluster cavity
  Born--Oppenheimer approximation for electronic strong coupling}},}\
  }\href@noop {} {\bibfield  {journal} {\bibinfo  {journal} {J. Chem. Phys.}\
  }\textbf {\bibinfo {volume} {159}} (\bibinfo {year} {2023})}\BibitemShut
  {NoStop}%
\bibitem [{\citenamefont {Castagnola}\ \emph
  {et~al.}(2024{\natexlab{a}})\citenamefont {Castagnola}, \citenamefont {Riso},
  \citenamefont {Barlini}, \citenamefont {Ronca},\ and\ \citenamefont
  {Koch}}]{castagnola2024polaritonic}%
  \BibitemOpen
  \bibfield  {author} {\bibinfo {author} {\bibfnamefont {M.}~\bibnamefont
  {Castagnola}}, \bibinfo {author} {\bibfnamefont {R.~R.}\ \bibnamefont
  {Riso}}, \bibinfo {author} {\bibfnamefont {A.}~\bibnamefont {Barlini}},
  \bibinfo {author} {\bibfnamefont {E.}~\bibnamefont {Ronca}}, \ and\ \bibinfo
  {author} {\bibfnamefont {H.}~\bibnamefont {Koch}},\ }\bibfield  {title}
  {\enquote {\bibinfo {title} {Polaritonic response theory for exact and
  approximate wave functions},}\ }\href@noop {} {\bibfield  {journal} {\bibinfo
   {journal} {{WIREs} Comput. Mol. Sci.}\ }\textbf {\bibinfo {volume} {14}},\
  \bibinfo {pages} {e1684} (\bibinfo {year} {2024}{\natexlab{a}})}\BibitemShut
  {NoStop}%
\bibitem [{\citenamefont {Fregoni}\ \emph {et~al.}(2021)\citenamefont
  {Fregoni}, \citenamefont {Haugland}, \citenamefont {Pipolo}, \citenamefont
  {Giovannini}, \citenamefont {Koch},\ and\ \citenamefont
  {Corni}}]{Fregoni2021}%
  \BibitemOpen
  \bibfield  {author} {\bibinfo {author} {\bibfnamefont {J.}~\bibnamefont
  {Fregoni}}, \bibinfo {author} {\bibfnamefont {T.~S.}\ \bibnamefont
  {Haugland}}, \bibinfo {author} {\bibfnamefont {S.}~\bibnamefont {Pipolo}},
  \bibinfo {author} {\bibfnamefont {T.}~\bibnamefont {Giovannini}}, \bibinfo
  {author} {\bibfnamefont {H.}~\bibnamefont {Koch}}, \ and\ \bibinfo {author}
  {\bibfnamefont {S.}~\bibnamefont {Corni}},\ }\bibfield  {title} {\enquote
  {\bibinfo {title} {{Strong Coupling between Localized Surface Plasmons and
  Molecules by Coupled Cluster Theory}},}\ }\href@noop {} {\bibfield  {journal}
  {\bibinfo  {journal} {Nano Letters}\ }\textbf {\bibinfo {volume} {21}},\
  \bibinfo {pages} {6664--6670} (\bibinfo {year} {2021})}\BibitemShut {NoStop}%
\bibitem [{\citenamefont {Romanelli}\ \emph {et~al.}(2023)\citenamefont
  {Romanelli}, \citenamefont {Riso}, \citenamefont {Haugland}, \citenamefont
  {Ronca}, \citenamefont {Corni},\ and\ \citenamefont {Koch}}]{Romanelli2023}%
  \BibitemOpen
  \bibfield  {author} {\bibinfo {author} {\bibfnamefont {M.}~\bibnamefont
  {Romanelli}}, \bibinfo {author} {\bibfnamefont {R.~R.}\ \bibnamefont {Riso}},
  \bibinfo {author} {\bibfnamefont {T.~S.}\ \bibnamefont {Haugland}}, \bibinfo
  {author} {\bibfnamefont {E.}~\bibnamefont {Ronca}}, \bibinfo {author}
  {\bibfnamefont {S.}~\bibnamefont {Corni}}, \ and\ \bibinfo {author}
  {\bibfnamefont {H.}~\bibnamefont {Koch}},\ }\bibfield  {title} {\enquote
  {\bibinfo {title} {{Effective Single-Mode Methodology for Strongly Coupled
  Multimode Molecular-Plasmon Nanosystems}},}\ }\href@noop {} {\bibfield
  {journal} {\bibinfo  {journal} {Nano Letters}\ }\textbf {\bibinfo {volume}
  {23}},\ \bibinfo {pages} {4938--4946} (\bibinfo {year} {2023})}\BibitemShut
  {NoStop}%
\bibitem [{\citenamefont {Schnack-Petersen}\ \emph {et~al.}(2022)\citenamefont
  {Schnack-Petersen}, \citenamefont {Koch}, \citenamefont {Coriani},\ and\
  \citenamefont {Kj{\o}nstad}}]{schnack2022efficient}%
  \BibitemOpen
  \bibfield  {author} {\bibinfo {author} {\bibfnamefont {A.~K.}\ \bibnamefont
  {Schnack-Petersen}}, \bibinfo {author} {\bibfnamefont {H.}~\bibnamefont
  {Koch}}, \bibinfo {author} {\bibfnamefont {S.}~\bibnamefont {Coriani}}, \
  and\ \bibinfo {author} {\bibfnamefont {E.~F.}\ \bibnamefont {Kj{\o}nstad}},\
  }\bibfield  {title} {\enquote {\bibinfo {title} {{Efficient implementation of
  molecular CCSD gradients with Cholesky-decomposed electron repulsion
  integrals}},}\ }\href@noop {} {\bibfield  {journal} {\bibinfo  {journal} {J.
  Chem. Phys.}\ }\textbf {\bibinfo {volume} {156}},\ \bibinfo {pages} {244111}
  (\bibinfo {year} {2022})}\BibitemShut {NoStop}%
\bibitem [{\citenamefont {Lexander}\ \emph
  {et~al.}(2025{\natexlab{a}})\citenamefont {Lexander}, \citenamefont
  {M.~Trabski}, \citenamefont {Bianchi}, \citenamefont {Kj{\o}nstad},
  \citenamefont {Haugland},\ and\ \citenamefont
  {Koch}}]{lexander2025exploring}%
  \BibitemOpen
  \bibfield  {author} {\bibinfo {author} {\bibfnamefont {M.~T.}\ \bibnamefont
  {Lexander}}, \bibinfo {author} {\bibfnamefont {J.~H.}\ \bibnamefont
  {M.~Trabski}}, \bibinfo {author} {\bibfnamefont {A.}~\bibnamefont {Bianchi}},
  \bibinfo {author} {\bibfnamefont {E.~F.}\ \bibnamefont {Kj{\o}nstad}},
  \bibinfo {author} {\bibfnamefont {T.~S.}\ \bibnamefont {Haugland}}, \ and\
  \bibinfo {author} {\bibfnamefont {H.}~\bibnamefont {Koch}},\ }\bibfield
  {title} {\enquote {\bibinfo {title} {Exploring molecular equilibrium
  geometries in static and quantized fields},}\ }\href@noop {} {\bibfield
  {journal} {\bibinfo  {journal} {J. Chem. Theory Comput.}\ } (\bibinfo {year}
  {2025}{\natexlab{a}})}\BibitemShut {NoStop}%
\bibitem [{\citenamefont {Barlini}\ \emph {et~al.}(2025)\citenamefont
  {Barlini}, \citenamefont {Bianchi}, \citenamefont {M.~Trabski}, \citenamefont
  {Bloino},\ and\ \citenamefont {Koch}}]{barlini2025}%
  \BibitemOpen
  \bibfield  {author} {\bibinfo {author} {\bibfnamefont {A.}~\bibnamefont
  {Barlini}}, \bibinfo {author} {\bibfnamefont {A.}~\bibnamefont {Bianchi}},
  \bibinfo {author} {\bibfnamefont {J.~H.}\ \bibnamefont {M.~Trabski}},
  \bibinfo {author} {\bibfnamefont {J.}~\bibnamefont {Bloino}}, \ and\ \bibinfo
  {author} {\bibfnamefont {H.}~\bibnamefont {Koch}},\ }\bibfield  {title}
  {\enquote {\bibinfo {title} {Cavity field-driven symmetry breaking and
  modulation of vibrational properties: Insights from the analytical qed-hf
  hessian},}\ }\href@noop {} {\bibfield  {journal} {\bibinfo  {journal} {J.
  Chem. Theory Comput.}\ } (\bibinfo {year} {2025})}\BibitemShut {NoStop}%
\bibitem [{\citenamefont {Lexander}\ \emph {et~al.}(2024)\citenamefont
  {Lexander}, \citenamefont {Angelico}, \citenamefont {Kj{\o}nstad},\ and\
  \citenamefont {Koch}}]{lexander2024analytical}%
  \BibitemOpen
  \bibfield  {author} {\bibinfo {author} {\bibfnamefont {M.~T.}\ \bibnamefont
  {Lexander}}, \bibinfo {author} {\bibfnamefont {S.}~\bibnamefont {Angelico}},
  \bibinfo {author} {\bibfnamefont {E.~F.}\ \bibnamefont {Kj{\o}nstad}}, \ and\
  \bibinfo {author} {\bibfnamefont {H.}~\bibnamefont {Koch}},\ }\bibfield
  {title} {\enquote {\bibinfo {title} {Analytical evaluation of ground state
  gradients in quantum electrodynamics coupled cluster theory},}\ }\href@noop
  {} {\bibfield  {journal} {\bibinfo  {journal} {J. Chem. Theory Comput.}\
  }\textbf {\bibinfo {volume} {20}},\ \bibinfo {pages} {8876--8885} (\bibinfo
  {year} {2024})}\BibitemShut {NoStop}%
\bibitem [{\citenamefont {Skeidsvoll}, \citenamefont {Balbi},\ and\
  \citenamefont {Koch}(2020)}]{skeidsvoll2020time}%
  \BibitemOpen
  \bibfield  {author} {\bibinfo {author} {\bibfnamefont {A.~S.}\ \bibnamefont
  {Skeidsvoll}}, \bibinfo {author} {\bibfnamefont {A.}~\bibnamefont {Balbi}}, \
  and\ \bibinfo {author} {\bibfnamefont {H.}~\bibnamefont {Koch}},\ }\bibfield
  {title} {\enquote {\bibinfo {title} {{Time-dependent coupled-cluster theory
  for ultrafast transient-absorption spectroscopy}},}\ }\href@noop {}
  {\bibfield  {journal} {\bibinfo  {journal} {Phys. Rev. A}\ }\textbf {\bibinfo
  {volume} {102}},\ \bibinfo {pages} {023115} (\bibinfo {year}
  {2020})}\BibitemShut {NoStop}%
\bibitem [{\citenamefont {Skeidsvoll}\ \emph {et~al.}(2022)\citenamefont
  {Skeidsvoll}, \citenamefont {Moitra}, \citenamefont {Balbi}, \citenamefont
  {Paul}, \citenamefont {Coriani},\ and\ \citenamefont
  {Koch}}]{skeidsvoll2022}%
  \BibitemOpen
  \bibfield  {author} {\bibinfo {author} {\bibfnamefont {A.~S.}\ \bibnamefont
  {Skeidsvoll}}, \bibinfo {author} {\bibfnamefont {T.}~\bibnamefont {Moitra}},
  \bibinfo {author} {\bibfnamefont {A.}~\bibnamefont {Balbi}}, \bibinfo
  {author} {\bibfnamefont {A.~C.}\ \bibnamefont {Paul}}, \bibinfo {author}
  {\bibfnamefont {S.}~\bibnamefont {Coriani}}, \ and\ \bibinfo {author}
  {\bibfnamefont {H.}~\bibnamefont {Koch}},\ }\bibfield  {title} {\enquote
  {\bibinfo {title} {Simulating weak-field attosecond processes with a
  {L}anczos reduced basis approach to time-dependent equation-of-motion
  coupled-cluster theory},}\ }\href {\doibase 10.1103/PhysRevA.105.023103}
  {\bibfield  {journal} {\bibinfo  {journal} {Phys. Rev. A}\ }\textbf {\bibinfo
  {volume} {105}},\ \bibinfo {pages} {023103} (\bibinfo {year}
  {2022})}\BibitemShut {NoStop}%
\bibitem [{\citenamefont {Tsuchimochi}\ and\ \citenamefont
  {Scuseria}(2010)}]{tsuchimochi2010communication}%
  \BibitemOpen
  \bibfield  {author} {\bibinfo {author} {\bibfnamefont {T.}~\bibnamefont
  {Tsuchimochi}}\ and\ \bibinfo {author} {\bibfnamefont {G.~E.}\ \bibnamefont
  {Scuseria}},\ }\bibfield  {title} {\enquote {\bibinfo {title} {Communication:
  {ROHF} theory made simple},}\ }\href@noop {} {\bibfield  {journal} {\bibinfo
  {journal} {J. Chem. Phys.}\ }\textbf {\bibinfo {volume} {133}},\ \bibinfo
  {pages} {141102} (\bibinfo {year} {2010})}\BibitemShut {NoStop}%
\bibitem [{\citenamefont {Marrazzini}\ \emph {et~al.}(2021)\citenamefont
  {Marrazzini}, \citenamefont {Giovannini}, \citenamefont {Scavino},
  \citenamefont {Egidi}, \citenamefont {Cappelli},\ and\ \citenamefont
  {Koch}}]{marrazzini2021multilevel}%
  \BibitemOpen
  \bibfield  {author} {\bibinfo {author} {\bibfnamefont {G.}~\bibnamefont
  {Marrazzini}}, \bibinfo {author} {\bibfnamefont {T.}~\bibnamefont
  {Giovannini}}, \bibinfo {author} {\bibfnamefont {M.}~\bibnamefont {Scavino}},
  \bibinfo {author} {\bibfnamefont {F.}~\bibnamefont {Egidi}}, \bibinfo
  {author} {\bibfnamefont {C.}~\bibnamefont {Cappelli}}, \ and\ \bibinfo
  {author} {\bibfnamefont {H.}~\bibnamefont {Koch}},\ }\bibfield  {title}
  {\enquote {\bibinfo {title} {Multilevel density functional theory},}\
  }\href@noop {} {\bibfield  {journal} {\bibinfo  {journal} {J. Chem. Theory
  Comput.}\ }\textbf {\bibinfo {volume} {17}},\ \bibinfo {pages} {791--803}
  (\bibinfo {year} {2021})}\BibitemShut {NoStop}%
\bibitem [{\citenamefont {Giovannini}\ \emph {et~al.}(2023)\citenamefont
  {Giovannini}, \citenamefont {Marrazzini}, \citenamefont {Scavino},
  \citenamefont {Koch},\ and\ \citenamefont
  {Cappelli}}]{giovannini2023integrated}%
  \BibitemOpen
  \bibfield  {author} {\bibinfo {author} {\bibfnamefont {T.}~\bibnamefont
  {Giovannini}}, \bibinfo {author} {\bibfnamefont {G.}~\bibnamefont
  {Marrazzini}}, \bibinfo {author} {\bibfnamefont {M.}~\bibnamefont {Scavino}},
  \bibinfo {author} {\bibfnamefont {H.}~\bibnamefont {Koch}}, \ and\ \bibinfo
  {author} {\bibfnamefont {C.}~\bibnamefont {Cappelli}},\ }\bibfield  {title}
  {\enquote {\bibinfo {title} {Integrated multiscale multilevel approach to
  open shell molecular systems},}\ }\href@noop {} {\bibfield  {journal}
  {\bibinfo  {journal} {J. Chem. Theory Comput.}\ }\textbf {\bibinfo {volume}
  {19}},\ \bibinfo {pages} {1446--1456} (\bibinfo {year} {2023})}\BibitemShut
  {NoStop}%
\bibitem [{\citenamefont {Lebedev}\ and\ \citenamefont
  {Laikov}(1999)}]{lebedev1999quadrature}%
  \BibitemOpen
  \bibfield  {author} {\bibinfo {author} {\bibfnamefont {V.~I.}\ \bibnamefont
  {Lebedev}}\ and\ \bibinfo {author} {\bibfnamefont {D.~N.}\ \bibnamefont
  {Laikov}},\ }\bibfield  {title} {\enquote {\bibinfo {title} {A quadrature
  formula for the sphere of the 131st algebraic order of accuracy},}\ }in\
  \href@noop {} {\emph {\bibinfo {booktitle} {Doklady Mathematics}}},\
  Vol.~\bibinfo {volume} {59}\ (\bibinfo {organization} {Pleiades Publishing} \bibinfo {year} {1999})\
  pp.\ \bibinfo {pages} {477--481}\BibitemShut {NoStop}%
\bibitem [{\citenamefont {Krack}\ and\ \citenamefont
  {K{\"o}ster}(1998)}]{krack1998adaptive}%
  \BibitemOpen
  \bibfield  {author} {\bibinfo {author} {\bibfnamefont {M.}~\bibnamefont
  {Krack}}\ and\ \bibinfo {author} {\bibfnamefont {A.~M.}\ \bibnamefont
  {K{\"o}ster}},\ }\bibfield  {title} {\enquote {\bibinfo {title} {An adaptive
  numerical integrator for molecular integrals},}\ }\href@noop {} {\bibfield
  {journal} {\bibinfo  {journal} {J. Chem. Phys.}\ }\textbf {\bibinfo {volume}
  {108}},\ \bibinfo {pages} {3226--3234} (\bibinfo {year} {1998})}\BibitemShut
  {NoStop}%
\bibitem [{\citenamefont {Sun}(2015)}]{libcint}%
  \BibitemOpen
  \bibfield  {author} {\bibinfo {author} {\bibfnamefont {Q.}~\bibnamefont
  {Sun}},\ }\bibfield  {title} {\enquote {\bibinfo {title} {Libcint: An
  efficient general integral library for gaussian basis functions},}\ }\href
  {\doibase 10.1002/jcc.23981} {\bibfield  {journal} {\bibinfo  {journal} {J.
  Comput. Chem.}\ }\textbf {\bibinfo {volume} {36}},\ \bibinfo {pages}
  {1664--1671} (\bibinfo {year} {2015})}\BibitemShut {NoStop}%
\bibitem [{\citenamefont {Valeyev}\ \emph {et~al.}(2025)\citenamefont
  {Valeyev}, \citenamefont {Surjuse}, \citenamefont {Burns}, \citenamefont
  {Abbott}, \citenamefont {Kawashima}, \citenamefont {Seewald}, \citenamefont
  {Misiewicz}, \citenamefont {Lewis}, \citenamefont {Calvin}, \citenamefont
  {Dullea}, \citenamefont {Peng}, \citenamefont {Asadchev}, \citenamefont
  {Panyala}, \citenamefont {Nishimra}, \citenamefont {powellsr}, \citenamefont
  {mclement1}, \citenamefont {samslattery}, \citenamefont {jfermann},
  \citenamefont {Lehtola}, \citenamefont {Herbst}, \citenamefont
  {Mejia-Rodriguez}, \citenamefont {Mitchell}, \citenamefont {Čertík},
  \citenamefont {Ellerbrock}, \citenamefont {Whitfield}, \citenamefont
  {Wingate}, \citenamefont {Wiedemann}, \citenamefont {Bosia}, \citenamefont
  {Banck},\ and\ \citenamefont {Willow}}]{libint}%
  \BibitemOpen
  \bibfield  {author} {\bibinfo {author} {\bibfnamefont {E.}~\bibnamefont
  {Valeyev}}, \bibinfo {author} {\bibfnamefont {K.}~\bibnamefont {Surjuse}},
  \bibinfo {author} {\bibfnamefont {L.~A.}\ \bibnamefont {Burns}}, \bibinfo
  {author} {\bibfnamefont {A.}~\bibnamefont {Abbott}}, \bibinfo {author}
  {\bibfnamefont {E.}~\bibnamefont {Kawashima}}, \bibinfo {author}
  {\bibfnamefont {P.}~\bibnamefont {Seewald}}, \bibinfo {author} {\bibfnamefont
  {J.}~\bibnamefont {Misiewicz}}, \bibinfo {author} {\bibfnamefont
  {D.}~\bibnamefont {Lewis}}, \bibinfo {author} {\bibfnamefont
  {J.}~\bibnamefont {Calvin}}, \bibinfo {author} {\bibfnamefont
  {J.}~\bibnamefont {Dullea}}, \bibinfo {author} {\bibfnamefont
  {C.}~\bibnamefont {Peng}}, \bibinfo {author} {\bibfnamefont {A.}~\bibnamefont
  {Asadchev}}, \bibinfo {author} {\bibfnamefont {A.}~\bibnamefont {Panyala}},
  \bibinfo {author} {\bibfnamefont {K.}~\bibnamefont {Nishimra}}, \bibinfo
  {author} {\bibnamefont {powellsr}}, \bibinfo {author} {\bibnamefont
  {mclement1}}, \bibinfo {author} {\bibnamefont {samslattery}}, \bibinfo
  {author} {\bibnamefont {jfermann}}, \bibinfo {author} {\bibfnamefont
  {S.}~\bibnamefont {Lehtola}}, \bibinfo {author} {\bibfnamefont {M.~F.}\
  \bibnamefont {Herbst}}, \bibinfo {author} {\bibfnamefont {D.}~\bibnamefont
  {Mejia-Rodriguez}}, \bibinfo {author} {\bibfnamefont {E.}~\bibnamefont
  {Mitchell}}, \bibinfo {author} {\bibfnamefont {O.}~\bibnamefont {Čertík}},
  \bibinfo {author} {\bibfnamefont {R.}~\bibnamefont {Ellerbrock}}, \bibinfo
  {author} {\bibfnamefont {J.~D.}\ \bibnamefont {Whitfield}}, \bibinfo {author}
  {\bibfnamefont {A.}~\bibnamefont {Wingate}}, \bibinfo {author} {\bibfnamefont
  {B.~M.}\ \bibnamefont {Wiedemann}}, \bibinfo {author} {\bibfnamefont
  {F.}~\bibnamefont {Bosia}}, \bibinfo {author} {\bibfnamefont
  {M.}~\bibnamefont {Banck}}, \ and\ \bibinfo {author} {\bibfnamefont {S.~Y.}\
  \bibnamefont {Willow}},\ }\href {\doibase 10.5281/zenodo.15420235} {\enquote
  {\bibinfo {title} {evaleev/libint: 2.11.1},}\ } (\bibinfo {year}
  {2025})\BibitemShut {NoStop}%
\bibitem [{\citenamefont {Bianchi}(2025)}]{bianchi2025}%
  \BibitemOpen
  \bibfield  {author} {\bibinfo {author} {\bibfnamefont {A.}~\bibnamefont
  {Bianchi}},\ }\href {\doibase 10.5281/zenodo.17025111} {\enquote {\bibinfo
  {title} {Phasedint},}\ } (\bibinfo {year} {2025})\BibitemShut {NoStop}%
\bibitem [{\citenamefont {Di~Remigio}\ \emph {et~al.}(2019)\citenamefont
  {Di~Remigio}, \citenamefont {Steindal}, \citenamefont {Mozgawa},
  \citenamefont {Weijo}, \citenamefont {Cao},\ and\ \citenamefont
  {Frediani}}]{di2019pcmsolver}%
  \BibitemOpen
  \bibfield  {author} {\bibinfo {author} {\bibfnamefont {R.}~\bibnamefont
  {Di~Remigio}}, \bibinfo {author} {\bibfnamefont {A.~H.}\ \bibnamefont
  {Steindal}}, \bibinfo {author} {\bibfnamefont {K.}~\bibnamefont {Mozgawa}},
  \bibinfo {author} {\bibfnamefont {V.}~\bibnamefont {Weijo}}, \bibinfo
  {author} {\bibfnamefont {H.}~\bibnamefont {Cao}}, \ and\ \bibinfo {author}
  {\bibfnamefont {L.}~\bibnamefont {Frediani}},\ }\bibfield  {title} {\enquote
  {\bibinfo {title} {{PCMSolver}: An open-source library for solvation
  modeling},}\ }\href@noop {} {\bibfield  {journal} {\bibinfo  {journal} {Int.
  J. Quantum Chem.}\ }\textbf {\bibinfo {volume} {119}},\ \bibinfo {pages}
  {e25685} (\bibinfo {year} {2019})}\BibitemShut {NoStop}%
\bibitem [{\citenamefont {Lehtola}\ \emph {et~al.}(2018)\citenamefont
  {Lehtola}, \citenamefont {Steigemann}, \citenamefont {Oliveira},\ and\
  \citenamefont {Marques}}]{libxc}%
  \BibitemOpen
  \bibfield  {author} {\bibinfo {author} {\bibfnamefont {S.}~\bibnamefont
  {Lehtola}}, \bibinfo {author} {\bibfnamefont {C.}~\bibnamefont {Steigemann}},
  \bibinfo {author} {\bibfnamefont {M.~J.~T.}\ \bibnamefont {Oliveira}}, \ and\
  \bibinfo {author} {\bibfnamefont {M.~A.~L.}\ \bibnamefont {Marques}},\
  }\bibfield  {title} {\enquote {\bibinfo {title} {{Recent developments in
  libxc—A comprehensive library of functionals for density functional
  theory}},}\ }\href@noop {} {\bibfield  {journal} {\bibinfo  {journal}
  {SoftwareX}\ }\textbf {\bibinfo {volume} {7}},\ \bibinfo {pages} {1--5}
  (\bibinfo {year} {2018})}\BibitemShut {NoStop}%
\bibitem [{\citenamefont {Marques}, \citenamefont {Oliveira},\ and\
  \citenamefont {Burnus}(2012)}]{marques2012libxc}%
  \BibitemOpen
  \bibfield  {author} {\bibinfo {author} {\bibfnamefont {M.~A.}\ \bibnamefont
  {Marques}}, \bibinfo {author} {\bibfnamefont {M.~J.}\ \bibnamefont
  {Oliveira}}, \ and\ \bibinfo {author} {\bibfnamefont {T.}~\bibnamefont
  {Burnus}},\ }\bibfield  {title} {\enquote {\bibinfo {title} {Libxc: A library
  of exchange and correlation functionals for density functional theory},}\
  }\href@noop {} {\bibfield  {journal} {\bibinfo  {journal} {Comput. Phys.
  Commun.}\ }\textbf {\bibinfo {volume} {183}},\ \bibinfo {pages} {2272--2281}
  (\bibinfo {year} {2012})}\BibitemShut {NoStop}%
\bibitem [{\citenamefont {Fletcher}(2013)}]{fletcher2013practical}%
  \BibitemOpen
  \bibfield  {author} {\bibinfo {author} {\bibfnamefont {R.}~\bibnamefont
  {Fletcher}},\ }\href@noop {} {\emph {\bibinfo {title} {{Practical methods of
  optimization}}}}\ (\bibinfo  {publisher} {John Wiley \& Sons},\ \bibinfo
  {year} {2013})\BibitemShut {NoStop}%
\bibitem [{\citenamefont {J{\o}rgensen}, \citenamefont {Swanstr{\o}m},\ and\
  \citenamefont {Yeager}(1983)}]{Jorgensen:1983aa}%
  \BibitemOpen
  \bibfield  {author} {\bibinfo {author} {\bibfnamefont {P.}~\bibnamefont
  {J{\o}rgensen}}, \bibinfo {author} {\bibfnamefont {P.}~\bibnamefont
  {Swanstr{\o}m}}, \ and\ \bibinfo {author} {\bibfnamefont {D.~L.}\
  \bibnamefont {Yeager}},\ }\bibfield  {title} {\enquote {\bibinfo {title}
  {Guaranteed convergence in ground state multiconfigurational
  self‐consistent field calculations},}\ }\href {\doibase 10.1063/1.444508}
  {\bibfield  {journal} {\bibinfo  {journal} {J. Chem. Phys.}\ }\textbf
  {\bibinfo {volume} {78}},\ \bibinfo {pages} {347--356} (\bibinfo {year}
  {1983})}\BibitemShut {NoStop}%
\bibitem [{\citenamefont {H{\o}yvik}, \citenamefont {Jansik},\ and\
  \citenamefont {J{\o}rgensen}(2012)}]{hoyvik2012trust}%
  \BibitemOpen
  \bibfield  {author} {\bibinfo {author} {\bibfnamefont {I.-M.}\ \bibnamefont
  {H{\o}yvik}}, \bibinfo {author} {\bibfnamefont {B.}~\bibnamefont {Jansik}}, \
  and\ \bibinfo {author} {\bibfnamefont {P.}~\bibnamefont {J{\o}rgensen}},\
  }\bibfield  {title} {\enquote {\bibinfo {title} {{Trust region minimization
  of orbital localization functions}},}\ }\href@noop {} {\bibfield  {journal}
  {\bibinfo  {journal} {J. Chem. Theory Comput.}\ }\textbf {\bibinfo {volume}
  {8}},\ \bibinfo {pages} {3137--3146} (\bibinfo {year} {2012})}\BibitemShut
  {NoStop}%
\bibitem [{\citenamefont {Folkestad}\ \emph {et~al.}(2022)\citenamefont
  {Folkestad}, \citenamefont {Matveeva}, \citenamefont {H{\o}yvik},\ and\
  \citenamefont {Koch}}]{ER2022implementation}%
  \BibitemOpen
  \bibfield  {author} {\bibinfo {author} {\bibfnamefont {S.~D.}\ \bibnamefont
  {Folkestad}}, \bibinfo {author} {\bibfnamefont {R.}~\bibnamefont {Matveeva}},
  \bibinfo {author} {\bibfnamefont {I.-M.}\ \bibnamefont {H{\o}yvik}}, \ and\
  \bibinfo {author} {\bibfnamefont {H.}~\bibnamefont {Koch}},\ }\bibfield
  {title} {\enquote {\bibinfo {title} {{Implementation of occupied and virtual
  {Edmiston--Ruedenberg} orbitals using {Cholesky} decomposed integrals}},}\
  }\href@noop {} {\bibfield  {journal} {\bibinfo  {journal} {J. Chem. Theory
  Comput.}\ }\textbf {\bibinfo {volume} {18}},\ \bibinfo {pages} {4733--4744}
  (\bibinfo {year} {2022})}\BibitemShut {NoStop}%
\bibitem [{\citenamefont {Davidson}(1975)}]{DAVIDSON197587}%
  \BibitemOpen
  \bibfield  {author} {\bibinfo {author} {\bibfnamefont {E.~R.}\ \bibnamefont
  {Davidson}},\ }\bibfield  {title} {\enquote {\bibinfo {title} {The iterative
  calculation of a few of the lowest eigenvalues and corresponding eigenvectors
  of large real-symmetric matrices},}\ }\href {\doibase
  https://doi.org/10.1016/0021-9991(75)90065-0} {\bibfield  {journal} {\bibinfo
   {journal} {J. Comput. Phys.}\ }\textbf {\bibinfo {volume} {17}},\ \bibinfo
  {pages} {87 -- 94} (\bibinfo {year} {1975})}\BibitemShut {NoStop}%
\bibitem [{\citenamefont {Hirao}\ and\ \citenamefont
  {Nakatsuji}(1982)}]{hirao1982generalization}%
  \BibitemOpen
  \bibfield  {author} {\bibinfo {author} {\bibfnamefont {K.}~\bibnamefont
  {Hirao}}\ and\ \bibinfo {author} {\bibfnamefont {H.}~\bibnamefont
  {Nakatsuji}},\ }\bibfield  {title} {\enquote {\bibinfo {title} {{A
  generalization of the Davidson's method to large nonsymmetric eigenvalue
  problems}},}\ }\href@noop {} {\bibfield  {journal} {\bibinfo  {journal}
  {Journal of Computational Physics}\ }\textbf {\bibinfo {volume} {45}},\
  \bibinfo {pages} {246--254} (\bibinfo {year} {1982})}\BibitemShut {NoStop}%
\bibitem [{\citenamefont {Christiansen}, \citenamefont {Koch},\ and\
  \citenamefont {J{\o}rgensen}(1995)}]{christiansen1995second}%
  \BibitemOpen
  \bibfield  {author} {\bibinfo {author} {\bibfnamefont {O.}~\bibnamefont
  {Christiansen}}, \bibinfo {author} {\bibfnamefont {H.}~\bibnamefont {Koch}},
  \ and\ \bibinfo {author} {\bibfnamefont {P.}~\bibnamefont {J{\o}rgensen}},\
  }\bibfield  {title} {\enquote {\bibinfo {title} {{The second-order
  approximate coupled cluster singles and doubles model CC2}},}\ }\href@noop {}
  {\bibfield  {journal} {\bibinfo  {journal} {Chem. Phys. Lett.}\ }\textbf
  {\bibinfo {volume} {243}},\ \bibinfo {pages} {409--418} (\bibinfo {year}
  {1995})}\BibitemShut {NoStop}%
\bibitem [{\citenamefont {Koch}\ \emph {et~al.}(1997)\citenamefont {Koch},
  \citenamefont {Christiansen}, \citenamefont {Sanchez~de Mer{\'a}s},
  \citenamefont {Helgaker} \emph {et~al.}}]{koch1997cc3}%
  \BibitemOpen
  \bibfield  {author} {\bibinfo {author} {\bibfnamefont {H.}~\bibnamefont
  {Koch}}, \bibinfo {author} {\bibfnamefont {O.}~\bibnamefont {Christiansen}},
  \bibinfo {author} {\bibfnamefont {A.~M.}\ \bibnamefont {Sanchez~de
  Mer{\'a}s}}, \bibinfo {author} {\bibfnamefont {T.}~\bibnamefont {Helgaker}},
  \emph {et~al.},\ }\bibfield  {title} {\enquote {\bibinfo {title} {{The CC3
  model: An iterative coupled cluster approach including connected triples}},}\
  }\href@noop {} {\bibfield  {journal} {\bibinfo  {journal} {J. Chem. Phys.}\
  }\textbf {\bibinfo {volume} {106}},\ \bibinfo {pages} {1808--1818} (\bibinfo
  {year} {1997})}\BibitemShut {NoStop}%
\bibitem [{\citenamefont {Paul}, \citenamefont {Myhre},\ and\ \citenamefont
  {Koch}(2020)}]{paul2020new}%
  \BibitemOpen
  \bibfield  {author} {\bibinfo {author} {\bibfnamefont {A.~C.}\ \bibnamefont
  {Paul}}, \bibinfo {author} {\bibfnamefont {R.~H.}\ \bibnamefont {Myhre}}, \
  and\ \bibinfo {author} {\bibfnamefont {H.}~\bibnamefont {Koch}},\ }\bibfield
  {title} {\enquote {\bibinfo {title} {New and efficient implementation of
  {CC3}},}\ }\href@noop {} {\bibfield  {journal} {\bibinfo  {journal} {J. Chem.
  Theory Comput.}\ }\textbf {\bibinfo {volume} {17}},\ \bibinfo {pages}
  {117--126} (\bibinfo {year} {2020})}\BibitemShut {NoStop}%
\bibitem [{\citenamefont {Scuseria}, \citenamefont {Lee},\ and\ \citenamefont
  {Schaefer}(1986)}]{SCUSERIA1986236}%
  \BibitemOpen
  \bibfield  {author} {\bibinfo {author} {\bibfnamefont {G.~E.}\ \bibnamefont
  {Scuseria}}, \bibinfo {author} {\bibfnamefont {T.~J.}\ \bibnamefont {Lee}}, \
  and\ \bibinfo {author} {\bibfnamefont {H.~F.}\ \bibnamefont {Schaefer}},\
  }\bibfield  {title} {\enquote {\bibinfo {title} {Accelerating the convergence
  of the coupled-cluster approach: The use of the {DIIS} method},}\ }\href
  {\doibase https://doi.org/10.1016/0009-2614(86)80461-4} {\bibfield  {journal}
  {\bibinfo  {journal} {Chem. Phys. Lett.}\ }\textbf {\bibinfo {volume}
  {130}},\ \bibinfo {pages} {236 -- 239} (\bibinfo {year} {1986})}\BibitemShut
  {NoStop}%
\bibitem [{\citenamefont {Hamilton}\ and\ \citenamefont
  {Pulay}(1986)}]{hamilton1986direct}%
  \BibitemOpen
  \bibfield  {author} {\bibinfo {author} {\bibfnamefont {T.~P.}\ \bibnamefont
  {Hamilton}}\ and\ \bibinfo {author} {\bibfnamefont {P.}~\bibnamefont
  {Pulay}},\ }\bibfield  {title} {\enquote {\bibinfo {title} {{Direct inversion
  in the iterative subspace (DIIS) optimization of open-shell, excited-state,
  and small multiconfiguration SCF wave functions}},}\ }\href@noop {}
  {\bibfield  {journal} {\bibinfo  {journal} {J. Chem. Phys.}\ }\textbf
  {\bibinfo {volume} {84}},\ \bibinfo {pages} {5728--5734} (\bibinfo {year}
  {1986})}\BibitemShut {NoStop}%
\bibitem [{\citenamefont {Kj{\o}nstad}, \citenamefont {Folkestad},\ and\
  \citenamefont {Koch}(2020)}]{kjonstad2020accelerated}%
  \BibitemOpen
  \bibfield  {author} {\bibinfo {author} {\bibfnamefont {E.~F.}\ \bibnamefont
  {Kj{\o}nstad}}, \bibinfo {author} {\bibfnamefont {S.~D.}\ \bibnamefont
  {Folkestad}}, \ and\ \bibinfo {author} {\bibfnamefont {H.}~\bibnamefont
  {Koch}},\ }\bibfield  {title} {\enquote {\bibinfo {title} {Accelerated
  multimodel newton-type algorithms for faster convergence of ground and
  excited state coupled cluster equations},}\ }\href@noop {} {\bibfield
  {journal} {\bibinfo  {journal} {J. Chem. Phys.}\ }\textbf {\bibinfo {volume}
  {153}},\ \bibinfo {pages} {014104} (\bibinfo {year} {2020})}\BibitemShut
  {NoStop}%
\bibitem [{\citenamefont {Schnack-Petersen}\ \emph {et~al.}(2023)\citenamefont
  {Schnack-Petersen}, \citenamefont {Moitra}, \citenamefont {Folkestad},\ and\
  \citenamefont {Coriani}}]{schnack2023new}%
  \BibitemOpen
  \bibfield  {author} {\bibinfo {author} {\bibfnamefont {A.~K.}\ \bibnamefont
  {Schnack-Petersen}}, \bibinfo {author} {\bibfnamefont {T.}~\bibnamefont
  {Moitra}}, \bibinfo {author} {\bibfnamefont {S.~D.}\ \bibnamefont
  {Folkestad}}, \ and\ \bibinfo {author} {\bibfnamefont {S.}~\bibnamefont
  {Coriani}},\ }\bibfield  {title} {\enquote {\bibinfo {title} {New
  implementation of an equation-of-motion coupled-cluster damped-response
  framework with illustrative applications to resonant inelastic {X-ray}
  scattering},}\ }\href@noop {} {\bibfield  {journal} {\bibinfo  {journal} {J.
  Phys. Chem. A}\ }\textbf {\bibinfo {volume} {127}},\ \bibinfo {pages}
  {1775--1793} (\bibinfo {year} {2023})}\BibitemShut {NoStop}%
\bibitem [{\citenamefont {Bast}(2018)}]{runtest}%
  \BibitemOpen
  \bibfield  {author} {\bibinfo {author} {\bibfnamefont {R.}~\bibnamefont
  {Bast}},\ }\href {\doibase 10.5281/zenodo.1434751} {\enquote {\bibinfo
  {title} {Runtest},}\ } (\bibinfo {year} {2018}),\ \bibinfo {note}
  {https://doi.org/10.5281/zenodo.1434751}\BibitemShut {NoStop}%
\bibitem [{\citenamefont {Rilee}\ and\ \citenamefont
  {Clune}(2014)}]{rilee2014towards}%
  \BibitemOpen
  \bibfield  {author} {\bibinfo {author} {\bibfnamefont {M.}~\bibnamefont
  {Rilee}}\ and\ \bibinfo {author} {\bibfnamefont {T.}~\bibnamefont {Clune}},\
  }\bibfield  {title} {\enquote {\bibinfo {title} {Towards test driven
  development for computational science with pfunit},}\ }in\ \href@noop {}
  {\emph {\bibinfo {booktitle} {2014 Second International Workshop on Software
  Engineering for High Performance Computing in Computational Science and
  Engineering}}}\ (\bibinfo {organization} {IEEE},\ \bibinfo {year} {2014})\
  pp.\ \bibinfo {pages} {20--27}\BibitemShut {NoStop}%
\bibitem [{\citenamefont {Bloch}\ \emph {et~al.}(2022)\citenamefont {Bloch},
  \citenamefont {Cavalleri}, \citenamefont {Galitski}, \citenamefont {Hafezi},\
  and\ \citenamefont {Rubio}}]{bloch2022strongly}%
  \BibitemOpen
  \bibfield  {author} {\bibinfo {author} {\bibfnamefont {J.}~\bibnamefont
  {Bloch}}, \bibinfo {author} {\bibfnamefont {A.}~\bibnamefont {Cavalleri}},
  \bibinfo {author} {\bibfnamefont {V.}~\bibnamefont {Galitski}}, \bibinfo
  {author} {\bibfnamefont {M.}~\bibnamefont {Hafezi}}, \ and\ \bibinfo {author}
  {\bibfnamefont {A.}~\bibnamefont {Rubio}},\ }\bibfield  {title} {\enquote
  {\bibinfo {title} {Strongly correlated electron--photon systems},}\
  }\href@noop {} {\bibfield  {journal} {\bibinfo  {journal} {Nature}\ }\textbf
  {\bibinfo {volume} {606}},\ \bibinfo {pages} {41--48} (\bibinfo {year}
  {2022})}\BibitemShut {NoStop}%
\bibitem [{\citenamefont {Fregoni}, \citenamefont {Garcia-Vidal},\ and\
  \citenamefont {Feist}(2022)}]{fregoni2022theoretical}%
  \BibitemOpen
  \bibfield  {author} {\bibinfo {author} {\bibfnamefont {J.}~\bibnamefont
  {Fregoni}}, \bibinfo {author} {\bibfnamefont {F.~J.}\ \bibnamefont
  {Garcia-Vidal}}, \ and\ \bibinfo {author} {\bibfnamefont {J.}~\bibnamefont
  {Feist}},\ }\bibfield  {title} {\enquote {\bibinfo {title} {Theoretical
  challenges in polaritonic chemistry},}\ }\href@noop {} {\bibfield  {journal}
  {\bibinfo  {journal} {ACS photonics}\ }\textbf {\bibinfo {volume} {9}},\
  \bibinfo {pages} {1096--1107} (\bibinfo {year} {2022})}\BibitemShut {NoStop}%
\bibitem [{\citenamefont {Ruggenthaler}, \citenamefont {Sidler},\ and\
  \citenamefont {Rubio}(2023)}]{ruggenthaler2023understanding}%
  \BibitemOpen
  \bibfield  {author} {\bibinfo {author} {\bibfnamefont {M.}~\bibnamefont
  {Ruggenthaler}}, \bibinfo {author} {\bibfnamefont {D.}~\bibnamefont
  {Sidler}}, \ and\ \bibinfo {author} {\bibfnamefont {A.}~\bibnamefont
  {Rubio}},\ }\bibfield  {title} {\enquote {\bibinfo {title} {Understanding
  polaritonic chemistry from ab initio quantum electrodynamics},}\ }\href@noop
  {} {\bibfield  {journal} {\bibinfo  {journal} {Chem. Rev.}\ }\textbf
  {\bibinfo {volume} {123}},\ \bibinfo {pages} {11191--11229} (\bibinfo {year}
  {2023})}\BibitemShut {NoStop}%
\bibitem [{\citenamefont {Feist}, \citenamefont {Galego},\ and\ \citenamefont
  {Garcia-Vidal}(2018)}]{feist2018polaritonic}%
  \BibitemOpen
  \bibfield  {author} {\bibinfo {author} {\bibfnamefont {J.}~\bibnamefont
  {Feist}}, \bibinfo {author} {\bibfnamefont {J.}~\bibnamefont {Galego}}, \
  and\ \bibinfo {author} {\bibfnamefont {F.~J.}\ \bibnamefont {Garcia-Vidal}},\
  }\bibfield  {title} {\enquote {\bibinfo {title} {Polaritonic chemistry with
  organic molecules},}\ }\href@noop {} {\bibfield  {journal} {\bibinfo
  {journal} {ACS Photonics}\ }\textbf {\bibinfo {volume} {5}},\ \bibinfo
  {pages} {205--216} (\bibinfo {year} {2018})}\BibitemShut {NoStop}%
\bibitem [{\citenamefont {Schafer}\ \emph {et~al.}(2024)\citenamefont
  {Schafer}, \citenamefont {Fojt}, \citenamefont {Lindgren},\ and\
  \citenamefont {Erhart}}]{schafer2024machine}%
  \BibitemOpen
  \bibfield  {author} {\bibinfo {author} {\bibfnamefont {C.}~\bibnamefont
  {Schafer}}, \bibinfo {author} {\bibfnamefont {J.}~\bibnamefont {Fojt}},
  \bibinfo {author} {\bibfnamefont {E.}~\bibnamefont {Lindgren}}, \ and\
  \bibinfo {author} {\bibfnamefont {P.}~\bibnamefont {Erhart}},\ }\bibfield
  {title} {\enquote {\bibinfo {title} {Machine learning for polaritonic
  chemistry: Accessing chemical kinetics},}\ }\href@noop {} {\bibfield
  {journal} {\bibinfo  {journal} {J. Am. Chem. Soc.}\ }\textbf {\bibinfo
  {volume} {146}},\ \bibinfo {pages} {5402--5413} (\bibinfo {year}
  {2024})}\BibitemShut {NoStop}%
\bibitem [{\citenamefont {George}\ and\ \citenamefont
  {Singh}(2023)}]{george2023polaritonic}%
  \BibitemOpen
  \bibfield  {author} {\bibinfo {author} {\bibfnamefont {J.}~\bibnamefont
  {George}}\ and\ \bibinfo {author} {\bibfnamefont {J.}~\bibnamefont {Singh}},\
  }\bibfield  {title} {\enquote {\bibinfo {title} {Polaritonic chemistry:
  Band-selective control of chemical reactions by vibrational strong
  coupling},}\ }\href@noop {} {\bibfield  {journal} {\bibinfo  {journal} {ACS
  Catalysis}\ }\textbf {\bibinfo {volume} {13}},\ \bibinfo {pages} {2631--2636}
  (\bibinfo {year} {2023})}\BibitemShut {NoStop}%
\bibitem [{\citenamefont {Ebbesen}, \citenamefont {Rubio},\ and\ \citenamefont
  {Scholes}(2023)}]{ebbesen2023introduction}%
  \BibitemOpen
  \bibfield  {author} {\bibinfo {author} {\bibfnamefont {T.~W.}\ \bibnamefont
  {Ebbesen}}, \bibinfo {author} {\bibfnamefont {A.}~\bibnamefont {Rubio}}, \
  and\ \bibinfo {author} {\bibfnamefont {G.~D.}\ \bibnamefont {Scholes}},\
  }\bibfield  {title} {\enquote {\bibinfo {title} {Introduction: polaritonic
  chemistry},}\ }\href@noop {} {\bibfield  {journal} {\bibinfo  {journal}
  {Chemical Reviews}\ }\textbf {\bibinfo {volume} {123}},\ \bibinfo {pages}
  {12037--12038} (\bibinfo {year} {2023})}\BibitemShut {NoStop}%
\bibitem [{\citenamefont {Campos-Gonzalez-Angulo}, \citenamefont {Ribeiro},\
  and\ \citenamefont {Yuen-Zhou}(2019)}]{campos2019resonant}%
  \BibitemOpen
  \bibfield  {author} {\bibinfo {author} {\bibfnamefont {J.~A.}\ \bibnamefont
  {Campos-Gonzalez-Angulo}}, \bibinfo {author} {\bibfnamefont {R.~F.}\
  \bibnamefont {Ribeiro}}, \ and\ \bibinfo {author} {\bibfnamefont
  {J.}~\bibnamefont {Yuen-Zhou}},\ }\bibfield  {title} {\enquote {\bibinfo
  {title} {Resonant catalysis of thermally activated chemical reactions with
  vibrational polaritons},}\ }\href@noop {} {\bibfield  {journal} {\bibinfo
  {journal} {Nat. Commun.}\ }\textbf {\bibinfo {volume} {10}},\ \bibinfo
  {pages} {4685} (\bibinfo {year} {2019})}\BibitemShut {NoStop}%
\bibitem [{\citenamefont {Wellnitz}, \citenamefont {Pupillo},\ and\
  \citenamefont {Schachenmayer}(2022)}]{wellnitz2022disorder}%
  \BibitemOpen
  \bibfield  {author} {\bibinfo {author} {\bibfnamefont {D.}~\bibnamefont
  {Wellnitz}}, \bibinfo {author} {\bibfnamefont {G.}~\bibnamefont {Pupillo}}, \
  and\ \bibinfo {author} {\bibfnamefont {J.}~\bibnamefont {Schachenmayer}},\
  }\bibfield  {title} {\enquote {\bibinfo {title} {Disorder enhanced
  vibrational entanglement and dynamics in polaritonic chemistry},}\
  }\href@noop {} {\bibfield  {journal} {\bibinfo  {journal} {Communications
  Physics}\ }\textbf {\bibinfo {volume} {5}},\ \bibinfo {pages} {120} (\bibinfo
  {year} {2022})}\BibitemShut {NoStop}%
\bibitem [{\citenamefont {Nagarajan}, \citenamefont {Thomas},\ and\
  \citenamefont {Ebbesen}(2021)}]{nagarajan2021chemistry}%
  \BibitemOpen
  \bibfield  {author} {\bibinfo {author} {\bibfnamefont {K.}~\bibnamefont
  {Nagarajan}}, \bibinfo {author} {\bibfnamefont {A.}~\bibnamefont {Thomas}}, \
  and\ \bibinfo {author} {\bibfnamefont {T.~W.}\ \bibnamefont {Ebbesen}},\
  }\bibfield  {title} {\enquote {\bibinfo {title} {Chemistry under vibrational
  strong coupling},}\ }\href@noop {} {\bibfield  {journal} {\bibinfo  {journal}
  {J. Am. Chem. Soc.}\ }\textbf {\bibinfo {volume} {143}},\ \bibinfo {pages}
  {16877--16889} (\bibinfo {year} {2021})}\BibitemShut {NoStop}%
\bibitem [{\citenamefont {Garcia-Vidal}, \citenamefont {Ciuti},\ and\
  \citenamefont {Ebbesen}(2021)}]{garcia2021manipulating}%
  \BibitemOpen
  \bibfield  {author} {\bibinfo {author} {\bibfnamefont {F.~J.}\ \bibnamefont
  {Garcia-Vidal}}, \bibinfo {author} {\bibfnamefont {C.}~\bibnamefont {Ciuti}},
  \ and\ \bibinfo {author} {\bibfnamefont {T.~W.}\ \bibnamefont {Ebbesen}},\
  }\bibfield  {title} {\enquote {\bibinfo {title} {Manipulating matter by
  strong coupling to vacuum fields},}\ }\href@noop {} {\bibfield  {journal}
  {\bibinfo  {journal} {Science}\ }\textbf {\bibinfo {volume} {373}},\ \bibinfo
  {pages} {eabd0336} (\bibinfo {year} {2021})}\BibitemShut {NoStop}%
\bibitem [{\citenamefont {Castagnola}\ \emph
  {et~al.}(2024{\natexlab{b}})\citenamefont {Castagnola}, \citenamefont
  {Lexander}, \citenamefont {Ronca},\ and\ \citenamefont
  {Koch}}]{castagnola2024strong}%
  \BibitemOpen
  \bibfield  {author} {\bibinfo {author} {\bibfnamefont {M.}~\bibnamefont
  {Castagnola}}, \bibinfo {author} {\bibfnamefont {M.~T.}\ \bibnamefont
  {Lexander}}, \bibinfo {author} {\bibfnamefont {E.}~\bibnamefont {Ronca}}, \
  and\ \bibinfo {author} {\bibfnamefont {H.}~\bibnamefont {Koch}},\ }\bibfield
  {title} {\enquote {\bibinfo {title} {Strong coupling electron-photon
  dynamics: A real-time investigation of energy redistribution in molecular
  polaritons},}\ }\href@noop {} {\bibfield  {journal} {\bibinfo  {journal}
  {Phys. Rev. Res.}\ }\textbf {\bibinfo {volume} {6}},\ \bibinfo {pages}
  {033283} (\bibinfo {year} {2024}{\natexlab{b}})}\BibitemShut {NoStop}%
\bibitem [{\citenamefont {Sandik}\ \emph {et~al.}(2024)\citenamefont {Sandik},
  \citenamefont {Feist}, \citenamefont {Garc{\'\i}a-Vidal},\ and\ \citenamefont
  {Schwartz}}]{sandik2024cavity}%
  \BibitemOpen
  \bibfield  {author} {\bibinfo {author} {\bibfnamefont {G.}~\bibnamefont
  {Sandik}}, \bibinfo {author} {\bibfnamefont {J.}~\bibnamefont {Feist}},
  \bibinfo {author} {\bibfnamefont {F.~J.}\ \bibnamefont {Garc{\'\i}a-Vidal}},
  \ and\ \bibinfo {author} {\bibfnamefont {T.}~\bibnamefont {Schwartz}},\
  }\bibfield  {title} {\enquote {\bibinfo {title} {Cavity-enhanced energy
  transport in molecular systems},}\ }\href@noop {} {\bibfield  {journal}
  {\bibinfo  {journal} {Nature Materials}\ ,\ \bibinfo {pages} {1--12}}
  (\bibinfo {year} {2024})}\BibitemShut {NoStop}%
\bibitem [{\citenamefont {Chikkaraddy}\ \emph {et~al.}(2016)\citenamefont
  {Chikkaraddy}, \citenamefont {De~Nijs}, \citenamefont {Benz}, \citenamefont
  {Barrow}, \citenamefont {Scherman}, \citenamefont {Rosta}, \citenamefont
  {Demetriadou}, \citenamefont {Fox}, \citenamefont {Hess},\ and\ \citenamefont
  {Baumberg}}]{chikkaraddy2016single}%
  \BibitemOpen
  \bibfield  {author} {\bibinfo {author} {\bibfnamefont {R.}~\bibnamefont
  {Chikkaraddy}}, \bibinfo {author} {\bibfnamefont {B.}~\bibnamefont
  {De~Nijs}}, \bibinfo {author} {\bibfnamefont {F.}~\bibnamefont {Benz}},
  \bibinfo {author} {\bibfnamefont {S.~J.}\ \bibnamefont {Barrow}}, \bibinfo
  {author} {\bibfnamefont {O.~A.}\ \bibnamefont {Scherman}}, \bibinfo {author}
  {\bibfnamefont {E.}~\bibnamefont {Rosta}}, \bibinfo {author} {\bibfnamefont
  {A.}~\bibnamefont {Demetriadou}}, \bibinfo {author} {\bibfnamefont
  {P.}~\bibnamefont {Fox}}, \bibinfo {author} {\bibfnamefont {O.}~\bibnamefont
  {Hess}}, \ and\ \bibinfo {author} {\bibfnamefont {J.~J.}\ \bibnamefont
  {Baumberg}},\ }\bibfield  {title} {\enquote {\bibinfo {title}
  {Single-molecule strong coupling at room temperature in plasmonic
  nanocavities},}\ }\href@noop {} {\bibfield  {journal} {\bibinfo  {journal}
  {Nature}\ }\textbf {\bibinfo {volume} {535}},\ \bibinfo {pages} {127--130}
  (\bibinfo {year} {2016})}\BibitemShut {NoStop}%
\bibitem [{\citenamefont {Santhosh}\ \emph {et~al.}(2016)\citenamefont
  {Santhosh}, \citenamefont {Bitton}, \citenamefont {Chuntonov},\ and\
  \citenamefont {Haran}}]{santhosh2016vacuum}%
  \BibitemOpen
  \bibfield  {author} {\bibinfo {author} {\bibfnamefont {K.}~\bibnamefont
  {Santhosh}}, \bibinfo {author} {\bibfnamefont {O.}~\bibnamefont {Bitton}},
  \bibinfo {author} {\bibfnamefont {L.}~\bibnamefont {Chuntonov}}, \ and\
  \bibinfo {author} {\bibfnamefont {G.}~\bibnamefont {Haran}},\ }\bibfield
  {title} {\enquote {\bibinfo {title} {Vacuum rabi splitting in a plasmonic
  cavity at the single quantum emitter limit},}\ }\href@noop {} {\bibfield
  {journal} {\bibinfo  {journal} {Nat. Commun.}\ }\textbf {\bibinfo {volume}
  {7}},\ \bibinfo {pages} {ncomms11823} (\bibinfo {year} {2016})}\BibitemShut
  {NoStop}%
\bibitem [{\citenamefont {Wu}\ \emph {et~al.}(2023)\citenamefont {Wu},
  \citenamefont {Battie}, \citenamefont {Genet}, \citenamefont {Ebbesen},
  \citenamefont {Decher},\ and\ \citenamefont {Pauly}}]{wu2023bottom}%
  \BibitemOpen
  \bibfield  {author} {\bibinfo {author} {\bibfnamefont {W.}~\bibnamefont
  {Wu}}, \bibinfo {author} {\bibfnamefont {Y.}~\bibnamefont {Battie}}, \bibinfo
  {author} {\bibfnamefont {C.}~\bibnamefont {Genet}}, \bibinfo {author}
  {\bibfnamefont {T.~W.}\ \bibnamefont {Ebbesen}}, \bibinfo {author}
  {\bibfnamefont {G.}~\bibnamefont {Decher}}, \ and\ \bibinfo {author}
  {\bibfnamefont {M.}~\bibnamefont {Pauly}},\ }\bibfield  {title} {\enquote
  {\bibinfo {title} {Bottom-up tunable broadband semi-reflective chiral
  mirrors},}\ }\href@noop {} {\bibfield  {journal} {\bibinfo  {journal}
  {Advanced Optical Materials}\ }\textbf {\bibinfo {volume} {11}},\ \bibinfo
  {pages} {2202831} (\bibinfo {year} {2023})}\BibitemShut {NoStop}%
\bibitem [{\citenamefont {Thomas}\ \emph {et~al.}(2019)\citenamefont {Thomas},
  \citenamefont {Lethuillier-Karl}, \citenamefont {Nagarajan}, \citenamefont
  {Vergauwe}, \citenamefont {George}, \citenamefont {Chervy}, \citenamefont
  {Shalabney}, \citenamefont {Devaux}, \citenamefont {Genet}, \citenamefont
  {Moran} \emph {et~al.}}]{thomas2019tilting}%
  \BibitemOpen
  \bibfield  {author} {\bibinfo {author} {\bibfnamefont {A.}~\bibnamefont
  {Thomas}}, \bibinfo {author} {\bibfnamefont {L.}~\bibnamefont
  {Lethuillier-Karl}}, \bibinfo {author} {\bibfnamefont {K.}~\bibnamefont
  {Nagarajan}}, \bibinfo {author} {\bibfnamefont {R.~M.}\ \bibnamefont
  {Vergauwe}}, \bibinfo {author} {\bibfnamefont {J.}~\bibnamefont {George}},
  \bibinfo {author} {\bibfnamefont {T.}~\bibnamefont {Chervy}}, \bibinfo
  {author} {\bibfnamefont {A.}~\bibnamefont {Shalabney}}, \bibinfo {author}
  {\bibfnamefont {E.}~\bibnamefont {Devaux}}, \bibinfo {author} {\bibfnamefont
  {C.}~\bibnamefont {Genet}}, \bibinfo {author} {\bibfnamefont
  {J.}~\bibnamefont {Moran}},  \emph {et~al.},\ }\bibfield  {title} {\enquote
  {\bibinfo {title} {Tilting a ground-state reactivity landscape by vibrational
  strong coupling},}\ }\href@noop {} {\bibfield  {journal} {\bibinfo  {journal}
  {Science}\ }\textbf {\bibinfo {volume} {363}},\ \bibinfo {pages} {615--619}
  (\bibinfo {year} {2019})}\BibitemShut {NoStop}%
\bibitem [{\citenamefont {Hutchison}\ \emph {et~al.}(2012)\citenamefont
  {Hutchison}, \citenamefont {Schwartz}, \citenamefont {Genet}, \citenamefont
  {Devaux},\ and\ \citenamefont {Ebbesen}}]{hutchison2012modifying}%
  \BibitemOpen
  \bibfield  {author} {\bibinfo {author} {\bibfnamefont {J.~A.}\ \bibnamefont
  {Hutchison}}, \bibinfo {author} {\bibfnamefont {T.}~\bibnamefont {Schwartz}},
  \bibinfo {author} {\bibfnamefont {C.}~\bibnamefont {Genet}}, \bibinfo
  {author} {\bibfnamefont {E.}~\bibnamefont {Devaux}}, \ and\ \bibinfo {author}
  {\bibfnamefont {T.~W.}\ \bibnamefont {Ebbesen}},\ }\bibfield  {title}
  {\enquote {\bibinfo {title} {Modifying chemical landscapes by coupling to
  vacuum fields},}\ }\href@noop {} {\bibfield  {journal} {\bibinfo  {journal}
  {Angew. Chem. Int. Ed.}\ }\textbf {\bibinfo {volume} {51}},\ \bibinfo {pages}
  {1592--1596} (\bibinfo {year} {2012})}\BibitemShut {NoStop}%
\bibitem [{\citenamefont {Thomas}\ \emph {et~al.}(2016)\citenamefont {Thomas},
  \citenamefont {George}, \citenamefont {Shalabney}, \citenamefont {Dryzhakov},
  \citenamefont {Varma}, \citenamefont {Moran}, \citenamefont {Chervy},
  \citenamefont {Zhong}, \citenamefont {Devaux}, \citenamefont {Genet} \emph
  {et~al.}}]{thomas2016ground}%
  \BibitemOpen
  \bibfield  {author} {\bibinfo {author} {\bibfnamefont {A.}~\bibnamefont
  {Thomas}}, \bibinfo {author} {\bibfnamefont {J.}~\bibnamefont {George}},
  \bibinfo {author} {\bibfnamefont {A.}~\bibnamefont {Shalabney}}, \bibinfo
  {author} {\bibfnamefont {M.}~\bibnamefont {Dryzhakov}}, \bibinfo {author}
  {\bibfnamefont {S.~J.}\ \bibnamefont {Varma}}, \bibinfo {author}
  {\bibfnamefont {J.}~\bibnamefont {Moran}}, \bibinfo {author} {\bibfnamefont
  {T.}~\bibnamefont {Chervy}}, \bibinfo {author} {\bibfnamefont
  {X.}~\bibnamefont {Zhong}}, \bibinfo {author} {\bibfnamefont
  {E.}~\bibnamefont {Devaux}}, \bibinfo {author} {\bibfnamefont
  {C.}~\bibnamefont {Genet}},  \emph {et~al.},\ }\bibfield  {title} {\enquote
  {\bibinfo {title} {Ground-state chemical reactivity under vibrational
  coupling to the vacuum electromagnetic field},}\ }\href@noop {} {\bibfield
  {journal} {\bibinfo  {journal} {Angew. Chem.}\ }\textbf {\bibinfo {volume}
  {128}},\ \bibinfo {pages} {11634--11638} (\bibinfo {year}
  {2016})}\BibitemShut {NoStop}%
\bibitem [{\citenamefont {Ahn}\ \emph {et~al.}(2023)\citenamefont {Ahn},
  \citenamefont {Triana}, \citenamefont {Recabal}, \citenamefont {Herrera},\
  and\ \citenamefont {Simpkins}}]{ahn2023modification}%
  \BibitemOpen
  \bibfield  {author} {\bibinfo {author} {\bibfnamefont {W.}~\bibnamefont
  {Ahn}}, \bibinfo {author} {\bibfnamefont {J.~F.}\ \bibnamefont {Triana}},
  \bibinfo {author} {\bibfnamefont {F.}~\bibnamefont {Recabal}}, \bibinfo
  {author} {\bibfnamefont {F.}~\bibnamefont {Herrera}}, \ and\ \bibinfo
  {author} {\bibfnamefont {B.~S.}\ \bibnamefont {Simpkins}},\ }\bibfield
  {title} {\enquote {\bibinfo {title} {Modification of ground-state chemical
  reactivity via light--matter coherence in infrared cavities},}\ }\href@noop
  {} {\bibfield  {journal} {\bibinfo  {journal} {Science}\ }\textbf {\bibinfo
  {volume} {380}},\ \bibinfo {pages} {1165--1168} (\bibinfo {year}
  {2023})}\BibitemShut {NoStop}%
\bibitem [{\citenamefont {Imperatore}, \citenamefont {Asbury},\ and\
  \citenamefont {Giebink}(2021)}]{imperatore2021reproducibility}%
  \BibitemOpen
  \bibfield  {author} {\bibinfo {author} {\bibfnamefont {M.~V.}\ \bibnamefont
  {Imperatore}}, \bibinfo {author} {\bibfnamefont {J.~B.}\ \bibnamefont
  {Asbury}}, \ and\ \bibinfo {author} {\bibfnamefont {N.~C.}\ \bibnamefont
  {Giebink}},\ }\bibfield  {title} {\enquote {\bibinfo {title} {Reproducibility
  of cavity-enhanced chemical reaction rates in the vibrational strong coupling
  regime},}\ }\href@noop {} {\bibfield  {journal} {\bibinfo  {journal} {J.
  Chem. Phys.}\ }\textbf {\bibinfo {volume} {154}},\ \bibinfo {pages} {191103}
  (\bibinfo {year} {2021})}\BibitemShut {NoStop}%
\bibitem [{\citenamefont {Ruggenthaler}\ \emph {et~al.}(2014)\citenamefont
  {Ruggenthaler}, \citenamefont {Flick}, \citenamefont {Pellegrini},
  \citenamefont {Appel}, \citenamefont {Tokatly},\ and\ \citenamefont
  {Rubio}}]{ruggenthaler2014quantum}%
  \BibitemOpen
  \bibfield  {author} {\bibinfo {author} {\bibfnamefont {M.}~\bibnamefont
  {Ruggenthaler}}, \bibinfo {author} {\bibfnamefont {J.}~\bibnamefont {Flick}},
  \bibinfo {author} {\bibfnamefont {C.}~\bibnamefont {Pellegrini}}, \bibinfo
  {author} {\bibfnamefont {H.}~\bibnamefont {Appel}}, \bibinfo {author}
  {\bibfnamefont {I.~V.}\ \bibnamefont {Tokatly}}, \ and\ \bibinfo {author}
  {\bibfnamefont {A.}~\bibnamefont {Rubio}},\ }\bibfield  {title} {\enquote
  {\bibinfo {title} {Quantum-electrodynamical density-functional theory:
  Bridging quantum optics and electronic-structure theory},}\ }\href@noop {}
  {\bibfield  {journal} {\bibinfo  {journal} {Phys. Rev. A}\ }\textbf {\bibinfo
  {volume} {90}},\ \bibinfo {pages} {012508} (\bibinfo {year}
  {2014})}\BibitemShut {NoStop}%
\bibitem [{\citenamefont {Flick}\ \emph {et~al.}(2015)\citenamefont {Flick},
  \citenamefont {Ruggenthaler}, \citenamefont {Appel},\ and\ \citenamefont
  {Rubio}}]{flick2015kohn}%
  \BibitemOpen
  \bibfield  {author} {\bibinfo {author} {\bibfnamefont {J.}~\bibnamefont
  {Flick}}, \bibinfo {author} {\bibfnamefont {M.}~\bibnamefont {Ruggenthaler}},
  \bibinfo {author} {\bibfnamefont {H.}~\bibnamefont {Appel}}, \ and\ \bibinfo
  {author} {\bibfnamefont {A.}~\bibnamefont {Rubio}},\ }\bibfield  {title}
  {\enquote {\bibinfo {title} {Kohn--sham approach to quantum electrodynamical
  density-functional theory: Exact time-dependent effective potentials in real
  space},}\ }\href@noop {} {\bibfield  {journal} {\bibinfo  {journal}
  {Proceedings of the National Academy of Sciences}\ }\textbf {\bibinfo
  {volume} {112}},\ \bibinfo {pages} {15285--15290} (\bibinfo {year}
  {2015})}\BibitemShut {NoStop}%
\bibitem [{\citenamefont {Lu}\ \emph {et~al.}(2024)\citenamefont {Lu},
  \citenamefont {Ruggenthaler}, \citenamefont {Tancogne-Dejean}, \citenamefont
  {Latini}, \citenamefont {Penz},\ and\ \citenamefont
  {Rubio}}]{lu2024electron}%
  \BibitemOpen
  \bibfield  {author} {\bibinfo {author} {\bibfnamefont {I.-T.}\ \bibnamefont
  {Lu}}, \bibinfo {author} {\bibfnamefont {M.}~\bibnamefont {Ruggenthaler}},
  \bibinfo {author} {\bibfnamefont {N.}~\bibnamefont {Tancogne-Dejean}},
  \bibinfo {author} {\bibfnamefont {S.}~\bibnamefont {Latini}}, \bibinfo
  {author} {\bibfnamefont {M.}~\bibnamefont {Penz}}, \ and\ \bibinfo {author}
  {\bibfnamefont {A.}~\bibnamefont {Rubio}},\ }\bibfield  {title} {\enquote
  {\bibinfo {title} {Electron-photon exchange-correlation approximation for
  quantum-electrodynamical density-functional theory},}\ }\href@noop {}
  {\bibfield  {journal} {\bibinfo  {journal} {Phys. Rev. A}\ }\textbf {\bibinfo
  {volume} {109}},\ \bibinfo {pages} {052823} (\bibinfo {year}
  {2024})}\BibitemShut {NoStop}%
\bibitem [{\citenamefont {Balbi}, \citenamefont {Skeidsvoll},\ and\
  \citenamefont {Koch}(2023)}]{balbi2023coupled}%
  \BibitemOpen
  \bibfield  {author} {\bibinfo {author} {\bibfnamefont {A.}~\bibnamefont
  {Balbi}}, \bibinfo {author} {\bibfnamefont {A.~S.}\ \bibnamefont
  {Skeidsvoll}}, \ and\ \bibinfo {author} {\bibfnamefont {H.}~\bibnamefont
  {Koch}},\ }\bibfield  {title} {\enquote {\bibinfo {title} {Coupled cluster
  simulation of impulsive stimulated x-ray raman scattering},}\ }\href@noop {}
  {\bibfield  {journal} {\bibinfo  {journal} {J. Phys. Chem. A}\ }\textbf
  {\bibinfo {volume} {127}},\ \bibinfo {pages} {8676--8684} (\bibinfo {year}
  {2023})}\BibitemShut {NoStop}%
\bibitem [{\citenamefont {Skeidsvoll}\ and\ \citenamefont
  {Koch}(2023)}]{Skeidsvoll2023Rabi}%
  \BibitemOpen
  \bibfield  {author} {\bibinfo {author} {\bibfnamefont {A.~S.}\ \bibnamefont
  {Skeidsvoll}}\ and\ \bibinfo {author} {\bibfnamefont {H.}~\bibnamefont
  {Koch}},\ }\bibfield  {title} {\enquote {\bibinfo {title} {Comparing
  real-time coupled-cluster methods through simulation of collective {R}abi
  oscillations},}\ }\href {\doibase 10.1103/PhysRevA.108.033116} {\bibfield
  {journal} {\bibinfo  {journal} {Phys. Rev. A}\ }\textbf {\bibinfo {volume}
  {108}},\ \bibinfo {pages} {033116} (\bibinfo {year} {2023})}\BibitemShut
  {NoStop}%
\bibitem [{\citenamefont {Dormand}\ and\ \citenamefont
  {Prince}(1980)}]{DORMAND198019}%
  \BibitemOpen
  \bibfield  {author} {\bibinfo {author} {\bibfnamefont {J.}~\bibnamefont
  {Dormand}}\ and\ \bibinfo {author} {\bibfnamefont {P.}~\bibnamefont
  {Prince}},\ }\bibfield  {title} {\enquote {\bibinfo {title} {A family of
  embedded runge-kutta formulae},}\ }\href {\doibase
  https://doi.org/10.1016/0771-050X(80)90013-3} {\bibfield  {journal} {\bibinfo
   {journal} {J. Comput. Appl. Math.}\ }\textbf {\bibinfo {volume} {6}},\
  \bibinfo {pages} {19--26} (\bibinfo {year} {1980})}\BibitemShut {NoStop}%
\bibitem [{Hai(1993)}]{Hairer1993}%
  \BibitemOpen
  \enquote {\bibinfo {title} {Runge-kutta and extrapolation methods},}\ in\
  \href {\doibase 10.1007/978-3-540-78862-1_2} {\emph {\bibinfo {booktitle}
  {Solving Ordinary Differential Equations I: Nonstiff Problems}}}\ (\bibinfo
  {publisher} {Springer Berlin Heidelberg},\ \bibinfo {address} {Berlin,
  Heidelberg},\ \bibinfo {year} {1993})\ pp.\ \bibinfo {pages}
  {129--353}\BibitemShut {NoStop}%
\bibitem [{\citenamefont {Pettersen}\ \emph {et~al.}(2004)\citenamefont
  {Pettersen}, \citenamefont {Goddard}, \citenamefont {Huang}, \citenamefont
  {Couch}, \citenamefont {Greenblatt}, \citenamefont {Meng},\ and\
  \citenamefont {Ferrin}}]{Pettersen2004Chimera}%
  \BibitemOpen
  \bibfield  {author} {\bibinfo {author} {\bibfnamefont {E.~F.}\ \bibnamefont
  {Pettersen}}, \bibinfo {author} {\bibfnamefont {T.~D.}\ \bibnamefont
  {Goddard}}, \bibinfo {author} {\bibfnamefont {C.~C.}\ \bibnamefont {Huang}},
  \bibinfo {author} {\bibfnamefont {G.~S.}\ \bibnamefont {Couch}}, \bibinfo
  {author} {\bibfnamefont {D.~M.}\ \bibnamefont {Greenblatt}}, \bibinfo
  {author} {\bibfnamefont {E.~C.}\ \bibnamefont {Meng}}, \ and\ \bibinfo
  {author} {\bibfnamefont {T.~E.}\ \bibnamefont {Ferrin}},\ }\bibfield  {title}
  {\enquote {\bibinfo {title} {{UCSF Chimera--A} visualization system for
  exploratory research and analysis},}\ }\href {\doibase 10.1002/jcc.20084}
  {\bibfield  {journal} {\bibinfo  {journal} {J. Comput. Chem.}\ }\textbf
  {\bibinfo {volume} {25}},\ \bibinfo {pages} {1605--1612} (\bibinfo {year}
  {2004})},\ \Eprint
  {http://arxiv.org/abs/https://onlinelibrary.wiley.com/doi/pdf/10.1002/jcc.20084}
  {https://onlinelibrary.wiley.com/doi/pdf/10.1002/jcc.20084} \BibitemShut
  {NoStop}%
\bibitem [{\citenamefont {Hariharan}\ and\ \citenamefont
  {Pople}(1973)}]{Hariharan1973basis}%
  \BibitemOpen
  \bibfield  {author} {\bibinfo {author} {\bibfnamefont {P.~C.}\ \bibnamefont
  {Hariharan}}\ and\ \bibinfo {author} {\bibfnamefont {J.~A.}\ \bibnamefont
  {Pople}},\ }\bibfield  {title} {\enquote {\bibinfo {title} {The influence of
  polarization functions on molecular orbital hydrogenation energies},}\ }\href
  {\doibase 10.1007/BF00533485} {\bibfield  {journal} {\bibinfo  {journal}
  {Theor. Chim. Acta}\ }\textbf {\bibinfo {volume} {28}},\ \bibinfo {pages}
  {213--222} (\bibinfo {year} {1973})}\BibitemShut {NoStop}%
\bibitem [{\citenamefont {Bakken}\ and\ \citenamefont
  {Helgaker}(2002)}]{bakken2002efficient}%
  \BibitemOpen
  \bibfield  {author} {\bibinfo {author} {\bibfnamefont {V.}~\bibnamefont
  {Bakken}}\ and\ \bibinfo {author} {\bibfnamefont {T.}~\bibnamefont
  {Helgaker}},\ }\bibfield  {title} {\enquote {\bibinfo {title} {The efficient
  optimization of molecular geometries using redundant internal coordinates},}\
  }\href@noop {} {\bibfield  {journal} {\bibinfo  {journal} {J. Chem. Phys.}\
  }\textbf {\bibinfo {volume} {117}},\ \bibinfo {pages} {9160--9174} (\bibinfo
  {year} {2002})}\BibitemShut {NoStop}%
\bibitem [{\citenamefont {Wang}\ and\ \citenamefont
  {Song}(2016)}]{wang2016geometry}%
  \BibitemOpen
  \bibfield  {author} {\bibinfo {author} {\bibfnamefont {L.-P.}\ \bibnamefont
  {Wang}}\ and\ \bibinfo {author} {\bibfnamefont {C.}~\bibnamefont {Song}},\
  }\bibfield  {title} {\enquote {\bibinfo {title} {Geometry optimization made
  simple with translation and rotation coordinates},}\ }\href@noop {}
  {\bibfield  {journal} {\bibinfo  {journal} {J. Chem. Phys.}\ }\textbf
  {\bibinfo {volume} {144}},\ \bibinfo {pages} {214108} (\bibinfo {year}
  {2016})}\BibitemShut {NoStop}%
\bibitem [{\citenamefont {Coriani}\ and\ \citenamefont
  {Koch}(2015)}]{Coriani2015}%
  \BibitemOpen
  \bibfield  {author} {\bibinfo {author} {\bibfnamefont {S.}~\bibnamefont
  {Coriani}}\ and\ \bibinfo {author} {\bibfnamefont {H.}~\bibnamefont {Koch}},\
  }\bibfield  {title} {\enquote {\bibinfo {title} {Communication: X-ray
  absorption spectra and core-ionization potentials within a core-valence
  separated coupled cluster framework},}\ }\href {\doibase 10.1063/1.4935712}
  {\bibfield  {journal} {\bibinfo  {journal} {J. Chem. Phys.}\ }\textbf
  {\bibinfo {volume} {143}},\ \bibinfo {pages} {181103} (\bibinfo {year}
  {2015})},\ \Eprint {http://arxiv.org/abs/https://doi.org/10.1063/1.4935712}
  {https://doi.org/10.1063/1.4935712} \BibitemShut {NoStop}%
\bibitem [{\citenamefont {Coriani}\ and\ \citenamefont
  {Koch}(2016)}]{Coriani2016erratum}%
  \BibitemOpen
  \bibfield  {author} {\bibinfo {author} {\bibfnamefont {S.}~\bibnamefont
  {Coriani}}\ and\ \bibinfo {author} {\bibfnamefont {H.}~\bibnamefont {Koch}},\
  }\bibfield  {title} {\enquote {\bibinfo {title} {Erratum: “communication:
  X-ray absorption spectra and core-ionization potentials within a core-valence
  separated coupled cluster framework” {[J. Chem. Phys. 143, 181103
  (2015)]}},}\ }\href {\doibase 10.1063/1.4964714} {\bibfield  {journal}
  {\bibinfo  {journal} {J. Chem. Phys.}\ }\textbf {\bibinfo {volume} {145}},\
  \bibinfo {pages} {149901} (\bibinfo {year} {2016})},\ \Eprint
  {http://arxiv.org/abs/https://doi.org/10.1063/1.4964714}
  {https://doi.org/10.1063/1.4964714} \BibitemShut {NoStop}%
\bibitem [{\citenamefont {Moitra}\ \emph {et~al.}(2022)\citenamefont {Moitra},
  \citenamefont {Paul}, \citenamefont {Decleva}, \citenamefont {Koch},\ and\
  \citenamefont {Coriani}}]{moitra2022multi}%
  \BibitemOpen
  \bibfield  {author} {\bibinfo {author} {\bibfnamefont {T.}~\bibnamefont
  {Moitra}}, \bibinfo {author} {\bibfnamefont {A.~C.}\ \bibnamefont {Paul}},
  \bibinfo {author} {\bibfnamefont {P.}~\bibnamefont {Decleva}}, \bibinfo
  {author} {\bibfnamefont {H.}~\bibnamefont {Koch}}, \ and\ \bibinfo {author}
  {\bibfnamefont {S.}~\bibnamefont {Coriani}},\ }\bibfield  {title} {\enquote
  {\bibinfo {title} {Multi-electron excitation contributions towards primary
  and satellite states in the photoelectron spectrum},}\ }\href {\doibase
  10.1039/D1CP04695K} {\bibfield  {journal} {\bibinfo  {journal} {Phys. Chem.
  Chem. Phys.}\ }\textbf {\bibinfo {volume} {24}},\ \bibinfo {pages}
  {8329--8343} (\bibinfo {year} {2022})}\BibitemShut {NoStop}%
\bibitem [{\citenamefont {Curchod}\ and\ \citenamefont
  {Mart{\'\i}nez}(2018)}]{curchod2018ab}%
  \BibitemOpen
  \bibfield  {author} {\bibinfo {author} {\bibfnamefont {B.~F.}\ \bibnamefont
  {Curchod}}\ and\ \bibinfo {author} {\bibfnamefont {T.~J.}\ \bibnamefont
  {Mart{\'\i}nez}},\ }\bibfield  {title} {\enquote {\bibinfo {title} {Ab initio
  nonadiabatic quantum molecular dynamics},}\ }\href@noop {} {\bibfield
  {journal} {\bibinfo  {journal} {Chemical reviews}\ }\textbf {\bibinfo
  {volume} {118}},\ \bibinfo {pages} {3305--3336} (\bibinfo {year}
  {2018})}\BibitemShut {NoStop}%
\bibitem [{\citenamefont {H{\o}yvik}, \citenamefont {Myhre},\ and\
  \citenamefont {Koch}(2017)}]{hoyvik2017correlated}%
  \BibitemOpen
  \bibfield  {author} {\bibinfo {author} {\bibfnamefont {I.-M.}\ \bibnamefont
  {H{\o}yvik}}, \bibinfo {author} {\bibfnamefont {R.~H.}\ \bibnamefont
  {Myhre}}, \ and\ \bibinfo {author} {\bibfnamefont {H.}~\bibnamefont {Koch}},\
  }\bibfield  {title} {\enquote {\bibinfo {title} {Correlated natural
  transition orbitals for core excitation energies in multilevel coupled
  cluster models},}\ }\href@noop {} {\bibfield  {journal} {\bibinfo  {journal}
  {J. Chem. Phys.}\ }\textbf {\bibinfo {volume} {146}},\ \bibinfo {pages}
  {144109} (\bibinfo {year} {2017})}\BibitemShut {NoStop}%
\bibitem [{\citenamefont {Folkestad}\ and\ \citenamefont
  {Koch}(2019)}]{folkestad2019multilevel}%
  \BibitemOpen
  \bibfield  {author} {\bibinfo {author} {\bibfnamefont {S.~D.}\ \bibnamefont
  {Folkestad}}\ and\ \bibinfo {author} {\bibfnamefont {H.}~\bibnamefont
  {Koch}},\ }\bibfield  {title} {\enquote {\bibinfo {title} {The multilevel
  {CC2} and {CCSD} methods with correlated natural transition orbitals},}\
  }\href@noop {} {\bibfield  {journal} {\bibinfo  {journal} {J. Chem. Theor.
  Comput.}\ } (\bibinfo {year} {2019})}\BibitemShut {NoStop}%
\bibitem [{\citenamefont {Senn}\ and\ \citenamefont
  {Thiel}(2009)}]{senn2009qm}%
  \BibitemOpen
  \bibfield  {author} {\bibinfo {author} {\bibfnamefont {H.~M.}\ \bibnamefont
  {Senn}}\ and\ \bibinfo {author} {\bibfnamefont {W.}~\bibnamefont {Thiel}},\
  }\bibfield  {title} {\enquote {\bibinfo {title} {{QM/MM} methods for
  biomolecular systems},}\ }\href@noop {} {\bibfield  {journal} {\bibinfo
  {journal} {Angew. Chem. Int. Ed.}\ }\textbf {\bibinfo {volume} {48}},\
  \bibinfo {pages} {1198--1229} (\bibinfo {year} {2009})}\BibitemShut {NoStop}%
\bibitem [{\citenamefont {Tomasi}, \citenamefont {Mennucci},\ and\
  \citenamefont {Cammi}(2005)}]{tomasi2005quantum}%
  \BibitemOpen
  \bibfield  {author} {\bibinfo {author} {\bibfnamefont {J.}~\bibnamefont
  {Tomasi}}, \bibinfo {author} {\bibfnamefont {B.}~\bibnamefont {Mennucci}}, \
  and\ \bibinfo {author} {\bibfnamefont {R.}~\bibnamefont {Cammi}},\ }\bibfield
   {title} {\enquote {\bibinfo {title} {Quantum mechanical continuum solvation
  models},}\ }\href@noop {} {\bibfield  {journal} {\bibinfo  {journal} {Chem.
  Rev.}\ }\textbf {\bibinfo {volume} {105}},\ \bibinfo {pages} {2999--3094}
  (\bibinfo {year} {2005})}\BibitemShut {NoStop}%
\bibitem [{\citenamefont {Warshel}\ and\ \citenamefont
  {Karplus}(1972)}]{warshel1972calculation}%
  \BibitemOpen
  \bibfield  {author} {\bibinfo {author} {\bibfnamefont {A.}~\bibnamefont
  {Warshel}}\ and\ \bibinfo {author} {\bibfnamefont {M.}~\bibnamefont
  {Karplus}},\ }\bibfield  {title} {\enquote {\bibinfo {title} {Calculation of
  ground and excited state potential surfaces of conjugated molecules. i.
  formulation and parametrization},}\ }\href@noop {} {\bibfield  {journal}
  {\bibinfo  {journal} {J. Am. Chem. Soc.}\ }\textbf {\bibinfo {volume} {94}},\
  \bibinfo {pages} {5612--5625} (\bibinfo {year} {1972})}\BibitemShut {NoStop}%
\bibitem [{\citenamefont {Warshel}\ and\ \citenamefont
  {Levitt}(1976)}]{warshel1976theoretical}%
  \BibitemOpen
  \bibfield  {author} {\bibinfo {author} {\bibfnamefont {A.}~\bibnamefont
  {Warshel}}\ and\ \bibinfo {author} {\bibfnamefont {M.}~\bibnamefont
  {Levitt}},\ }\bibfield  {title} {\enquote {\bibinfo {title} {Theoretical
  studies of enzymic reactions: dielectric, electrostatic and steric
  stabilization of the carbonium ion in the reaction of lysozyme},}\
  }\href@noop {} {\bibfield  {journal} {\bibinfo  {journal} {J. Mol. Biol.}\
  }\textbf {\bibinfo {volume} {103}},\ \bibinfo {pages} {227--249} (\bibinfo
  {year} {1976})}\BibitemShut {NoStop}%
\bibitem [{\citenamefont {Wesolowski}\ and\ \citenamefont
  {Warshel}(1993)}]{wesolowski1993frozen}%
  \BibitemOpen
  \bibfield  {author} {\bibinfo {author} {\bibfnamefont {T.~A.}\ \bibnamefont
  {Wesolowski}}\ and\ \bibinfo {author} {\bibfnamefont {A.}~\bibnamefont
  {Warshel}},\ }\bibfield  {title} {\enquote {\bibinfo {title} {Frozen density
  functional approach for ab initio calculations of solvated molecules},}\
  }\href@noop {} {\bibfield  {journal} {\bibinfo  {journal} {J. Phys. Chem.}\
  }\textbf {\bibinfo {volume} {97}},\ \bibinfo {pages} {8050--8053} (\bibinfo
  {year} {1993})}\BibitemShut {NoStop}%
\bibitem [{\citenamefont {Tomasi}\ \emph {et~al.}(2002)\citenamefont {Tomasi},
  \citenamefont {Cammi}, \citenamefont {Mennucci}, \citenamefont {Cappelli},\
  and\ \citenamefont {Corni}}]{tomasi2002molecular}%
  \BibitemOpen
  \bibfield  {author} {\bibinfo {author} {\bibfnamefont {J.}~\bibnamefont
  {Tomasi}}, \bibinfo {author} {\bibfnamefont {R.}~\bibnamefont {Cammi}},
  \bibinfo {author} {\bibfnamefont {B.}~\bibnamefont {Mennucci}}, \bibinfo
  {author} {\bibfnamefont {C.}~\bibnamefont {Cappelli}}, \ and\ \bibinfo
  {author} {\bibfnamefont {S.}~\bibnamefont {Corni}},\ }\bibfield  {title}
  {\enquote {\bibinfo {title} {Molecular properties in solution described with
  a continuum solvation model},}\ }\href@noop {} {\bibfield  {journal}
  {\bibinfo  {journal} {Phys. Chem. Chem. Phys.}\ }\textbf {\bibinfo {volume}
  {4}},\ \bibinfo {pages} {5697--5712} (\bibinfo {year} {2002})}\BibitemShut
  {NoStop}%
\bibitem [{\citenamefont {Cappelli}(2016)}]{cappelli2016integrated}%
  \BibitemOpen
  \bibfield  {author} {\bibinfo {author} {\bibfnamefont {C.}~\bibnamefont
  {Cappelli}},\ }\bibfield  {title} {\enquote {\bibinfo {title} {Integrated
  {QM}/polarizable {MM}/continuum approaches to model chiroptical properties of
  strongly interacting solute-solvent systems},}\ }\href@noop {} {\bibfield
  {journal} {\bibinfo  {journal} {Int. J. Quantum Chem.}\ }\textbf {\bibinfo
  {volume} {116}},\ \bibinfo {pages} {1532--1542} (\bibinfo {year}
  {2016})}\BibitemShut {NoStop}%
\bibitem [{\citenamefont {Giovannini}, \citenamefont {Egidi},\ and\
  \citenamefont {Cappelli}(2020{\natexlab{a}})}]{giovannini2020molecular}%
  \BibitemOpen
  \bibfield  {author} {\bibinfo {author} {\bibfnamefont {T.}~\bibnamefont
  {Giovannini}}, \bibinfo {author} {\bibfnamefont {F.}~\bibnamefont {Egidi}}, \
  and\ \bibinfo {author} {\bibfnamefont {C.}~\bibnamefont {Cappelli}},\
  }\bibfield  {title} {\enquote {\bibinfo {title} {Molecular spectroscopy of
  aqueous solutions: a theoretical perspective},}\ }\href@noop {} {\bibfield
  {journal} {\bibinfo  {journal} {Chem. Soc. Rev.}\ }\textbf {\bibinfo {volume}
  {49}},\ \bibinfo {pages} {5664--5677} (\bibinfo {year}
  {2020}{\natexlab{a}})}\BibitemShut {NoStop}%
\bibitem [{\citenamefont {Giovannini}, \citenamefont {Egidi},\ and\
  \citenamefont {Cappelli}(2020{\natexlab{b}})}]{giovannini2020theory}%
  \BibitemOpen
  \bibfield  {author} {\bibinfo {author} {\bibfnamefont {T.}~\bibnamefont
  {Giovannini}}, \bibinfo {author} {\bibfnamefont {F.}~\bibnamefont {Egidi}}, \
  and\ \bibinfo {author} {\bibfnamefont {C.}~\bibnamefont {Cappelli}},\
  }\bibfield  {title} {\enquote {\bibinfo {title} {Theory and algorithms for
  chiroptical properties and spectroscopies of aqueous systems},}\ }\href@noop
  {} {\bibfield  {journal} {\bibinfo  {journal} {Phys. Chem. Chem. Phys.}\
  }\textbf {\bibinfo {volume} {22}},\ \bibinfo {pages} {22864--22879} (\bibinfo
  {year} {2020}{\natexlab{b}})}\BibitemShut {NoStop}%
\bibitem [{\citenamefont {Folkestad}\ \emph
  {et~al.}(2024{\natexlab{a}})\citenamefont {Folkestad}, \citenamefont {Paul},
  \citenamefont {Paul}, \citenamefont {Coriani}, \citenamefont {Odelius},
  \citenamefont {Iannuzzi},\ and\ \citenamefont
  {Koch}}]{folkestad2024understanding}%
  \BibitemOpen
  \bibfield  {author} {\bibinfo {author} {\bibfnamefont {S.~D.}\ \bibnamefont
  {Folkestad}}, \bibinfo {author} {\bibfnamefont {A.~C.}\ \bibnamefont {Paul}},
  \bibinfo {author} {\bibfnamefont {R.}~\bibnamefont {Paul}}, \bibinfo {author}
  {\bibfnamefont {S.}~\bibnamefont {Coriani}}, \bibinfo {author} {\bibfnamefont
  {M.}~\bibnamefont {Odelius}}, \bibinfo {author} {\bibfnamefont
  {M.}~\bibnamefont {Iannuzzi}}, \ and\ \bibinfo {author} {\bibfnamefont
  {H.}~\bibnamefont {Koch}},\ }\bibfield  {title} {\enquote {\bibinfo {title}
  {Understanding x-ray absorption in liquid water using triple excitations in
  multilevel coupled cluster theory},}\ }\href@noop {} {\bibfield  {journal}
  {\bibinfo  {journal} {Nat. Commun.}\ }\textbf {\bibinfo {volume} {15}},\
  \bibinfo {pages} {3551} (\bibinfo {year} {2024}{\natexlab{a}})}\BibitemShut
  {NoStop}%
\bibitem [{\citenamefont {Folkestad}\ \emph
  {et~al.}(2024{\natexlab{b}})\citenamefont {Folkestad}, \citenamefont {Paul},
  \citenamefont {Paul~née Matveeva}, \citenamefont {Reinholdt}, \citenamefont
  {Coriani}, \citenamefont {Odelius},\ and\ \citenamefont
  {Koch}}]{folkestad2024quantum}%
  \BibitemOpen
  \bibfield  {author} {\bibinfo {author} {\bibfnamefont {S.~D.}\ \bibnamefont
  {Folkestad}}, \bibinfo {author} {\bibfnamefont {A.~C.}\ \bibnamefont {Paul}},
  \bibinfo {author} {\bibfnamefont {R.}~\bibnamefont {Paul~nee Matveeva}},
  \bibinfo {author} {\bibfnamefont {P.}~\bibnamefont {Reinholdt}}, \bibinfo
  {author} {\bibfnamefont {S.}~\bibnamefont {Coriani}}, \bibinfo {author}
  {\bibfnamefont {M.}~\bibnamefont {Odelius}}, \ and\ \bibinfo {author}
  {\bibfnamefont {H.}~\bibnamefont {Koch}},\ }\bibfield  {title} {\enquote
  {\bibinfo {title} {Quantum mechanical versus polarizable embedding schemes: A
  study of the xray absorption spectra of aqueous ammonia and ammonium},}\
  }\href@noop {} {\bibfield  {journal} {\bibinfo  {journal} {J. Chem. Theory
  Comput.}\ }\textbf {\bibinfo {volume} {20}},\ \bibinfo {pages} {4161--4169}
  (\bibinfo {year} {2024}{\natexlab{b}})}\BibitemShut {NoStop}%
\bibitem [{\citenamefont {Giovannini}, \citenamefont {G{\'o}mez},\ and\
  \citenamefont {Cappelli}(2025)}]{giovannini2025modeling}%
  \BibitemOpen
  \bibfield  {author} {\bibinfo {author} {\bibfnamefont {T.}~\bibnamefont
  {Giovannini}}, \bibinfo {author} {\bibfnamefont {S.}~\bibnamefont
  {G{\'o}mez}}, \ and\ \bibinfo {author} {\bibfnamefont {C.}~\bibnamefont
  {Cappelli}},\ }\bibfield  {title} {\enquote {\bibinfo {title} {Modeling raman
  spectra in complex environments: From solutions to surface-enhanced raman
  scattering},}\ }\href@noop {} {\bibfield  {journal} {\bibinfo  {journal} {J.
  Phys. Chem. Lett.}\ }\textbf {\bibinfo {volume} {16}},\ \bibinfo {pages}
  {3106--3121} (\bibinfo {year} {2025})}\BibitemShut {NoStop}%
\bibitem [{\citenamefont {Folkestad}, \citenamefont {Kjønstad},\ and\
  \citenamefont {Koch}(2019)}]{Folkestad2019}%
  \BibitemOpen
  \bibfield  {author} {\bibinfo {author} {\bibfnamefont {S.~D.}\ \bibnamefont
  {Folkestad}}, \bibinfo {author} {\bibfnamefont {E.~F.}\ \bibnamefont
  {Kjønstad}}, \ and\ \bibinfo {author} {\bibfnamefont {H.}~\bibnamefont
  {Koch}},\ }\bibfield  {title} {\enquote {\bibinfo {title} {An efficient
  algorithm for {Cholesky} decomposition of electron repulsion integrals},}\
  }\href {\doibase 10.1063/1.5083802} {\bibfield  {journal} {\bibinfo
  {journal} {J. Chem. Phys.}\ }\textbf {\bibinfo {volume} {150}},\ \bibinfo
  {pages} {194112} (\bibinfo {year} {2019})}\BibitemShut {NoStop}%
\bibitem [{\citenamefont {Folkestad}\ \emph {et~al.}(2021)\citenamefont
  {Folkestad}, \citenamefont {Kj{\o}nstad}, \citenamefont {Goletto},\ and\
  \citenamefont {Koch}}]{folkestad2021multilevel}%
  \BibitemOpen
  \bibfield  {author} {\bibinfo {author} {\bibfnamefont {S.~D.}\ \bibnamefont
  {Folkestad}}, \bibinfo {author} {\bibfnamefont {E.~F.}\ \bibnamefont
  {Kj{\o}nstad}}, \bibinfo {author} {\bibfnamefont {L.}~\bibnamefont
  {Goletto}}, \ and\ \bibinfo {author} {\bibfnamefont {H.}~\bibnamefont
  {Koch}},\ }\bibfield  {title} {\enquote {\bibinfo {title} {{Multilevel CC2
  and CCSD in reduced orbital spaces: electronic excitations in large molecular
  systems}},}\ }\href@noop {} {\bibfield  {journal} {\bibinfo  {journal} {J.
  Chem. Theory Comput.}\ }\textbf {\bibinfo {volume} {17}},\ \bibinfo {pages}
  {714--726} (\bibinfo {year} {2021})}\BibitemShut {NoStop}%
\bibitem [{\citenamefont {Olsen}, \citenamefont {J{\o}rgensen},\ and\
  \citenamefont {Simons}(1990)}]{olsen1990passing}%
  \BibitemOpen
  \bibfield  {author} {\bibinfo {author} {\bibfnamefont {J.}~\bibnamefont
  {Olsen}}, \bibinfo {author} {\bibfnamefont {P.}~\bibnamefont {J{\o}rgensen}},
  \ and\ \bibinfo {author} {\bibfnamefont {J.}~\bibnamefont {Simons}},\
  }\bibfield  {title} {\enquote {\bibinfo {title} {Passing the one-billion
  limit in full configuration-interaction (fci) calculations},}\ }\href@noop {}
  {\bibfield  {journal} {\bibinfo  {journal} {Chemical Physics Letters}\
  }\textbf {\bibinfo {volume} {169}},\ \bibinfo {pages} {463--472} (\bibinfo
  {year} {1990})}\BibitemShut {NoStop}%
\bibitem [{\citenamefont {Myhre}\ and\ \citenamefont
  {Koch}(2016)}]{myhre2016multilevel}%
  \BibitemOpen
  \bibfield  {author} {\bibinfo {author} {\bibfnamefont {R.~H.}\ \bibnamefont
  {Myhre}}\ and\ \bibinfo {author} {\bibfnamefont {H.}~\bibnamefont {Koch}},\
  }\bibfield  {title} {\enquote {\bibinfo {title} {{The multilevel {CC3}
  coupled cluster model}},}\ }\href@noop {} {\bibfield  {journal} {\bibinfo
  {journal} {J. Chem. Phys.}\ }\textbf {\bibinfo {volume} {145}},\ \bibinfo
  {pages} {44111} (\bibinfo {year} {2016})}\BibitemShut {NoStop}%
\bibitem [{\citenamefont {eT-program
  developers}(2025)}]{geometries_and_ouputs}%
  \BibitemOpen
  \bibfield  {author} {\bibinfo {author} {\bibfnamefont {T.}~\bibnamefont
  {eT-program developers}},\ }\href {\doibase 10.5281/zenodo.17314764}
  {\enquote {\bibinfo {title} {Geometries and outputs for "et 2.0: An efficient
  open-source molecular electronic structure program"},}\ } (\bibinfo {year}
  {2025})\BibitemShut {NoStop}%
\bibitem [{\citenamefont {Kj{\o}nstad}\ and\ \citenamefont
  {Koch}(2023)}]{kjonstad2023communication}%
  \BibitemOpen
  \bibfield  {author} {\bibinfo {author} {\bibfnamefont {E.~F.}\ \bibnamefont
  {Kj{\o}nstad}}\ and\ \bibinfo {author} {\bibfnamefont {H.}~\bibnamefont
  {Koch}},\ }\bibfield  {title} {\enquote {\bibinfo {title} {Communication:
  Non-adiabatic derivative coupling elements for the coupled cluster singles
  and doubles model},}\ }\href@noop {} {\bibfield  {journal} {\bibinfo
  {journal} {J. Chem. Phys.}\ }\textbf {\bibinfo {volume} {158}},\ \bibinfo
  {pages} {161106} (\bibinfo {year} {2023})}\BibitemShut {NoStop}%
\bibitem [{\citenamefont {Kj{\o}nstad}\ and\ \citenamefont
  {Koch}(2017)}]{kjonstad2017resolving}%
  \BibitemOpen
  \bibfield  {author} {\bibinfo {author} {\bibfnamefont {E.~F.}\ \bibnamefont
  {Kj{\o}nstad}}\ and\ \bibinfo {author} {\bibfnamefont {H.}~\bibnamefont
  {Koch}},\ }\bibfield  {title} {\enquote {\bibinfo {title} {Resolving the
  notorious case of conical intersections for coupled cluster dynamics},}\
  }\href@noop {} {\bibfield  {journal} {\bibinfo  {journal} {J. Phys. Chem.
  Lett.}\ }\textbf {\bibinfo {volume} {8}},\ \bibinfo {pages} {4801--4807}
  (\bibinfo {year} {2017})}\BibitemShut {NoStop}%
\bibitem [{\citenamefont {Kj{\o}nstad}\ and\ \citenamefont
  {Koch}(2019)}]{kjonstad2019orbital}%
  \BibitemOpen
  \bibfield  {author} {\bibinfo {author} {\bibfnamefont {E.~F.}\ \bibnamefont
  {Kj{\o}nstad}}\ and\ \bibinfo {author} {\bibfnamefont {H.}~\bibnamefont
  {Koch}},\ }\bibfield  {title} {\enquote {\bibinfo {title} {An orbital
  invariant similarity constrained coupled cluster model},}\ }\href@noop {}
  {\bibfield  {journal} {\bibinfo  {journal} {J. Chem. Theory Comput.}\
  }\textbf {\bibinfo {volume} {15}},\ \bibinfo {pages} {5386--5397} (\bibinfo
  {year} {2019})}\BibitemShut {NoStop}%
\bibitem [{\citenamefont {Kj{\o}nstad}, \citenamefont {Angelico},\ and\
  \citenamefont {Koch}(2024)}]{kjonstad2024coupled}%
  \BibitemOpen
  \bibfield  {author} {\bibinfo {author} {\bibfnamefont {E.~F.}\ \bibnamefont
  {Kj{\o}nstad}}, \bibinfo {author} {\bibfnamefont {S.}~\bibnamefont
  {Angelico}}, \ and\ \bibinfo {author} {\bibfnamefont {H.}~\bibnamefont
  {Koch}},\ }\bibfield  {title} {\enquote {\bibinfo {title} {Coupled cluster
  theory for nonadiabatic dynamics: nuclear gradients and nonadiabatic
  couplings in similarity constrained coupled cluster theory},}\ }\href@noop {}
  {\bibfield  {journal} {\bibinfo  {journal} {J. Chem. Theory Comput.}\
  }\textbf {\bibinfo {volume} {20}},\ \bibinfo {pages} {7080--7092} (\bibinfo
  {year} {2024})}\BibitemShut {NoStop}%
\bibitem [{\citenamefont {Hait}\ \emph {et~al.}(2024)\citenamefont {Hait},
  \citenamefont {Lahana}, \citenamefont {Fajen}, \citenamefont {Paz},
  \citenamefont {Unzueta}, \citenamefont {Rana}, \citenamefont {Lu},
  \citenamefont {Wang}, \citenamefont {Kj{\o}nstad}, \citenamefont {Koch} \emph
  {et~al.}}]{hait2024prediction}%
  \BibitemOpen
  \bibfield  {author} {\bibinfo {author} {\bibfnamefont {D.}~\bibnamefont
  {Hait}}, \bibinfo {author} {\bibfnamefont {D.}~\bibnamefont {Lahana}},
  \bibinfo {author} {\bibfnamefont {O.~J.}\ \bibnamefont {Fajen}}, \bibinfo
  {author} {\bibfnamefont {A.~S.}\ \bibnamefont {Paz}}, \bibinfo {author}
  {\bibfnamefont {P.~A.}\ \bibnamefont {Unzueta}}, \bibinfo {author}
  {\bibfnamefont {B.}~\bibnamefont {Rana}}, \bibinfo {author} {\bibfnamefont
  {L.}~\bibnamefont {Lu}}, \bibinfo {author} {\bibfnamefont {Y.}~\bibnamefont
  {Wang}}, \bibinfo {author} {\bibfnamefont {E.~F.}\ \bibnamefont
  {Kj{\o}nstad}}, \bibinfo {author} {\bibfnamefont {H.}~\bibnamefont {Koch}},
  \emph {et~al.},\ }\bibfield  {title} {\enquote {\bibinfo {title} {Prediction
  of photodynamics of 200 nm excited cyclobutanone with linear response
  electronic structure and ab initio multiple spawning},}\ }\href@noop {}
  {\bibfield  {journal} {\bibinfo  {journal} {J. Chem. Phys.}\ }\textbf
  {\bibinfo {volume} {160}},\ \bibinfo {pages} {244101} (\bibinfo {year}
  {2024})}\BibitemShut {NoStop}%
\bibitem [{\citenamefont {Stoll}\ \emph {et~al.}(2025)\citenamefont {Stoll},
  \citenamefont {Angelico}, \citenamefont {Kj{\o}nstad},\ and\ \citenamefont
  {Koch}}]{stoll2025similarity}%
  \BibitemOpen
  \bibfield  {author} {\bibinfo {author} {\bibfnamefont {L.}~\bibnamefont
  {Stoll}}, \bibinfo {author} {\bibfnamefont {S.}~\bibnamefont {Angelico}},
  \bibinfo {author} {\bibfnamefont {E.~F.}\ \bibnamefont {Kj{\o}nstad}}, \ and\
  \bibinfo {author} {\bibfnamefont {H.}~\bibnamefont {Koch}},\ }\bibfield
  {title} {\enquote {\bibinfo {title} {{Similarity Constrained CC2 for
  Efficient Coupled Cluster Nonadiabatic Dynamics}},}\ }\href@noop {}
  {\bibfield  {journal} {\bibinfo  {journal} {arXiv preprint arXiv:2504.11157}\
  } (\bibinfo {year} {2025})}\BibitemShut {NoStop}%
\bibitem [{\citenamefont {Angelico}, \citenamefont {Kj{\o}nstad},\ and\
  \citenamefont {Koch}(2025)}]{angelico2025determining}%
  \BibitemOpen
  \bibfield  {author} {\bibinfo {author} {\bibfnamefont {S.}~\bibnamefont
  {Angelico}}, \bibinfo {author} {\bibfnamefont {E.~F.}\ \bibnamefont
  {Kj{\o}nstad}}, \ and\ \bibinfo {author} {\bibfnamefont {H.}~\bibnamefont
  {Koch}},\ }\bibfield  {title} {\enquote {\bibinfo {title} {Determining
  minimum energy conical intersections by enveloping the seam: exploring ground
  and excited state intersections in coupled cluster theory},}\ }\href@noop {}
  {\bibfield  {journal} {\bibinfo  {journal} {J. Phys. Chem. Lett.}\ }\textbf
  {\bibinfo {volume} {16}},\ \bibinfo {pages} {561--567} (\bibinfo {year}
  {2025})}\BibitemShut {NoStop}%
\bibitem [{\citenamefont {Rossi}\ \emph {et~al.}(2025)\citenamefont {Rossi},
  \citenamefont {Kj{\o}nstad}, \citenamefont {Angelico},\ and\ \citenamefont
  {Koch}}]{rossi2025generalized}%
  \BibitemOpen
  \bibfield  {author} {\bibinfo {author} {\bibfnamefont {F.}~\bibnamefont
  {Rossi}}, \bibinfo {author} {\bibfnamefont {E.~F.}\ \bibnamefont
  {Kj{\o}nstad}}, \bibinfo {author} {\bibfnamefont {S.}~\bibnamefont
  {Angelico}}, \ and\ \bibinfo {author} {\bibfnamefont {H.}~\bibnamefont
  {Koch}},\ }\bibfield  {title} {\enquote {\bibinfo {title} {Generalized
  coupled cluster theory for ground and excited state intersections},}\
  }\href@noop {} {\bibfield  {journal} {\bibinfo  {journal} {J. Phys. Chem.
  Lett.}\ }\textbf {\bibinfo {volume} {16}},\ \bibinfo {pages} {568--578}
  (\bibinfo {year} {2025})}\BibitemShut {NoStop}%
\bibitem [{\citenamefont {Rossi}\ and\ \citenamefont
  {Koch}(2025)}]{rossi2025convex}%
  \BibitemOpen
  \bibfield  {author} {\bibinfo {author} {\bibfnamefont {F.}~\bibnamefont
  {Rossi}}\ and\ \bibinfo {author} {\bibfnamefont {H.}~\bibnamefont {Koch}},\
  }\bibfield  {title} {\enquote {\bibinfo {title} {Convex hartree-fock theory:
  A simple framework for ground state conical intersections},}\ }\href@noop {}
  {\bibfield  {journal} {\bibinfo  {journal} {arXiv preprint arXiv:2508.21453}\
  } (\bibinfo {year} {2025})}\BibitemShut {NoStop}%
\bibitem [{\citenamefont {Smedsrud}(2024)}]{smedsrud2024coupled}%
  \BibitemOpen
  \bibfield  {author} {\bibinfo {author} {\bibfnamefont {W.~H.}\ \bibnamefont
  {Smedsrud}},\ }\emph {\bibinfo {title} {Coupled-Cluster Equation of Motion
  method (QED-CCSD-1) for Triplet Excitations in Optical Cavities}},\
  \href@noop {} {Master's thesis},\ \bibinfo  {school} {NTNU} (\bibinfo {year}
  {2024})\BibitemShut {NoStop}%
\bibitem [{\citenamefont {Castagnola}, \citenamefont {Lexander},\ and\
  \citenamefont {Koch}(2025)}]{castagnola2025}%
  \BibitemOpen
  \bibfield  {author} {\bibinfo {author} {\bibfnamefont {M.}~\bibnamefont
  {Castagnola}}, \bibinfo {author} {\bibfnamefont {M.~T.}\ \bibnamefont
  {Lexander}}, \ and\ \bibinfo {author} {\bibfnamefont {H.}~\bibnamefont
  {Koch}},\ }\bibfield  {title} {\enquote {\bibinfo {title} {{Realistic Ab
  Initio Predictions of Excimer Behavior under Collective Light-Matter Strong
  Coupling}},}\ }\href {\doibase 10.1103/PhysRevX.15.021040} {\bibfield
  {journal} {\bibinfo  {journal} {Phys. Rev. X}\ }\textbf {\bibinfo {volume}
  {15}},\ \bibinfo {pages} {021040} (\bibinfo {year} {2025})}\BibitemShut
  {NoStop}%
\bibitem [{\citenamefont {Barlini}\ \emph {et~al.}(2024)\citenamefont
  {Barlini}, \citenamefont {Bianchi}, \citenamefont {Ronca},\ and\
  \citenamefont {Koch}}]{barilini2024}%
  \BibitemOpen
  \bibfield  {author} {\bibinfo {author} {\bibfnamefont {A.}~\bibnamefont
  {Barlini}}, \bibinfo {author} {\bibfnamefont {A.}~\bibnamefont {Bianchi}},
  \bibinfo {author} {\bibfnamefont {E.}~\bibnamefont {Ronca}}, \ and\ \bibinfo
  {author} {\bibfnamefont {H.}~\bibnamefont {Koch}},\ }\bibfield  {title}
  {\enquote {\bibinfo {title} {{Theory of Magnetic Properties in Quantum
  Electrodynamics Environments: Application to Molecular Aromaticity}},}\
  }\href {\doibase 10.1021/acs.jctc.4c00195} {\bibfield  {journal} {\bibinfo
  {journal} {J. Chem. Theory Comput.}\ }\textbf {\bibinfo {volume} {20}},\
  \bibinfo {pages} {7841--7854} (\bibinfo {year} {2024})},\ \Eprint
  {http://arxiv.org/abs/https://doi.org/10.1021/acs.jctc.4c00195}
  {https://doi.org/10.1021/acs.jctc.4c00195} \BibitemShut {NoStop}%
\bibitem [{\citenamefont {Riso}\ \emph {et~al.}(2023)\citenamefont {Riso},
  \citenamefont {Grazioli}, \citenamefont {Ronca}, \citenamefont {Giovannini},\
  and\ \citenamefont {Koch}}]{Riso2023}%
  \BibitemOpen
  \bibfield  {author} {\bibinfo {author} {\bibfnamefont {R.~R.}\ \bibnamefont
  {Riso}}, \bibinfo {author} {\bibfnamefont {L.}~\bibnamefont {Grazioli}},
  \bibinfo {author} {\bibfnamefont {E.}~\bibnamefont {Ronca}}, \bibinfo
  {author} {\bibfnamefont {T.}~\bibnamefont {Giovannini}}, \ and\ \bibinfo
  {author} {\bibfnamefont {H.}~\bibnamefont {Koch}},\ }\bibfield  {title}
  {\enquote {\bibinfo {title} {Strong coupling in chiral cavities:
  Nonperturbative framework for enantiomer discrimination},}\ }\href {\doibase
  10.1103/PhysRevX.13.031002} {\bibfield  {journal} {\bibinfo  {journal} {Phys.
  Rev. X}\ }\textbf {\bibinfo {volume} {13}},\ \bibinfo {pages} {031002}
  (\bibinfo {year} {2023})}\BibitemShut {NoStop}%
\bibitem [{\citenamefont {Riso}\ \emph {et~al.}(2025)\citenamefont {Riso},
  \citenamefont {Castagnola}, \citenamefont {Ronca},\ and\ \citenamefont
  {Koch}}]{Riso_2025}%
  \BibitemOpen
  \bibfield  {author} {\bibinfo {author} {\bibfnamefont {R.~R.}\ \bibnamefont
  {Riso}}, \bibinfo {author} {\bibfnamefont {M.}~\bibnamefont {Castagnola}},
  \bibinfo {author} {\bibfnamefont {E.}~\bibnamefont {Ronca}}, \ and\ \bibinfo
  {author} {\bibfnamefont {H.}~\bibnamefont {Koch}},\ }\bibfield  {title}
  {\enquote {\bibinfo {title} {Chiral polaritonics: cavity-mediated
  enantioselective excitation condensation},}\ }\href {\doibase
  10.1088/1361-6633/ad9ed9} {\bibfield  {journal} {\bibinfo  {journal} {Reports
  on Progress in Physics}\ }\textbf {\bibinfo {volume} {88}},\ \bibinfo {pages}
  {027901} (\bibinfo {year} {2025})}\BibitemShut {NoStop}%
\bibitem [{\citenamefont {Alessandro}\ \emph {et~al.}(2025)\citenamefont
  {Alessandro}, \citenamefont {Castagnola}, \citenamefont {Koch},\ and\
  \citenamefont {Ronca}}]{Alessandro2025}%
  \BibitemOpen
  \bibfield  {author} {\bibinfo {author} {\bibfnamefont {R.}~\bibnamefont
  {Alessandro}}, \bibinfo {author} {\bibfnamefont {M.}~\bibnamefont
  {Castagnola}}, \bibinfo {author} {\bibfnamefont {H.}~\bibnamefont {Koch}}, \
  and\ \bibinfo {author} {\bibfnamefont {E.}~\bibnamefont {Ronca}},\ }\bibfield
   {title} {\enquote {\bibinfo {title} {{A Complete Active Space
  Self-Consistent Field Approach for Molecules in QED Environments}},}\ }\href
  {\doibase 10.1021/acs.jctc.5c00519} {\bibfield  {journal} {\bibinfo
  {journal} {J. Chem. Theory Comput.}\ }\textbf {\bibinfo {volume} {21}},\
  \bibinfo {pages} {6862--6873} (\bibinfo {year} {2025})},\ \bibinfo {note}
  {pMID: 40619708},\ \Eprint
  {http://arxiv.org/abs/https://doi.org/10.1021/acs.jctc.5c00519}
  {https://doi.org/10.1021/acs.jctc.5c00519} \BibitemShut {NoStop}%
\bibitem [{\citenamefont {Matveeva}, \citenamefont {Folkestad},\ and\
  \citenamefont {H{\o}yvik}(2023)}]{Matveeva:2023aa}%
  \BibitemOpen
  \bibfield  {author} {\bibinfo {author} {\bibfnamefont {R.}~\bibnamefont
  {Matveeva}}, \bibinfo {author} {\bibfnamefont {S.~D.}\ \bibnamefont
  {Folkestad}}, \ and\ \bibinfo {author} {\bibfnamefont {I.-M.}\ \bibnamefont
  {H{\o}yvik}},\ }\bibfield  {title} {\enquote {\bibinfo {title}
  {Particle-breaking hartree--fock theory for open molecular systems},}\ }\href
  {\doibase 10.1021/acs.jpca.2c07686} {\bibfield  {journal} {\bibinfo
  {journal} {J. Phys. Chem. A}\ }\textbf {\bibinfo {volume} {127}},\ \bibinfo
  {pages} {1329--1341} (\bibinfo {year} {2023})}\BibitemShut {NoStop}%
\bibitem [{\citenamefont {Paul~n{\'e}e Matveeva}\ \emph
  {et~al.}(2024)\citenamefont {Paul~n{\'e}e Matveeva}, \citenamefont
  {Folkestad}, \citenamefont {Sannes},\ and\ \citenamefont
  {H{\o}yvik}}]{Paul:2024aa}%
  \BibitemOpen
  \bibfield  {author} {\bibinfo {author} {\bibfnamefont {R.}~\bibnamefont
  {Paul~n{\'e}e Matveeva}}, \bibinfo {author} {\bibfnamefont {S.~D.}\
  \bibnamefont {Folkestad}}, \bibinfo {author} {\bibfnamefont {B.~S.}\
  \bibnamefont {Sannes}}, \ and\ \bibinfo {author} {\bibfnamefont {I.-M.}\
  \bibnamefont {H{\o}yvik}},\ }\bibfield  {title} {\enquote {\bibinfo {title}
  {Particle-breaking unrestricted hartree--fock theory for open molecular
  systems},}\ }\href {\doibase 10.1021/acs.jpca.3c07231} {\bibfield  {journal}
  {\bibinfo  {journal} {J. Phys. Chem. A}\ }\textbf {\bibinfo {volume} {128}},\
  \bibinfo {pages} {1533--1542} (\bibinfo {year} {2024})}\BibitemShut {NoStop}%
\bibitem [{\citenamefont {Pedersen}\ \emph {et~al.}(2025)\citenamefont
  {Pedersen}, \citenamefont {Sannes}, \citenamefont {Paul~n{\'e}e Matveeva},
  \citenamefont {Coriani},\ and\ \citenamefont {H{\o}yvik}}]{jacobsen:2025}%
  \BibitemOpen
  \bibfield  {author} {\bibinfo {author} {\bibfnamefont {J.}~\bibnamefont
  {Pedersen}}, \bibinfo {author} {\bibfnamefont {B.~S.}\ \bibnamefont
  {Sannes}}, \bibinfo {author} {\bibfnamefont {R.}~\bibnamefont {Paul~n{\'e}e
  Matveeva}}, \bibinfo {author} {\bibfnamefont {S.}~\bibnamefont {Coriani}}, \
  and\ \bibinfo {author} {\bibfnamefont {I.-M.}\ \bibnamefont {H{\o}yvik}},\
  }\bibfield  {title} {\enquote {\bibinfo {title} {Time-dependent
  particle-breaking hartree--fock model for electronically open molecules},}\
  }\href {\doibase 10.1021/acs.jpca.5c00810} {\bibfield  {journal} {\bibinfo
  {journal} {J. Phys. Chem. A}\ }\textbf {\bibinfo {volume} {129}},\ \bibinfo
  {pages} {4288--4300} (\bibinfo {year} {2025})}\BibitemShut {NoStop}%
\bibitem [{\citenamefont {Marques}\ and\ \citenamefont
  {Gross}(2004)}]{marques2004time}%
  \BibitemOpen
  \bibfield  {author} {\bibinfo {author} {\bibfnamefont {M.~A.}\ \bibnamefont
  {Marques}}\ and\ \bibinfo {author} {\bibfnamefont {E.~K.}\ \bibnamefont
  {Gross}},\ }\bibfield  {title} {\enquote {\bibinfo {title} {Time-dependent
  density functional theory},}\ }\href@noop {} {\bibfield  {journal} {\bibinfo
  {journal} {Annu. Rev. Phys. Chem.}\ }\textbf {\bibinfo {volume} {55}},\
  \bibinfo {pages} {427--455} (\bibinfo {year} {2004})}\BibitemShut {NoStop}%
\bibitem [{\citenamefont {Giovannini}, \citenamefont {Scavino},\ and\
  \citenamefont {Koch}(2024)}]{giovannini2024time}%
  \BibitemOpen
  \bibfield  {author} {\bibinfo {author} {\bibfnamefont {T.}~\bibnamefont
  {Giovannini}}, \bibinfo {author} {\bibfnamefont {M.}~\bibnamefont {Scavino}},
  \ and\ \bibinfo {author} {\bibfnamefont {H.}~\bibnamefont {Koch}},\
  }\bibfield  {title} {\enquote {\bibinfo {title} {Time-dependent multilevel
  density functional theory},}\ }\href@noop {} {\bibfield  {journal} {\bibinfo
  {journal} {J. Chem. Theory Comput.}\ }\textbf {\bibinfo {volume} {20}},\
  \bibinfo {pages} {3601--3612} (\bibinfo {year} {2024})}\BibitemShut {NoStop}%
\bibitem [{\citenamefont {Giovannini}\ and\ \citenamefont
  {Koch}(2020)}]{giovannini2020energy}%
  \BibitemOpen
  \bibfield  {author} {\bibinfo {author} {\bibfnamefont {T.}~\bibnamefont
  {Giovannini}}\ and\ \bibinfo {author} {\bibfnamefont {H.}~\bibnamefont
  {Koch}},\ }\bibfield  {title} {\enquote {\bibinfo {title} {Energy-based
  molecular orbital localization in a specific spatial region},}\ }\href@noop
  {} {\bibfield  {journal} {\bibinfo  {journal} {J. Chem. Theory Comput.}\
  }\textbf {\bibinfo {volume} {17}},\ \bibinfo {pages} {139--150} (\bibinfo
  {year} {2020})}\BibitemShut {NoStop}%
\bibitem [{\citenamefont {Giovannini}\ and\ \citenamefont
  {Koch}(2022)}]{giovannini2022fragment}%
  \BibitemOpen
  \bibfield  {author} {\bibinfo {author} {\bibfnamefont {T.}~\bibnamefont
  {Giovannini}}\ and\ \bibinfo {author} {\bibfnamefont {H.}~\bibnamefont
  {Koch}},\ }\bibfield  {title} {\enquote {\bibinfo {title} {Fragment localized
  molecular orbitals},}\ }\href@noop {} {\bibfield  {journal} {\bibinfo
  {journal} {J. Chem. Theory Comput.}\ }\textbf {\bibinfo {volume} {18}},\
  \bibinfo {pages} {4806--4813} (\bibinfo {year} {2022})}\BibitemShut {NoStop}%
\bibitem [{\citenamefont {Goletto}\ \emph {et~al.}(2022)\citenamefont
  {Goletto}, \citenamefont {G{\'o}mez}, \citenamefont {Andersen}, \citenamefont
  {Koch},\ and\ \citenamefont {Giovannini}}]{goletto2022linear}%
  \BibitemOpen
  \bibfield  {author} {\bibinfo {author} {\bibfnamefont {L.}~\bibnamefont
  {Goletto}}, \bibinfo {author} {\bibfnamefont {S.}~\bibnamefont {G{\'o}mez}},
  \bibinfo {author} {\bibfnamefont {J.~H.}\ \bibnamefont {Andersen}}, \bibinfo
  {author} {\bibfnamefont {H.}~\bibnamefont {Koch}}, \ and\ \bibinfo {author}
  {\bibfnamefont {T.}~\bibnamefont {Giovannini}},\ }\bibfield  {title}
  {\enquote {\bibinfo {title} {Linear response properties of solvated systems:
  a computational study},}\ }\href@noop {} {\bibfield  {journal} {\bibinfo
  {journal} {Phys. Chem. Chem. Phys.}\ }\textbf {\bibinfo {volume} {24}},\
  \bibinfo {pages} {27866--27878} (\bibinfo {year} {2022})}\BibitemShut
  {NoStop}%
\bibitem [{\citenamefont {Goletto}\ \emph {et~al.}(2021)\citenamefont
  {Goletto}, \citenamefont {Giovannini}, \citenamefont {Folkestad},\ and\
  \citenamefont {Koch}}]{goletto2021combining}%
  \BibitemOpen
  \bibfield  {author} {\bibinfo {author} {\bibfnamefont {L.}~\bibnamefont
  {Goletto}}, \bibinfo {author} {\bibfnamefont {T.}~\bibnamefont {Giovannini}},
  \bibinfo {author} {\bibfnamefont {S.~D.}\ \bibnamefont {Folkestad}}, \ and\
  \bibinfo {author} {\bibfnamefont {H.}~\bibnamefont {Koch}},\ }\bibfield
  {title} {\enquote {\bibinfo {title} {Combining multilevel hartree--fock and
  multilevel coupled cluster approaches with molecular mechanics: a study of
  electronic excitations in solutions},}\ }\href@noop {} {\bibfield  {journal}
  {\bibinfo  {journal} {Phys. Chem. Chem. Phys.}\ }\textbf {\bibinfo {volume}
  {23}},\ \bibinfo {pages} {4413--4425} (\bibinfo {year} {2021})}\BibitemShut
  {NoStop}%
\bibitem [{\citenamefont {Giovannini}(2024)}]{giovannini2024kohn}%
  \BibitemOpen
  \bibfield  {author} {\bibinfo {author} {\bibfnamefont {T.}~\bibnamefont
  {Giovannini}},\ }\bibfield  {title} {\enquote {\bibinfo {title} {Kohn--sham
  fragment energy decomposition analysis},}\ }\href@noop {} {\bibfield
  {journal} {\bibinfo  {journal} {J. Chem. Phys.}\ }\textbf {\bibinfo {volume}
  {161}} (\bibinfo {year} {2024})}\BibitemShut {NoStop}%
\bibitem [{\citenamefont {Lexander}\ \emph
  {et~al.}(2025{\natexlab{b}})\citenamefont {Lexander}, \citenamefont
  {Haugland}, \citenamefont {Rossi},\ and\ \citenamefont
  {Koch}}]{lexander2025spinadaptedsecondquantization}%
  \BibitemOpen
  \bibfield  {author} {\bibinfo {author} {\bibfnamefont {M.~T.}\ \bibnamefont
  {Lexander}}, \bibinfo {author} {\bibfnamefont {T.~S.}\ \bibnamefont
  {Haugland}}, \bibinfo {author} {\bibfnamefont {F.}~\bibnamefont {Rossi}}, \
  and\ \bibinfo {author} {\bibfnamefont {H.}~\bibnamefont {Koch}},\ }\bibfield
  {title} {\enquote {\bibinfo {title} {Spinadaptedsecondquantization. jl 1.0--a
  simple and pedagogical approach to symbolic quantum chemistry},}\ }\href@noop
  {} {\bibfield  {journal} {\bibinfo  {journal} {arXiv preprint
  arXiv:2508.16342}\ } (\bibinfo {year} {2025}{\natexlab{b}})}\BibitemShut
  {NoStop}%
\bibitem [{\citenamefont {Helgaker}, \citenamefont {J{\o}rgensen},\ and\
  \citenamefont {Olsen}(2014)}]{helgaker2014molecular}%
  \BibitemOpen
  \bibfield  {author} {\bibinfo {author} {\bibfnamefont {T.}~\bibnamefont
  {Helgaker}}, \bibinfo {author} {\bibfnamefont {P.}~\bibnamefont
  {J{\o}rgensen}}, \ and\ \bibinfo {author} {\bibfnamefont {J.}~\bibnamefont
  {Olsen}},\ }\href@noop {} {\emph {\bibinfo {title} {Molecular
  electronic-structure theory}}}\ (\bibinfo  {publisher} {John Wiley \& Sons},\
  \bibinfo {year} {2014})\BibitemShut {NoStop}%
\end{thebibliography}%

\end{document}